\documentclass[zpreprint,zbstpl]{zeus_paper}

\usepackage[english]{babel}

\chardef\usc=95
\chardef\til=126
\catcode`\@=11 
\DeclareRobustCommand\xdotspace{\futurelet\@let@token\@xdotspace}
\def\@xdotspace{%
  \ifx\@let@token.\else
  \ifx\@let@token\bgroup.\else
  \ifx\@let@token\egroup.\else
  \ifx\@let@token\/.\else
  \ifx\@let@token\ .\else
  \ifx\@let@token~.\else
  \ifx\@let@token!.\else
  \ifx\@let@token,.\else
  \ifx\@let@token:.\else
  \ifx\@let@token;.\else
  \ifx\@let@token?.\else
  \ifx\@let@token/.\else
  \ifx\@let@token'.\else
  \ifx\@let@token).\else
  \ifx\@let@token-.\else
  \ifx\@let@token\@xobeysp.\else
  \ifx\@let@token\space.\else
  \ifx\@let@token\@sptoken.\else
   .\space
   \fi\fi\fi\fi\fi\fi\fi\fi\fi\fi\fi\fi\fi\fi\fi\fi\fi\fi}
\catcode`\@=12 

\newcommand{\stru}[2]{%
   \relax\ifmmode\hbox{\vrule height#1 depth#2 width0pt}%
   \else\vrule height#1 depth#2 width0pt\fi}

\newcommand{\Ronum}[1]{\uppercase\expandafter{\romannumeral#1}}
\newcommand{\ronum}[1]{\expandafter{\romannumeral#1}}
\DeclareRobustCommand{\LaTeXZ}{%
  \LaTeX\kern-.05em4\kern-.1em
  {\raisebox{-0.2ex}{$\scriptstyle\text{ZEUS}$}}\xspace}

\DeclareMathAlphabet{\mathbf}{OT1}{cmr}{bx}{sl}
\newcommand{\eVdist}{\kern-0.06667em}

\newcommand{\pb}{\,\text{pb}}
\newcommand{\fb}{\,\text{fb}}

\newcommand{\Tesla}{\,\text{T}}

\newcommand{\slashfrac}[2]{%
  \raisebox{0.5ex}{\ensuremath #1}\kern-0.12em/\kern-0.08em
  \raisebox{-.8ex}{\ensuremath #2}}

\newcommand{\sqr}[3]{%
    {\vcenter{\hrule height.#3ex\hbox{\vrule width.#2ex height#1ex
     \kern#1ex\vrule width.#3ex}\hrule height.#2ex}}}

\catcode`\@=11 
\newcommand{\parenbar}{\mathpalette\p@renb@r}
\def\p@renb@r#1#2{\vbox{%
  \ifx#1\scriptscriptstyle \dimen@.7em\dimen@ii.2em\else
  \ifx#1\scriptstyle \dimen@.8em\dimen@ii.25em\else
  \dimen@1em\dimen@ii.4em\fi\fi \offinterlineskip
  \ialign{\hfill##\hfill\cr
    \vbox{\hrule width\dimen@ii}\cr
    \noalign{\vskip-.3ex}%
    \hbox to\dimen@{$\mathchar300\hfil\mathchar301$}\cr
    \noalign{\vskip-.3ex}%
    $#1#2$\cr}}}
\catcode`\@=12 

\newcommand{\IP}{{\rm I$\kern-0.01667em$P}\xspace}

\mathchardef\qsm=63
\mathchardef\pls=43
\mathchardef\mns=512
\mathchardef\plm=518
\mathchardef\eql=61
\mathchardef\smallleft=300
\mathchardef\smallright=301
\mathchardef\les=316
\mathchardef\gre=318
\mathchardef\leq=532
\mathchardef\grq=533

\catcode`\@=11 
\newcounter{pict@width}
\newcounter{pict@height}
\newlength{\pict@scale}
\newcommand{\psfigadd}[4]{%
\setcounter{pict@width}{1*\ratio{#2+\pict@scale/2}{\pict@scale}}
\setcounter{pict@height}{1*\ratio{#3+\pict@scale/2}{\pict@scale}}
\setlength{\unitlength}{\pict@scale}
\hbox to #2{\hspace{-\fill}\begin{picture}(\thepict@width,\thepict@height)
\put(0,0){\psfig{figure=#1,width=#2,height=#3,clip=}}
\SetScale{0.283466457}
\SetWidth{1.763889}
{#4}
\end{picture}}
}
\newcounter{pict@widthfst}
\newcounter{pict@widthscd}
\newcounter{pict@widthtot}
\newcommand{\psfigaddtwo}[7]{%
\setcounter{pict@widthfst}{1*\ratio{#2+\pict@scale/2}{\pict@scale}}
\setcounter{pict@widthscd}{1*\ratio{#2+#4+\pict@scale/2}{\pict@scale}}
\setcounter{pict@widthtot}{1*\ratio{#2+#4+#6+\pict@scale/2}{\pict@scale}}
\setcounter{pict@height}{1*\ratio{#3+\pict@scale/2}{\pict@scale}}
\setlength{\unitlength}{\pict@scale}
\hbox{\hspace{-\fill}\begin{picture}(\thepict@widthtot,\thepict@height)
\put(0,0){\psfig{figure=#1,width=#2,height=#3,clip=}}
\put(\thepict@widthscd,0){\psfig{figure=#5,width=#6,height=#3,clip=}}
\SetScale{0.283466457}
\SetWidth{1.763889}
{#7}
\end{picture}}
}
\newcommand{\psfigror}[4]{%
\setcounter{pict@width}{1*\ratio{#2+\pict@scale/2}{\pict@scale}}
\setcounter{pict@height}{1*\ratio{#3+\pict@scale/2}{\pict@scale}}
\setlength{\unitlength}{\pict@scale}
\hbox{\begin{picture}(\thepict@width,\thepict@height)
\put(0,\thepict@height){\psfig{figure=#1,width=#3,height=#2,clip=,angle=270}}
\SetScale{0.283466457}
\SetWidth{1.763889}
{#4}
\end{picture}}
}
\newcommand{\psfigrol}[4]{%
\setcounter{pict@width}{1*\ratio{#2+\pict@scale/2}{\pict@scale}}
\setcounter{pict@height}{1*\ratio{#3+\pict@scale/2}{\pict@scale}}
\setlength{\unitlength}{\pict@scale}
\hbox{\begin{picture}(\thepict@width,\thepict@height)
\put(0,0){\psfig{figure=#1,width=#3,height=#2,clip=,angle=90}}
\SetScale{0.283466457}
\SetWidth{1.763889}
{#4}
\end{picture}}
}
\catcode`\@=12 
\newlength\listtextwidth

\catcode`\@=11 
\newlength{\@tabfninsert}
\newlength{\@tabfnwidth}
\newcommand{\tabfootnote}[2]{%
  \setlength{\@tabfninsert}{0.8em}
  \setlength{\@tabfnwidth}{\textwidth}
  \addtolength{\@tabfnwidth}{-\@tabfninsert}
  \addtolength{\@tabfnwidth}{-0.4em}
  \noindent\makebox[\@tabfninsert][r]{\footnotesize$^{#1}$\hfil}\hfill%
  \parbox[t]{\@tabfnwidth}{\footnotesize #2\hfill}}
\catcode`\@=12 

\def\JHEP{JHEP}
\def\q2{Q^2}
\def\as{\alpha_s}
\def\qq{q\bar q}
\def\mw{M_W}
\def\asz{\as(\mz)}
\def\mz{M_Z}
\def\etjet{E_T^{\rm jet}}
\def\etajet{\eta^{\rm jet}}
\def\ptmiss{p_T^{\rm miss}}
\def\g2{GeV$^2$}
\def\kt{k_T}
\def\phijet{\phi^{\rm jet}}
\def\etj{E_T^{\rm jet1}}
\def\etjj{E_T^{\rm jet2}}
\def\etjjj{E_T^{\rm jet3}}
\def\etaj{\eta^{\rm jet1}}
\def\etajj{\eta^{\rm jet2}}
\def\etajjj{\eta^{\rm jet3}}
\def\oas{{\cal O}(\as)}
\def\oass{{\cal O}(\as^2)}
\def\seta{d\sigma/d\etajet}
\def\set{d\sigma/d\etjet}
\def\etabar{\overline{\eta}^{\rm jet}}
\def\etbar{\overline{E}_T^{\rm jet}}
\def\mj{M^{\rm jj}}
\def\m3j{M^{\rm 3j}}
\def\colab#1{#1 Coll.}
\def\etaphi{\eta-\phi}
\def\pb1{pb$^{-1}$}
\def\fb1{fb$^{-1}$}

\def\figdir{./}

\begin{document}
\prepnum{{DESY--08--024}}

\title{
        Multi-jet cross sections in charged current
                {\boldmath $e^{\pm}p$} scattering at HERA
}                                                       

\author{ZEUS Collaboration}
\date{February 2008}

\abstract{
Jet cross sections were measured in charged current deep inelastic
$e^{\pm}p$ scattering at high boson virtualities $\q2$ with the ZEUS
detector at HERA II using an integrated luminosity of $0.36$
\fb1. Differential cross sections are presented for inclusive-jet
production as functions of $\q2$, Bjorken $x$ and the jet transverse
energy and pseudorapidity. The dijet invariant mass cross section is
also presented. Observation of three- and four-jet events in
charged-current $e^{\pm}p$ processes is reported for the first
time. The predictions of next-to-leading-order (NLO) QCD calculations
are compared to the measurements. The measured inclusive-jet cross
sections are well described in shape and normalization by the NLO
predictions. The data have the potential to constrain the $u$ and $d$
valence quark distributions in the proton if included as input to
global fits.
}

\makezeustitle

\def\3{\ss}
\pagenumbering{Roman}

\begin{center}
{                      \Large  The ZEUS Collaboration              }
\end{center}

  S.~Chekanov,
  M.~Derrick,
  S.~Magill,
  B.~Musgrave,
  D.~Nicholass$^{   1}$,
  \mbox{J.~Repond},
  R.~Yoshida\\
  {\it Argonne National Laboratory, Argonne, Illinois 60439-4815,
    USA}~$^{n}$
\par \filbreak

  M.C.K.~Mattingly \\
  {\it Andrews University, Berrien Springs, Michigan 49104-0380, 
    USA}
\par \filbreak

  M.~Jechow,
  N.~Pavel~$^{\dagger}$\\
  {\it Institut f\"ur Physik der Humboldt-Universit\"at zu Berlin,
    Berlin, Germany}~$^{b}$
% ...

\newpage

$^{\    1}$ also affiliated with University College London, UK \\
% ...
$^{\dagger}$ deceased \\

\newpage

\begin{tabular}[h]{rp{14cm}}

$^{a}$ &  supported by the Natural Sciences and Engineering Research
Council of Canada (NSERC) \\
$^{b}$ &  supported by the German Federal Ministry for Education and
Research (BMBF), under contract numbers 05 HZ6PDA, 05 HZ6GUA, 05
HZ6VFA and 05 HZ4KHA\\
$^{c}$ &  supported in part by the MINERVA Gesellschaft f\"ur
Forschung GmbH, the Israel Science Foundation (grant no. 293/02-11.2)
and the U.S.-Israel Binational Science Foundation \\
$^{d}$ &  supported by the German-Israeli Foundation and the Israel
Science Foundation\\
$^{e}$ &  supported by the Italian National Institute for Nuclear
Physics (INFN) \\
$^{f}$ &  supported by the Japanese Ministry of Education, Culture,
Sports, Science and Technology (MEXT) and its grants for Scientific
Research\\
$^{g}$ &  supported by the Korean Ministry of Education and Korea
Science and Engineering Foundation\\
$^{h}$ &  supported by the Netherlands Foundation for Research on
Matter (FOM)\\
$^{i}$ &  supported by the Polish State Committee for Scientific
Research, grant no. 620/E-77/SPB/DESY/P-03/DZ 117/2003-2005 and grant
no. 1P03B07427/2004-2006\\
$^{j}$ &  partially supported by the German Federal Ministry for
Education and Research (BMBF)\\
$^{k}$ &  supported by RF Presidential grant N 8122.2006.2 for the
leading scientific schools and by the Russian Ministry of Education
and Science through its grant for Scientific Research on High Energy
Physics\\
$^{l}$ &  supported by the Spanish Ministry of Education and Science
through funds provided by CICYT\\
$^{m}$ &  supported by the Science and Technology Facilities Council,
UK\\
$^{n}$ &  supported by the US Department of Energy\\
$^{o}$ &  supported by the US National Science Foundation. Any
opinion, findings and conclusions or recommendations expressed in this
material are those of the authors and do not necessarily reflect the
views of the National Science Foundation.\\
$^{p}$ &  supported by the Polish Ministry of Science and Higher
Education as a scientific project (2006-2008)\\
$^{q}$ &  supported by FNRS and its associated funds (IISN and FRIA)
and by an Inter-University Attraction Poles Programme subsidised by
the Belgian Federal Science Policy Office\\
$^{r}$ &  supported by the Malaysian Ministry of Science, Technology
and Innovation/Akademi Sains Malaysia grant SAGA 66-02-03-0048\\

\end{tabular}

\newpage

\pagenumbering{arabic}
\pagestyle{plain}

\section{Introduction}
\begin{figure}[p]
\vfill
\setlength{\unitlength}{1.0cm}
\begin{picture} (18.0,17.0)
\put (0.0,10.0){\centerline{\epsfig{figure=\figdir 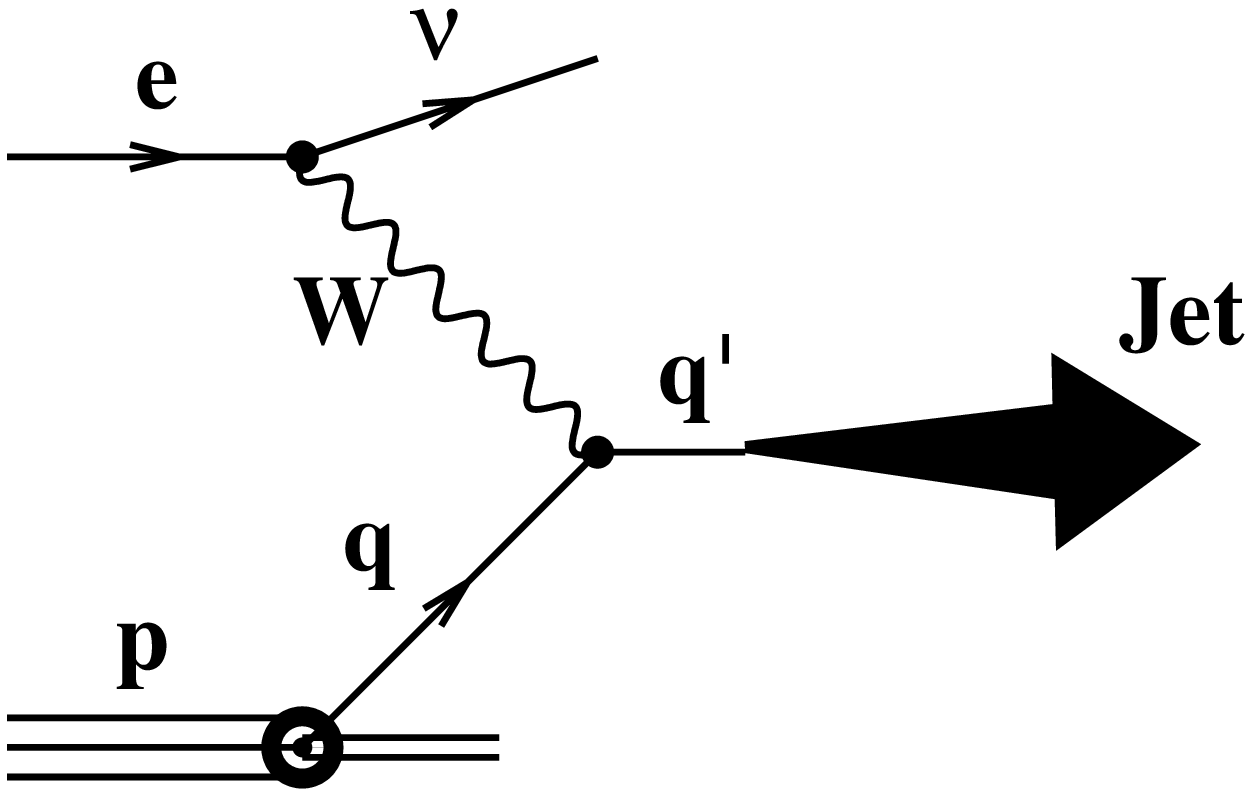,width=8cm}}}
\put (1.0,1.0){\epsfig{figure=\figdir 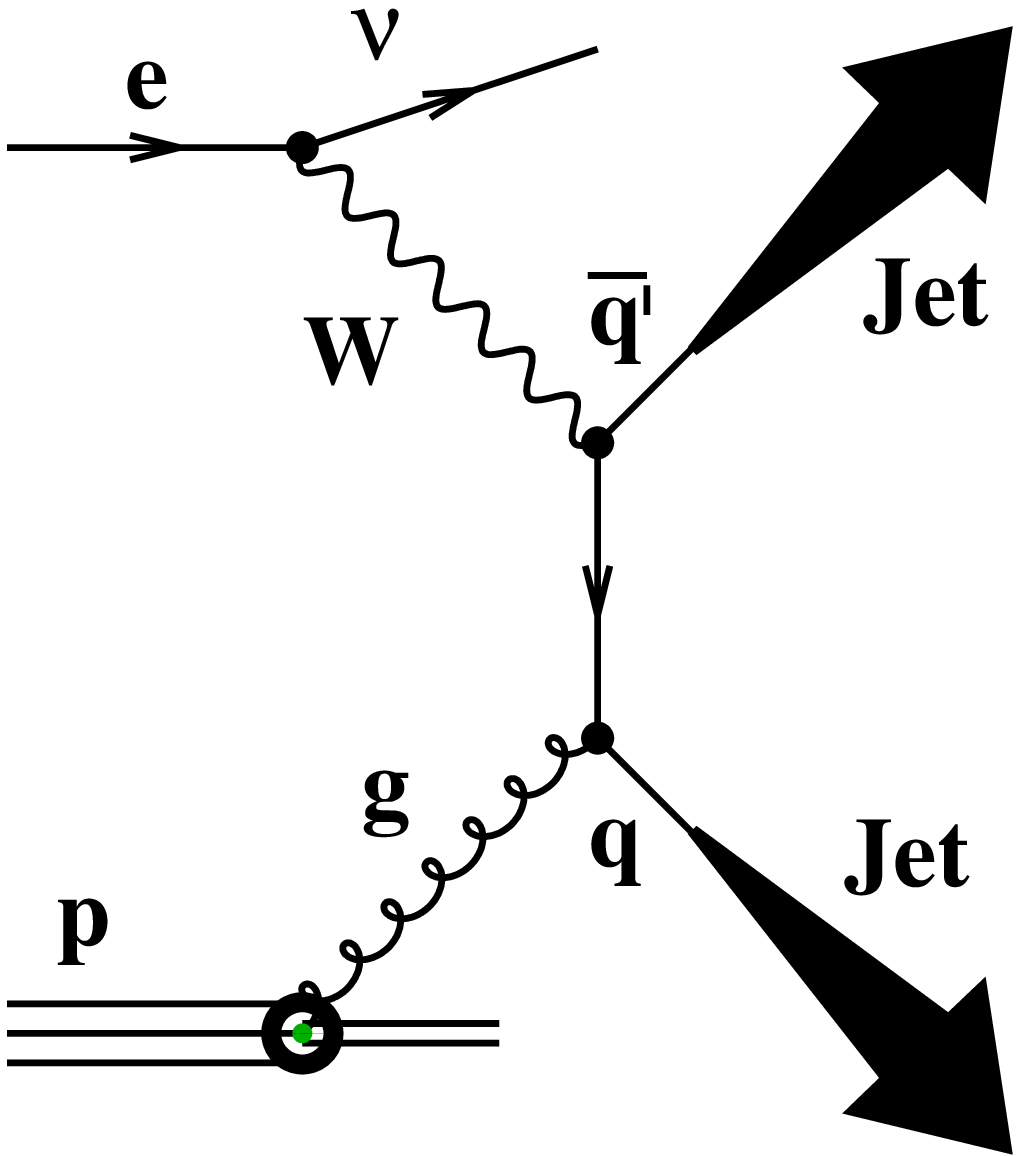,width=5cm}}
\put (8.0,1.0){\epsfig{figure=\figdir 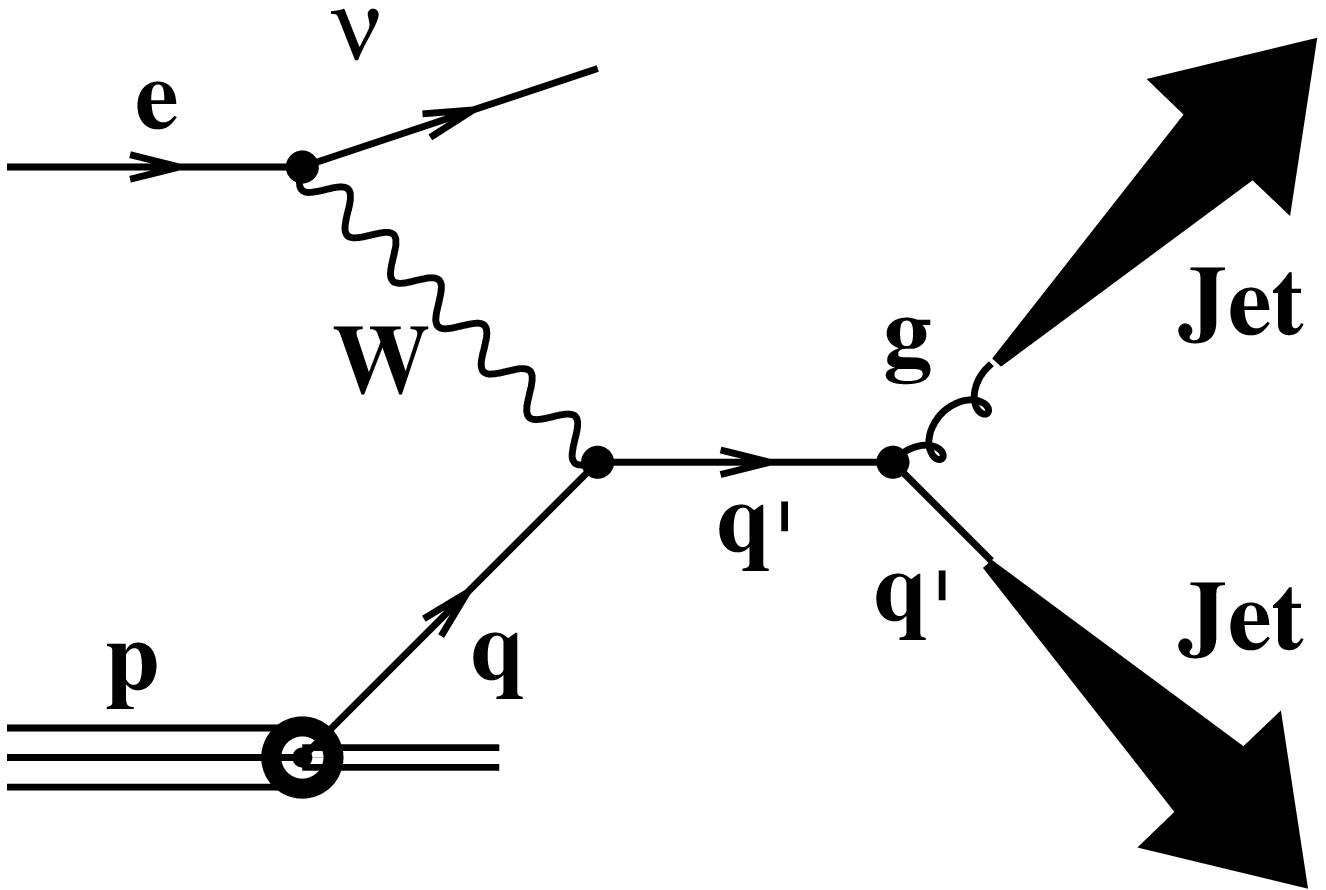,width=8cm}}
\put (6.3,9.4){\bf\small (a)}
\put (3.3,0.0){\bf\small (b)}
\put (12.3,0.0){\bf\small (c)}
\end{picture}
\caption
{\it 
Examples of jet production up to $\oas$ in CC DIS. Feynman diagrams
for: (a) quark-parton model, (b) boson-gluon fusion and (c)
QCD-Compton processes.
}
\label{fig1}
\vfill
\end{figure}

Jet production in charged-current (CC) deep inelastic $e^{\pm}p$
scattering (DIS) provides a testing ground for QCD and for the
electroweak sector of the Standard Model (SM). Up to leading order
(LO) in the strong coupling constant, $\as$, jet production in CC DIS
proceeds via the quark-parton model ($Wq\rightarrow q$,
Fig.~\ref{fig1}a), $W$-gluon fusion ($Wg\rightarrow \qq$,
Fig.~\ref{fig1}b) and the QCD-Compton ($Wq\rightarrow qg$,
Fig.~\ref{fig1}c) processes. Thus, differential cross sections for jet
production are sensitive to both the value of $\as$ and the mass of
the propagator, $\mw$, which are fundamental parameters of the
theory. Cross sections in CC DIS are also sensitive to the valence
flavor content of the proton, since the $W^{-(+)}$ couples to the 
up-like (down-like) quarks in the proton.

The large center-of-mass energy available at the HERA $e^{\pm}p$ collider
($\sqrt s=318$ GeV) extends the kinematic region for studying
CC DIS with respect to fixed-target neutrino scattering
experiments~\cite{zfp:c25:29,*zfp:c49:187,*zfp:c53:51,*zfp:c62:575} by
about two orders of magnitude in the virtuality of the exchanged
boson, $\q2$, and to lower values of the fraction of the proton
momentum carried by the struck parton, $x$. Measurements of the
CC DIS cross section at
HERA~\cite{pl:b324:241,*zfp:c67:565,*pl:b379:319,*prl:75:1006,zfp:c72:47,epj:c12:411}
demonstrated, at high $\q2$, the presence of a space-like propagator
with a finite mass, consistent with that of the $W$ boson. During
2002--2007, HERA provided longitudinally-polarized electron or
positron beams. Measurements of the fully-inclusive CC DIS cross
section for positive and negative values of the longitudinal
polarization of the beams were found to be in good agreement with
the predictions of the SM~\cite{pl:b634:173,pl:b637:210}.

At HERA, multijet structures were observed in CC
DIS~\cite{zfp:c72:47,epj:c19:429,epj:c31:149} at large $\q2$ and jet
shapes and subjet multiplicities were
measured~\cite{epj:c8:367,epj:c31:149} and compared with neutral 
current (NC) DIS processes~\cite{epj:c8:367,pl:b558:41}. The subjet
multiplicities were used to extract a value of
$\asz$~\cite{epj:c31:149}.

In this paper, measurements are presented of inclusive-jet and dijet
cross sections in CC $e^{\pm}p$ DIS in the laboratory
frame. Measurements of three-jet differential cross sections in CC DIS
were measured for the first time in $e^{\pm}p$ collisions. A small
sample of four-jet events was also observed in the data. The
measurements are presented as functions of $\q2$, $x$, the jet
transverse energy, $\etjet$, and pseudorapidity\footnote{The ZEUS
  coordinate system is a right-handed Cartesian system, with the $Z$
  axis pointing in the proton beam direction, referred to as the
  ``forward direction'', and the $X$ axis pointing left towards the
  center of HERA. The coordinate origin is at the nominal interaction
  point.}, $\etajet$, and as a function of the invariant mass of the
jet system in dijet and three-jet events. Predictions from
next-to-leading-order (NLO) QCD calculations are compared to the
measurements. Results for negative and positive
longitudinally-polarized electron and positron beams are also presented.

\section{Experimental set-up}
A detailed description of the ZEUS detector can be found
elsewhere~\cite{pl:b293:465,zeus:1993:bluebook}. A brief outline of
the components most relevant for this analysis is given below.

Charged particles were tracked in the central tracking detector
(CTD)~\cite{nim:a279:290,*npps:b32:181,*nim:a338:254}, which operated
in a magnetic field of $1.43\Tesla$ provided by a thin superconducting
solenoid. The CTD consisted of $72$~cylindrical drift-chamber
layers, organized in nine superlayers covering the
polar-angle region \mbox{$15^\circ<\theta<164^\circ$}.
In 2001, a silicon microvertex
detector (MVD)~\cite{nim:a581:656} was installed between
the beampipe and the inner radius of the CTD. The MVD was organized
into a barrel with three cylindrical layers and a forward section with
four planar layers perpendicular to the HERA beam direction. The
barrel contained 600 single-sided silicon strip sensors each having
512 strips of width 120 $\mu$m; the forward section contained 112
sensors each of which had 480 strips of width 120 $\mu$m. 
Charged-particle tracks were reconstructed online by using the ZEUS
global tracking trigger~\cite{nim:a580:1257}, which combined
information from the CTD and MVD. The online tracks were used to
reconstruct the interaction vertex and reject non-$ep$
background. Offline, the tracks used in this analysis were
reconstructed using information from the CTD and were used, in
addition to the vertex reconstruction, to cross-check the energy scale
of the calorimeter.

The high-resolution uranium--scintillator calorimeter
(CAL)~\cite{nim:a309:77,*nim:a309:101,*nim:a321:356,*nim:a336:23}
covered $99.7\%$ of the total solid angle and consisted of three parts:
the forward (FCAL), the barrel (BCAL) and the rear (RCAL) calorimeters. 
Each part was subdivided transversely into towers and longitudinally
into one electromagnetic section (EMC) and either one (in RCAL) or two
(in BCAL and FCAL) hadronic sections (HAC). The smallest subdivision
of the calorimeter is called a cell. Under test-beam conditions, the
CAL single-particle relative energy resolutions were
$\sigma(E)/E=0.18/\sqrt E$ for leptons and 
$\sigma(E)/E=0.35/\sqrt E$ for hadrons, with $E$ in GeV.

The luminosity was measured using the Bethe-Heitler reaction 
$ep\rightarrow e\gamma p$ by the luminosity detector which consisted of
two independent systems. In the first system, the photons were detected
by a lead-scintillator calorimeter placed in the HERA tunnel 107 m
from the interaction point in the lepton-beam direction. The system
used in previous ZEUS
publications~\cite{desy-92-066,*zfp:c63:391,*acpp:b32:2025} was
modified by the addition of active filters in order to suppress the
increased synchrotron radiation background of the upgraded HERA
collider. The second system was a magnetic spectrometer
arrangement~\cite{nim:a565:572}, which measured electron-positron pairs
from converted photons. The fractional uncertainty on the measured
luminosity was $3.5\%$.

The lepton beam in HERA became transversely polarized naturally
through the Sokolov-Ternov effect~\cite{sovpdo:8:1203}, with a 
build-up time of approximately 40 minutes. Spin rotators on either
side of the ZEUS detector changed the transverse polarization of the
beam into longitudinal polarization. The lepton-beam polarization was
measured using two independent polarimeters, the transverse
polarimeter (TPOL)~\cite{nim:a329:79,tech:pol2000prc} and the
longitudinal polarimeter (LPOL)~\cite{nim:a479:334,tech:pol2000prc}.
Both devices exploited the spin-dependent cross section for
Compton scattering of circularly polarized photons off leptons to
measure the beam polarization. The fractional uncertainty on the
measured polarization was $4.2\%$ and $3.6\%$ from TPOL and LPOL,
respectively.

\section{Data selection and jet search}
\label{selec}
The data were collected from 2004 to 2007, when HERA operated with
protons of energy $E_p=920$~GeV and electrons (positrons) of energy
$E_e=27.5$~GeV, and correspond to an integrated luminosity of
$180.0\pm 6.3$ ($178.5\pm 6.2$)~\pb1. Samples of negatively-
(positively-) polarized electron beams with an integrated luminosity
of $106.4$ ($73.6$) \pb1\ and luminosity-weighted average polarization
of $P_{e^-}^{\rm neg}=-0.27\pm 0.01$ ($P_{e^-}^{\rm pos}=+0.29\pm 0.01$)
were analyzed. For positrons, the samples analyzed were of $76.5$ and
$102.1$ \pb1\ with a luminosity-weighted average polarization of
$P_{e^+}^{\rm neg}=-0.37^{+0.01}_{-0.02}$ and 
$P_{e^+}^{\rm pos}=+0.32\pm 0.01$, respectively.

The main signature of a CC DIS event at HERA is the presence of large
$\ptmiss$ and large $E_T^{\rm tot}$, where $\ptmiss$ is the missing
transverse momentum arising from the energetic final-state neutrino
which escapes detection and $E_T^{\rm tot}$ is the total transverse
energy arising from the hard interaction. 
The variable $\ptmiss$ was reconstructed using the vectorial sum 
$\ptmiss=\sqrt{(\sum_ip_{X,i})^2+(\sum_ip_{Y,i})^2}$ and 
$E_T^{\rm tot}$ was reconstructed as the scalar sum
$E_T^{\rm tot}=\sum_iE_{T,i}$. In both cases, the sum runs over all
CAL cells. The online selection of the signal was based on the ZEUS
three-level trigger~\cite{zeus:1993:bluebook}. Two different trigger
selections were used. One trigger selection relied on large $\ptmiss$
and large $E_T^{\rm tot}$. The alternative trigger selection
additionally required the presence of at least one jet with transverse
energy above $8$ GeV. Events from CC DIS interactions were selected
offline using criteria similar to those of an earlier
publication~\cite{epj:c31:149}. The kinematic variables $\q2$, the
inelasticity, $y$, and $x$ were estimated using the method of
Jacquet-Blondel~\cite{proc:epfacility:1979:391}, which uses the
information from the hadronic energy flow of the event, and corrected
for detector effects as described
elsewhere~\cite{epj:c8:367,homer}. These estimators were reconstructed
as:

$$y_{\rm JB}=\frac{\sum_i(E_i-p_{Z,i})}{2\ E_e},\hspace{1cm} Q^2_{\rm
  JB}=\frac{(\ptmiss)^2}{1-y_{\rm JB}} \hspace{1cm} {\rm and} \hspace{1cm} x_{\rm JB}=\frac{Q^2_{\rm
    JB}}{s\ y_{\rm JB}},$$

where the sum runs over all CAL cells. The main selection criteria
were:

\begin{itemize}
 \item $\ptmiss>11$~GeV;
 \item $\ptmiss/E_T^{\rm tot} > 0.5$, to reject photoproduction and
   beam-gas background;
 \item the vertex position along the beam axis in the range
   $-35<Z_{\rm vtx}<33$~cm, consistent with an $e^{\pm}p$ interaction; 
 \item at least one track associated with the vertex, which had a
   polar angle between $15^{\circ}$ and $164^{\circ}$ and a 
   transverse momentum exceeding 0.15 GeV, to reject 
   non-$e^{\pm}p$ background;
 \item $|\Delta\varphi|<1$~rad, where $|\Delta\varphi|$ is the
   difference between the azimuth of the net transverse momentum as
   measured by the tracks associated with the vertex and the azimuth
   measured by the CAL. This requirement reduced the contamination
   from random coincidences of cosmic rays with $e^{\pm}p$
   interactions;
 \item $P_{T,\rm tracks}/\ptmiss>0.1$, where $P_{T,\rm tracks}$ is the net 
   transverse momentum of the tracks associated with the vertex (this
   condition was not applied if $\ptmiss>25$~GeV). This cut rejected
   events with additional energy deposits in the CAL not related to
   $e^{\pm}p$ interactions (mainly cosmic rays) and beam-related
   background in which $\ptmiss$ has a small polar angle;
 \item events were removed from the sample if there was an isolated
   electron or positron candidate with energy above $10$~GeV, to
   reject NC DIS events;
 \item events were rejected if $E_{\rm BHAC2}/E_{\rm BCAL}>0.5$ for
   $E_{\rm BCAL}>2$ GeV and $E_{\rm BHAC1}/E_{\rm BCAL}>0.85$ for
   $E_{\rm BCAL}>8$ GeV, where $E_{\rm BHAC1(BHAC2)}$ is the energy
   deposited in the first (second) HAC section of BCAL and 
   $E_{\rm BCAL}$ is the total energy deposition in BCAL. These
   requirements rejected beam-related background;
 \item tracking requirements were not applied if the highest-$\etjet$
   jet (see below) in the event had $\etajet>2$; in such a case, a
   tighter $\ptmiss$ cut of $20$~GeV was applied;
 \item $\q2>200$~\g2, to ensure high trigger efficiency;
 \item $y<0.9$, to avoid the degradation of the resolution in $\q2$
   near $y\sim 1$.
\end{itemize}

The selected events were visually inspected and a few remaining
non-$e^{\pm}p$ background events ($1.2\%$ of the final sample), mainly
cosmic-ray and halo-muon events, were removed.

Jets were identified in the pseudorapidity ($\eta$) - azimuth ($\phi$)
plane of the laboratory frame using the $\kt$ cluster
algorithm~\cite{np:b406:187} in the longitudinally invariant inclusive
mode~\cite{pr:d48:3160}. This algorithm combines objects with a small
relative distance $d_{ij}$,

  $$   d_{ij} = {\rm min}(E_{T,i},E_{T,j})^2 \cdot ((\eta_i-\eta_j)^2 +
              (\phi_i-\phi_j)^2), $$
where $E_{T,i}$, $\eta_i$ and $\phi_i$ are the transverse energy,
pseudorapidity and azimuth of object $i$. The axis of the jet was
defined according to the Snowmass
convention~\cite{proc:snowmass:1990:134}, where $\etajet$ ($\phijet$) 
is the transverse-energy weighted mean pseudorapidity (azimuth) of all
the objects belonging to that jet. The jets were reconstructed
using the CAL and corrected~\cite{pl:b547:164,homer} for detector
effects to yield jets of hadrons. Events with at least one jet in the
range $-1<\etajet<2.5$ were retained. The inclusive-jet sample
contained $N^{\rm neg}=5335$ ($870$) and $N^{\rm pos}=2122$ ($2284$)
jets with $\etjet>14$ GeV in the $e^-p$ ($e^+p$) data, where 
$N^{\rm neg(pos)}$ is the number of jets selected in events with
negatively-(positively-)polarized lepton beams. The dijet sample was
selected requiring the jets with the highest and second-highest
$\etjet$ to have $\etj>14$ and $\etjj>5$ GeV, respectively. The
three-jet sample was selected from the dijet sample by requiring 
the third-highest $\etjet$ jet to have $\etjjj>5$ GeV. The $e^-p$
($e^+p$) dijet and three-jet samples contained $1117$ ($464$) and
$109$ ($30$) events, respectively. Eleven events contained a fourth
jet with $\etjet>5$ GeV.

\section{Monte Carlo simulation}
\label{mc}

Samples of Monte Carlo (MC) events were generated to determine the
response of the detector and to evaluate the
correction factors necessary to obtain the hadron-level jet cross
sections. The hadron level is defined by those hadrons with lifetime
$\tau\geq 10$~ps. The generated events were passed through the {\sc
  Geant}~3.21-based~\cite{tech:cern-dd-ee-84-1} ZEUS detector- and
trigger-simulation programs~\cite{zeus:1993:bluebook}. They were
reconstructed and analyzed by the same program chain as the data.
The CC DIS events were generated using the {\sc Lepto}~6.5
program~\cite{cpc:101:108} interfaced to {\sc
  Heracles}~4.6.1~\cite{cpc:69:155,*spi:www:heracles} via {\sc
  Djangoh}~1.3~\cite{cpc:81:381,*spi:www:djangoh11}. The {\sc
  Heracles} program includes first-order electroweak radiative
corrections, vertex and propagator terms, and
two-boson exchange. The CTEQ5D~\cite{epj:c12:375} proton parton
distribution functions (PDFs) were used. The QCD radiation was
modeled with the color-dipole model
(CDM)~\cite{pl:b165:147,*pl:b175:453,*np:b306:746,*zfp:c43:625} by 
using the {\sc Ariadne}~4.08 program~\cite{cpc:71:15,*zfp:c65:285}
including the boson-gluon-fusion process. To study the systematic
effect of the modeling of parton showers in the correction of the
data, samples of events were generated using the {\sc Lepto} model
which is based on first-order QCD matrix elements and parton showers
(MEPS). For the generation of the {\sc Lepto}-MEPS
samples\footnote{The program {\sc Lepto} allows the generation of only
  QPM events when using the MEPS option for $e^-p$ collisions; thus,
  only positron samples were generated and used as an estimate of the
  systematic effect also for $e^-p$ collisions.},
the option for soft-color interactions was switched off. In both
cases, fragmentation into hadrons was performed using the Lund string
model~\cite{prep:97:31} as implemented in {\sc
  Jetset}~7.4~\cite{cpc:82:74,*cpc:135:238,cpc:39:347,*cpc:43:367}.

The photoproduction background was estimated using resolved and direct
samples generated using the program {\sc Herwig}
5.9~\cite{cpc:67:465,*jhep:0101:010}. After all the selection cuts
described in the previous section, the contribution from
photoproduction events to the inclusive-jet sample was estimated to be
smaller than $0.5\%$ overall and amounted to $\sim 2\%$ in the lowest
$\etjet$ bin. The NC DIS contamination was estimated to be smaller
than the photoproduction background. No background subtraction was
performed.

\begin{figure}[p]
\vfill
\setlength{\unitlength}{1.0cm}
\begin{picture} (18.0,17.0)
\put (0.0,0.0){\epsfig{figure=\figdir 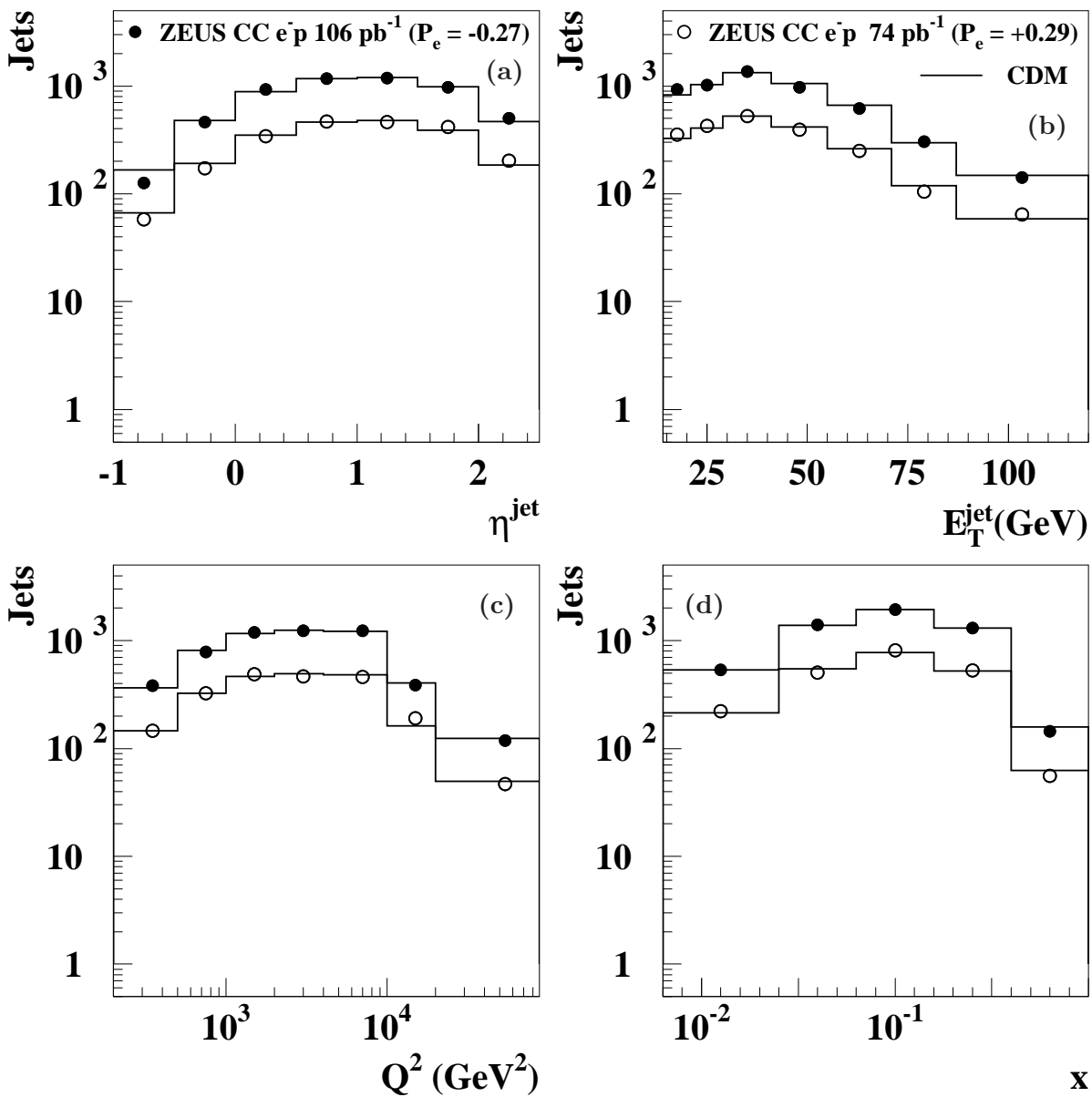,width=18cm}}
\put (7.4,15.2){\bf\small (a)}
\put (15.3,14.4){\bf\small (b)}
\put (7.3,7.4){\bf\small (c)}
\put (10.3,7.4){\bf\small (d)}
\end{picture}
\caption
{\it 
Detector-level data distributions for inclusive-jet production
with negative (dots) and positive (open circles) longitudinally
polarized electron beams with $\etjet>14$ GeV and $-1<\etajet<2.5$ in
the kinematic region given by $\q2>200$ \g2\ and $y<0.9$ as functions
of (a) $\etajet$, (b) $\etjet$, (c) $\q2$ and (d) Bjorken $x$. For
comparison, the distributions of the CDM Monte Carlo model normalized
to the number of jets in the data (solid histograms) are included.
}
\label{fig2}
\vfill
\end{figure}

\begin{figure}[p]
\vfill
\setlength{\unitlength}{1.0cm}
\begin{picture} (18.0,17.0)
\put (0.0,0.0){\epsfig{figure=\figdir 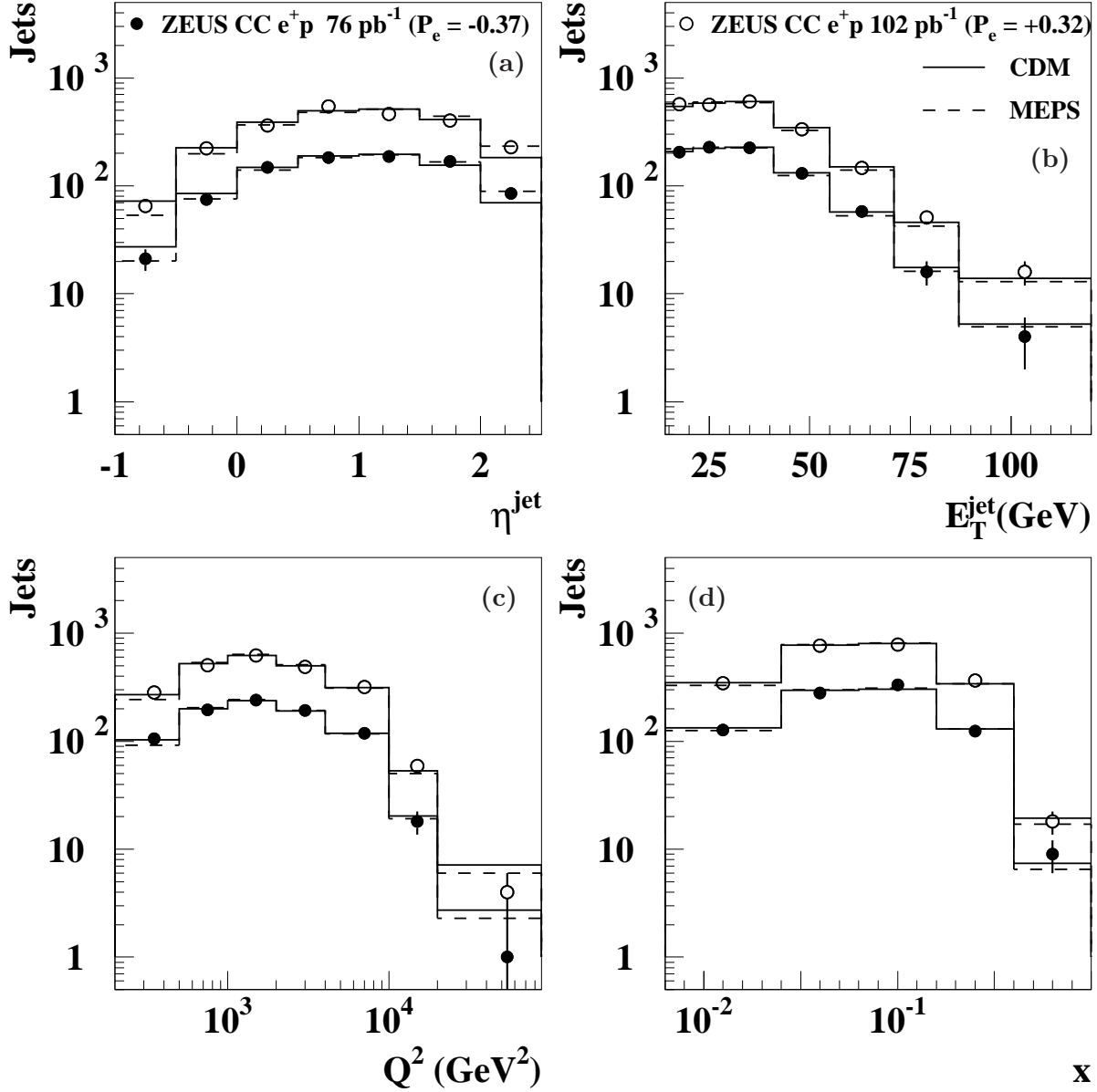,width=18cm}}
\put (7.4,15.2){\bf\small (a)}
\put (15.3,13.8){\bf\small (b)}
\put (7.3,7.4){\bf\small (c)}
\put (10.3,7.4){\bf\small (d)}
\end{picture}
\caption
{\it 
Detector-level data distributions for inclusive-jet production
with negative (dots) and positive (open circles) longitudinally
polarized positron beams with $\etjet>14$ GeV and $-1<\etajet<2.5$ in
the kinematic region given by $\q2>200$ \g2\ and $y<0.9$ as functions
of (a) $\etajet$, (b) $\etjet$, (c) $\q2$ and (d) Bjorken $x$. The
distributions of the {\sc Lepto}-MEPS Monte Carlo model are also
included (dashed histograms). Other details as in the caption to
Fig.~\ref{fig2}.
}
\label{fig3}
\vfill
\end{figure}

\begin{figure}[p]
\vfill
\setlength{\unitlength}{1.0cm}
\begin{picture} (18.0,17.0)
\put (0.0,0.0){\epsfig{figure=\figdir 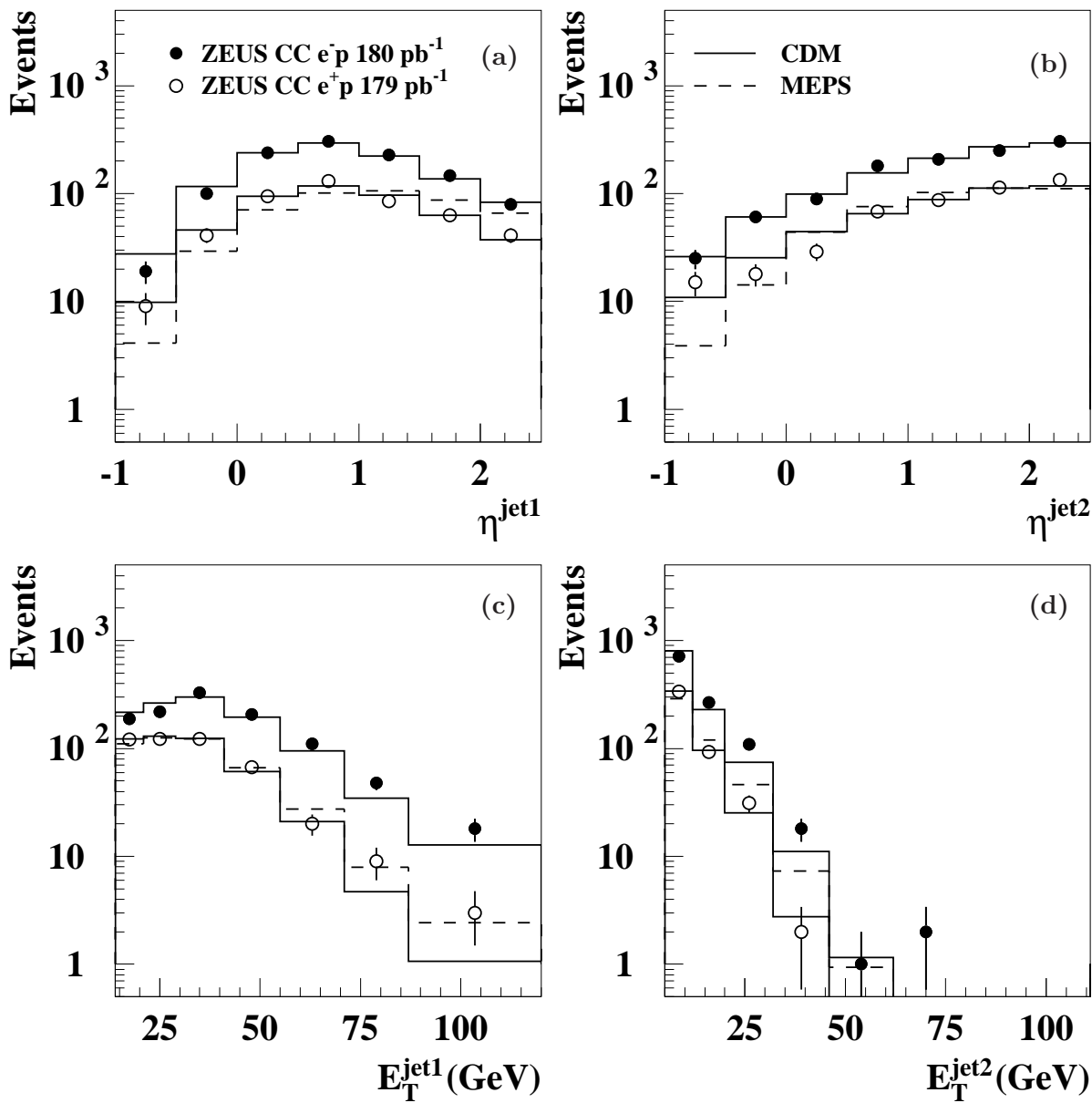,width=18cm}}
\put (7.3,15.4){\bf\small (a)}
\put (15.3,15.3){\bf\small (b)}
\put (7.3,7.4){\bf\small (c)}
\put (15.3,7.4){\bf\small (d)}
\end{picture}
\caption
{\it 
Detector-level data distributions for dijet production with electron
(dots) and positron (open circles) beams with $\etj>14$ GeV, $\etjj>5$
GeV and $-1<\etajet<2.5$ in the kinematic region given by $\q2>200$
\g2\ and $y<0.9$ as functions of (a) $\etaj$, (b) $\etajj$, (c) $\etj$
and (d) $\etjj$. Other details as in the caption to Fig.~\ref{fig2}.
}
\label{fig4}
\vfill
\end{figure}

The jet search was performed on the MC events using the energy
measured in the CAL cells in the same way as for the data. The
MC samples provided a good description of the measured
distributions of the kinematic and jet variables. The data
distributions as functions of $\etajet$, $\etjet$, $\q2$ and $x$ for
the inclusive-jet samples are shown in Figs.~\ref{fig2} ($e^-p$) and
\ref{fig3} ($e^+p$) separately for negatively- and
positively-polarized electron and positron beams. Figure~\ref{fig4}
shows the $\etaj$, $\etajj$, $\etj$ and $\etjj$ distributions for the
dijet sample. These figures also show the MC simulations normalized to
the number of jets in the data.

The same jet algorithm was also applied to the final-state hadrons to
obtain the predictions at the hadron level. To correct the data to the
hadron level, multiplicative correction factors, defined as ratios
of the measured quantities for jets (events) of hadrons over the same
quantity for jets (events) at detector level, were estimated by using
the CDM samples and applied to the inclusive-jet (dijet and three-jet)
data distributions. The samples of {\sc Lepto}-MEPS were used as an
estimation of the uncertainty on the modeling of the parton shower.

Parton-level predictions were also obtained by applying the jet
algorithm to the MC-generated partons. These predictions were used to
correct the fixed-order QCD calculations (see Section~\ref{nlo}) to
the hadron level. The program {\sc Heracles} was used to correct
the predicted cross sections to the electroweak Born level evaluated
using the fine structure constant $\alpha=1/137.035999$, the
Fermi coupling constant $G_F=1.1664\cdot 10^{-5}$~GeV$^{-2}$ and the
mass of the $Z$ boson $M_Z=91.1876$ GeV to determine the electroweak
parameters.

\section{Fixed-order QCD calculations}
\label{nlo}
Fixed-order QCD calculations were obtained using the program
{\sc Mepjet}~\cite{pl:b380:205}, which employs the phase-space slicing
method~\cite{pr:d46:1980}. This is the only available program
providing fixed-order QCD calculations for jet production in CC
DIS. The jet algorithm described in Section~\ref{selec} was
also applied to the partons in the events generated by {\sc Mepjet} to
compute the predictions for the jet cross sections. The calculations
were performed in the $\overline{\rm MS}$ scheme. The calculations are
$\oas$ ($\oass$) for inclusive-jet (dijet and three-jet) production;
this means that for inclusive-jet and dijet cross sections, the
predictions are NLO whereas those for three-jet cross sections are
only LO. The number of flavors was set to five and the renormalization
$(\mu_R)$ and factorization $(\mu_F)$ scales were chosen to be
$\mu_R=\mu_F=Q$. The calculations were performed using the
ZEUS-S~\cite{pr:d67:012007} parameterizations of the proton
PDFs. Alternative calculations were performed using the
CTEQ6~\cite{jhep:0207:012,*jhep:0310:046} and
MRST2001~\cite{epj:c28:455} sets of proton PDFs. The cross sections 
were evaluated using the same values for $\alpha$, $G_F$ and $\mz$ as
in the MC simulations (Section~\ref{mc}). The mass of the $W$ boson was
fixed to $80.40$ GeV~\cite{jp:g33:1}. The strong coupling constant was
calculated at two loops with $\Lambda^{(5)}_{\overline{\rm
    MS}}=226$~MeV, corresponding to $\asz=0.118$.

Since the measurements correspond to jets of hadrons whereas the QCD
calculations correspond to jet of partons, the predictions were
corrected to the hadron level using MC simulations. The
multiplicative correction factor ($C_{\rm had}$) is defined as the
ratio of the cross sections for jets of hadrons to the same quantity
for jets of partons, estimated using the MC programs described in
Section~\ref{mc}. The ratios obtained with the CDM model were taken as
the default corrections, whereas those from the {\sc Lepto}-MEPS model
were used as an estimation of the effect of the parton shower.
The value of $C_{\rm had}$ typically differs from unity by less than
$5\%$, $10\%$ and $30\%$ for the inclusive-jet, dijet and three-jet
cross sections, respectively.

The following sources of uncertainty in the theoretical predictions were
considered:
\begin{itemize}
  \item the uncertainty on the NLO QCD calculations due to terms
    beyond NLO, estimated by varying $\mu_R$ between $Q/2$ and
    $2Q$, was typically below $\pm 2\%$ for the inclusive-jet cross
    sections and below $\pm 5\%$ for the dijet cross sections. For the
    LO calculations of the three-jet cross sections the uncertainty
    was $\approx \pm 30\%$. For the three-jet cross sections, this
    uncertainty is dominant. Thus, no other theoretical uncertainty
    was taken into account for the three-jet cross sections;
  \item the uncertainty on the NLO QCD calculations due to those on the 
    proton PDFs was estimated by repeating the calculations using 22
    additional sets from the ZEUS-S analysis, which takes into
    account the statistical and correlated systematic experimental
    uncertainties of each data set used in the determination of the
    proton PDFs. The resulting uncertainty in the inclusive-jet $e^-p$
    ($e^+p$) cross sections was below $\pm 2\ (4)\%$, except in the
    high-$\etjet$, high-$\q2$ and high-$x$ regions where it reached
    $\pm 4\ (10)\%$. The resulting uncertainty in the dijet $e^-p$
    ($e^+p$) cross sections was below $\pm 5\ (5)\%$, except in the
    high-$\etjet$, high-$\q2$ and high-$x$ regions where it reached
    $\pm 7\ (15)\%$;
  \item the uncertainty on the NLO QCD calculations due to that on
    $\asz$ was estimated by repeating the calculations using two
    additional sets of proton PDFs, for which different values of
    $\asz$ were assumed in the fits. The difference between
    the calculations using these various sets was scaled by a factor
    such as to reflect the uncertainty on the current world average of
    $\as$~\cite{ppnp:58:351}. The resulting uncertainty in the
    cross sections was below $\pm 1\%$;
  \item the uncertainty from the modeling of the QCD cascade was
    estimated as the difference between the hadronization corrections
    obtained using the {\sc Ariadne} and {\sc Lepto-MEPS} models. The
    resulting uncertainty on the inclusive-jet and dijet cross
    sections was typically below $1\%$;
  \item the uncertainty of the calculations due to the value of
    $\mu_F$ was estimated by repeating the calculations with
    $\mu_F=Q/2$ and $2Q$. The effect was negligible.
\end{itemize}
 
The total theoretical uncertainty was obtained by adding the
individual uncertainties listed above in quadrature.

\section{Systematic uncertainties}
\label{expunc}
The following sources of systematic uncertainty were considered for
the measurements of the jet cross sections~\cite{homer}; 
values in percentage of the effects for the integrated inclusive-jet,
dijet and three-jet cross sections are shown in parentheses:
\begin{itemize}
  \item the uncertainty in the absolute energy scale of the jets was
    estimated by studying the differences between data and MC
    simulation in single-jet NC DIS
    events~\cite{proc:calor:2002:767,homer} by comparing the transverse
    energy imbalance between the scattered electron or positron and the
    jet. The uncertainty was found to be $\pm 1\%$ for $\etjet<75$ GeV
    and $\pm 2\%$ for $\etjet>75$ GeV ($0.5\%$, $2\%$, $4\%$);
  \item the uncertainty in the reconstruction of the kinematic
    variables due to the uncertainty in the absolute energy scale of
    the CAL was estimated by varying $\ptmiss$ and $y_{\rm JB}$ as
    measured with the CAL by $\pm 1\%$ for $\ptmiss<75$ GeV and
    $_{-1}^{+3}\%$ for $\ptmiss>75$ GeV in the MC samples ($0.2\%$,
    $0.3\%$, $0.4\%$);
  \item the differences in the results obtained by using either CDM or
    {\sc Lepto}-MEPS to correct the data for detector effects were
    taken as systematic uncertainties due to the modeling of the
    parton shower ($0.7\%$, $7\%$, $6\%$);
  \item the selection cut of $\ptmiss>11$ GeV was changed to $10$ GeV
    and $12$ GeV in data and MC events ($0.2\%$, $0.1\%$, below
    $0.1\%$);
  \item the uncertainty in the simulation of the vertex position was
    estimated by changing the selection cut to $-24<Z_{\rm vtx}<22$~cm
    in data and MC events ($2\%$, $2\%$, $2\%$);
  \item the uncertainty in the simulation of the trigger was estimated
    to be negligible by using two different trigger selections in data
    and MC events (see Section~\ref{selec}).
\end{itemize}

The experimental uncertainties are dominated by the statistical
uncertainty of the data, except for the inclusive-jet differential
cross sections at high $\etjet$ and high $\q2$, where the uncertainty
coming from the modeling of the parton shower is large. The
systematic uncertainties not associated with the absolute energy 
scale were added in quadrature to the statistical uncertainties and
are shown in the figures as error bars. The uncertainties due to the
absolute energy scale of the jets and the CAL were added linearly, due
to the large bin-to-bin correlation, and are shown separately in the
tables. In addition, there was an overall normalization uncertainty of
$3.5\%$ from the luminosity determination and $3-5\%$ uncertainty on
the polarization measurement.

\section{Results}
\label{results}

\subsection{Polarized inclusive-jet cross sections}

Differential inclusive-jet cross sections were measured in the
kinematic regime $\q2>200$ \g2\ and $y<0.9$. The cross sections were
determined for jets with $\etjet>14$ GeV and $-1<\etajet<2.5$. 

\begin{figure}[p]
\vfill
\setlength{\unitlength}{1.0cm}
\begin{picture} (18.0,17.0)
\put (0.0,11.0){\centerline{\epsfig{figure=\figdir 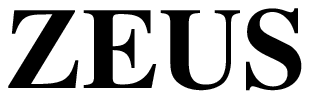,width=10cm}}}
\put (-1.0,9.5){\epsfig{figure=\figdir 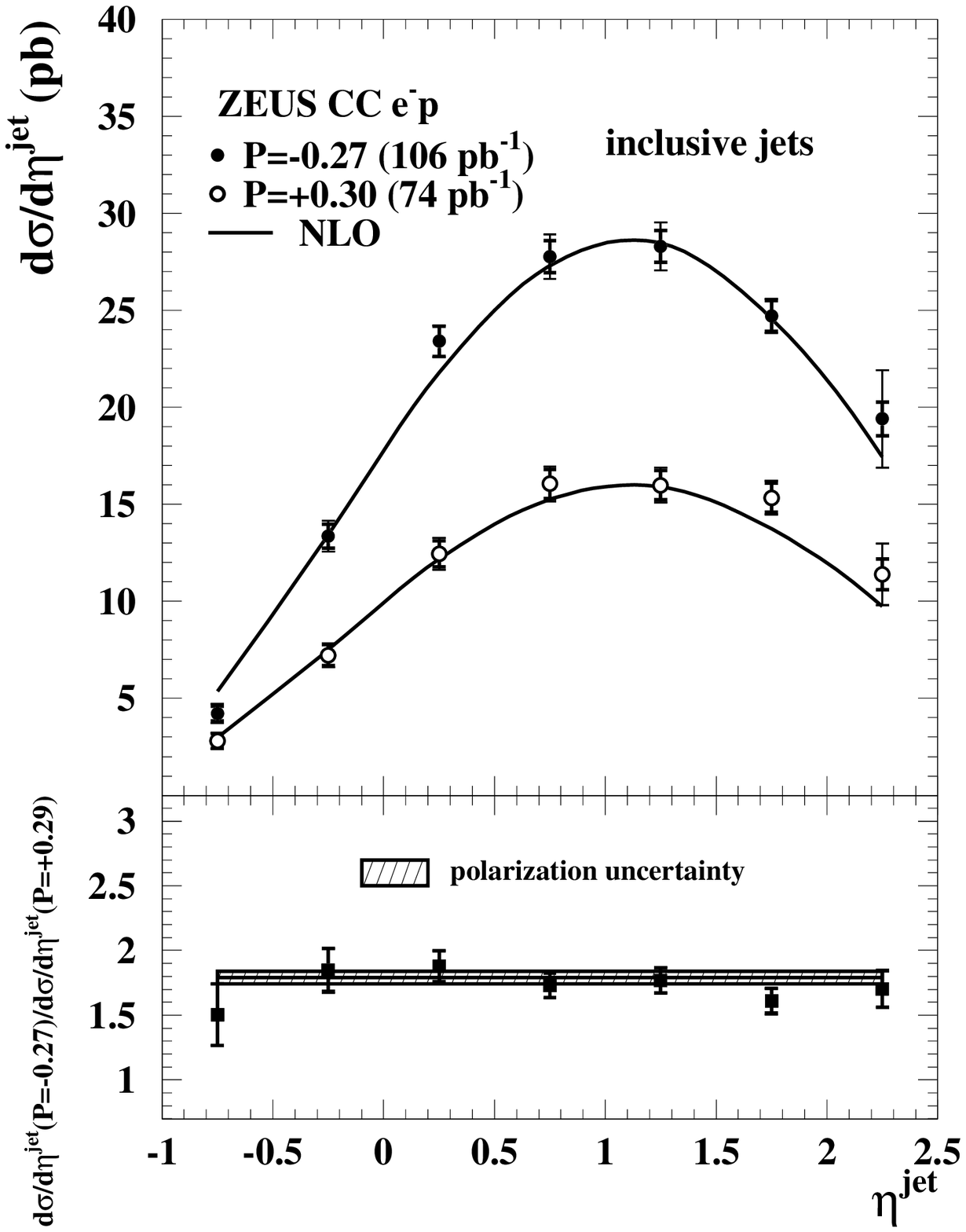,width=10cm}}
\put (8.0,9.5){\epsfig{figure=\figdir 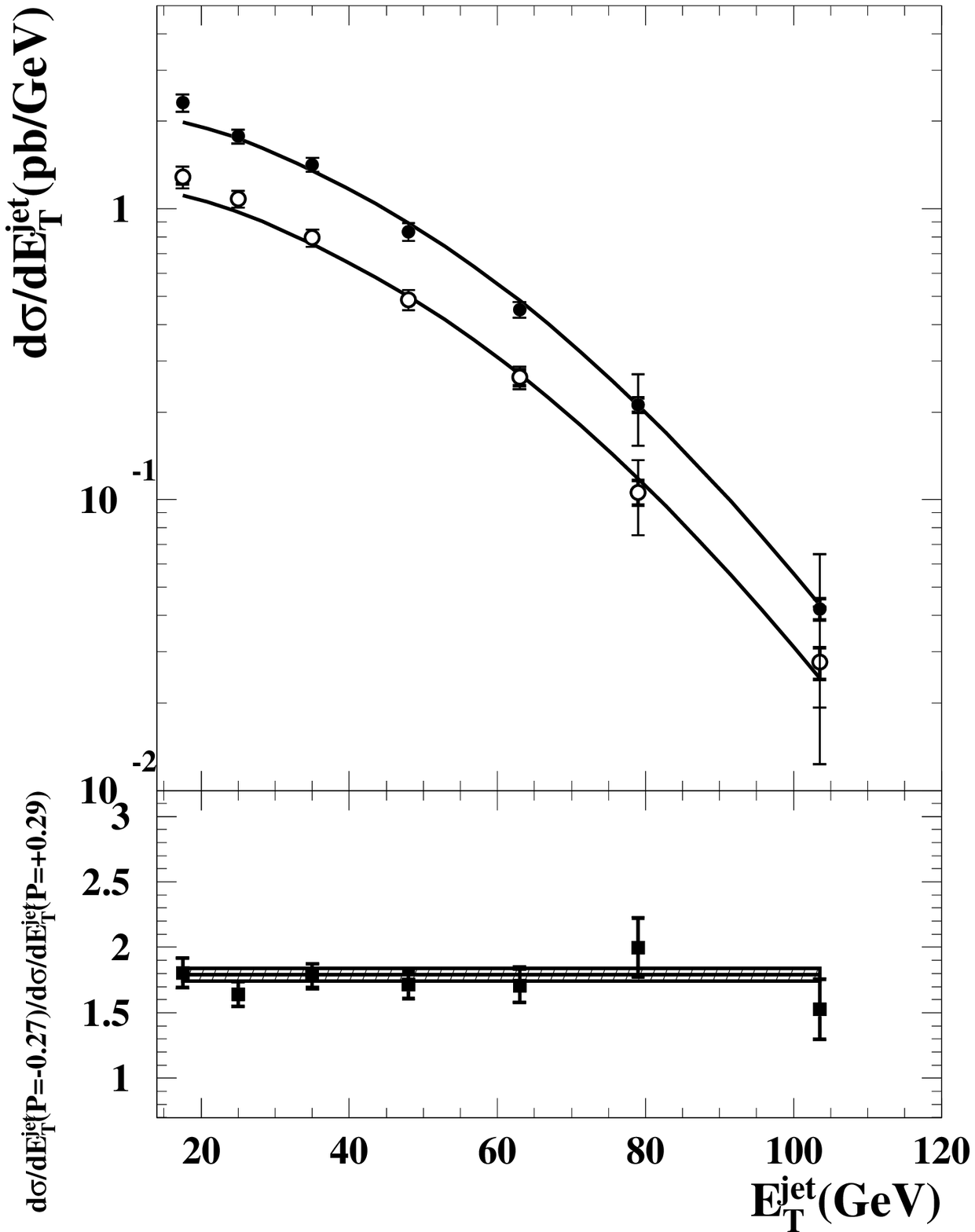,width=10cm}}
\put (-1.0,-0.5){\epsfig{figure=\figdir 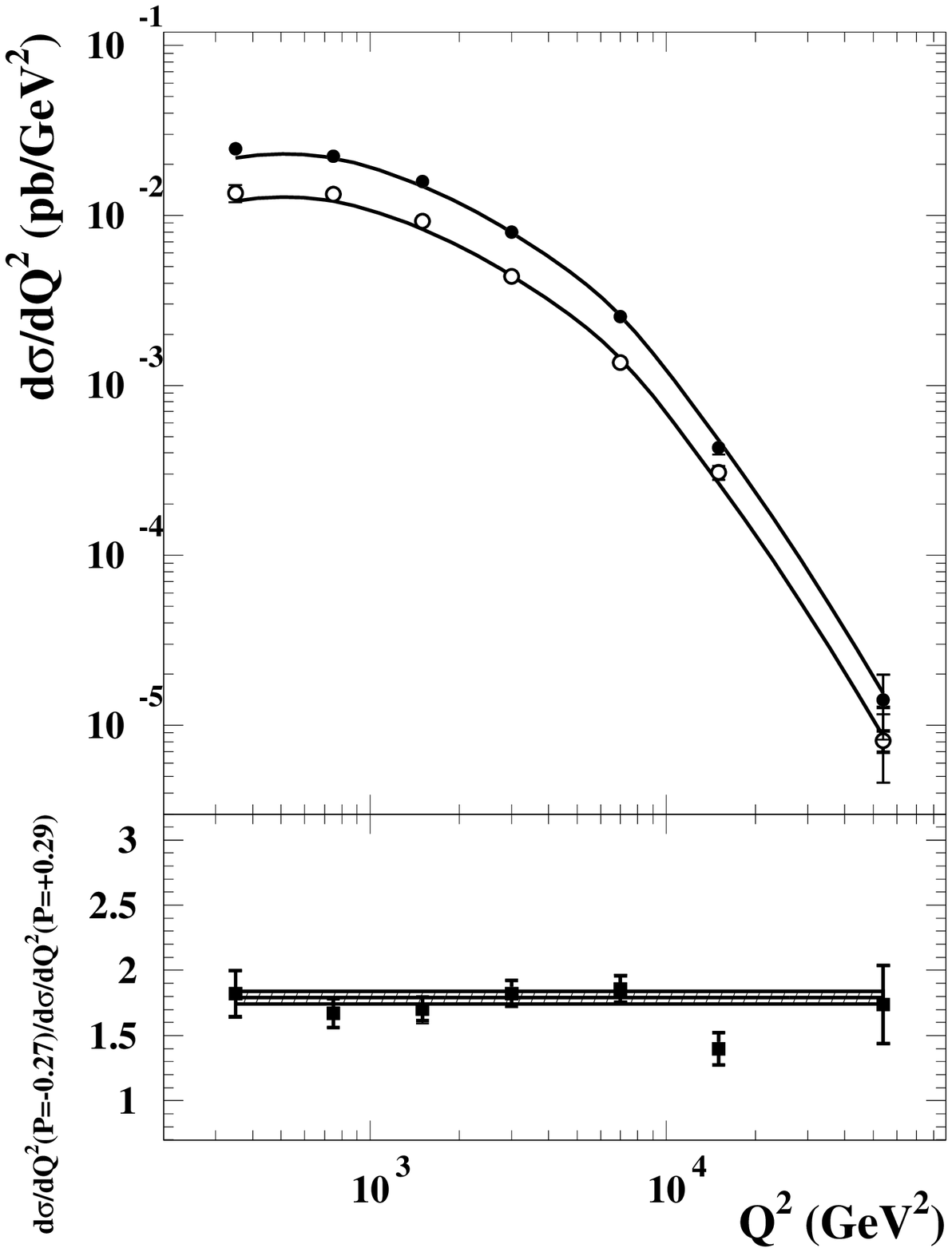,width=10cm}}
\put (8.0,-0.5){\epsfig{figure=\figdir 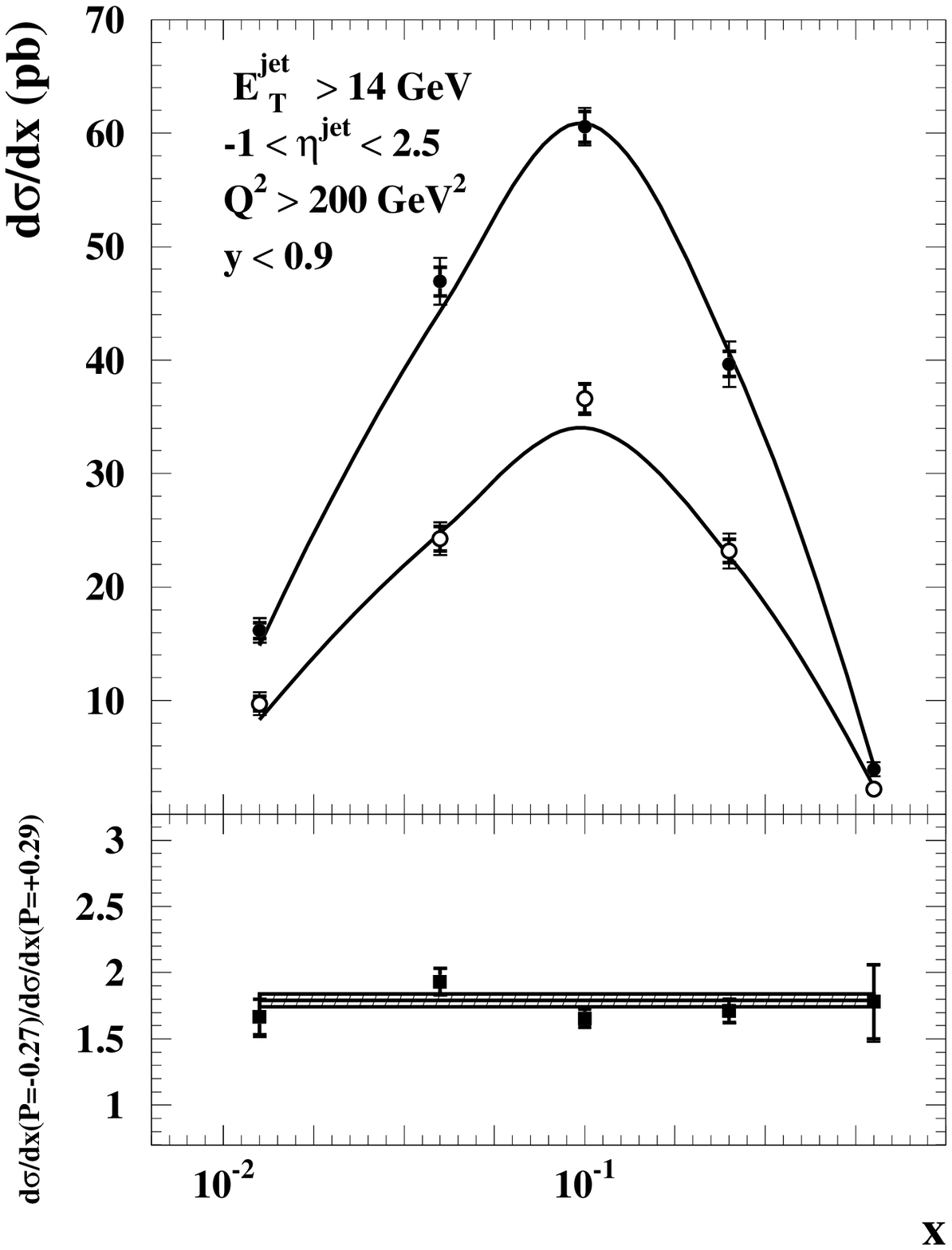,width=10cm}}
\put (6.3,18.4){\bf\small (a)}
\put (15.3,18.4){\bf\small (b)}
\put (6.3,8.4){\bf\small (c)}
\put (15.3,8.4){\bf\small (d)}
\end{picture}
\caption
{\it 
Measured inclusive-jet cross sections in CC $e^-p$ DIS for jets with 
$\etjet>14$ GeV and $-1<\etajet<2.5$ in the kinematic regime given by
$\q2>200$~\g2\ and $y<0.9$ as functions of (a) $\etajet$, (b)
$\etjet$, (c) $\q2$ and (d) Bjorken $x$ for negative (dots) and
positive (open circles) longitudinally polarized electron beams. The
data points are plotted at the bin centres. The inner error
bars represent the statistical uncertainties of the data, and the
outer error bars show the statistical and uncorrelated systematic 
uncertainties added in
quadrature. For comparison, the predictions of NLO QCD (solid lines)
are included. The lower parts of the figures display the ratio of the
cross sections for negatively- and positively-polarized electron beams 
(squares) and the prediction (solid line) for the measured polarizations.
The hatched band displays the uncertainty due to the polarization
measurement.
}
\label{fig5}
\vfill
\end{figure}

\begin{figure}[p]
\vfill
\setlength{\unitlength}{1.0cm}
\begin{picture} (18.0,17.0)
\put (0.0,11.0){\centerline{\epsfig{figure=\figdir zeus.eps,width=10cm}}}
\put (-1.0,9.5){\epsfig{figure=\figdir 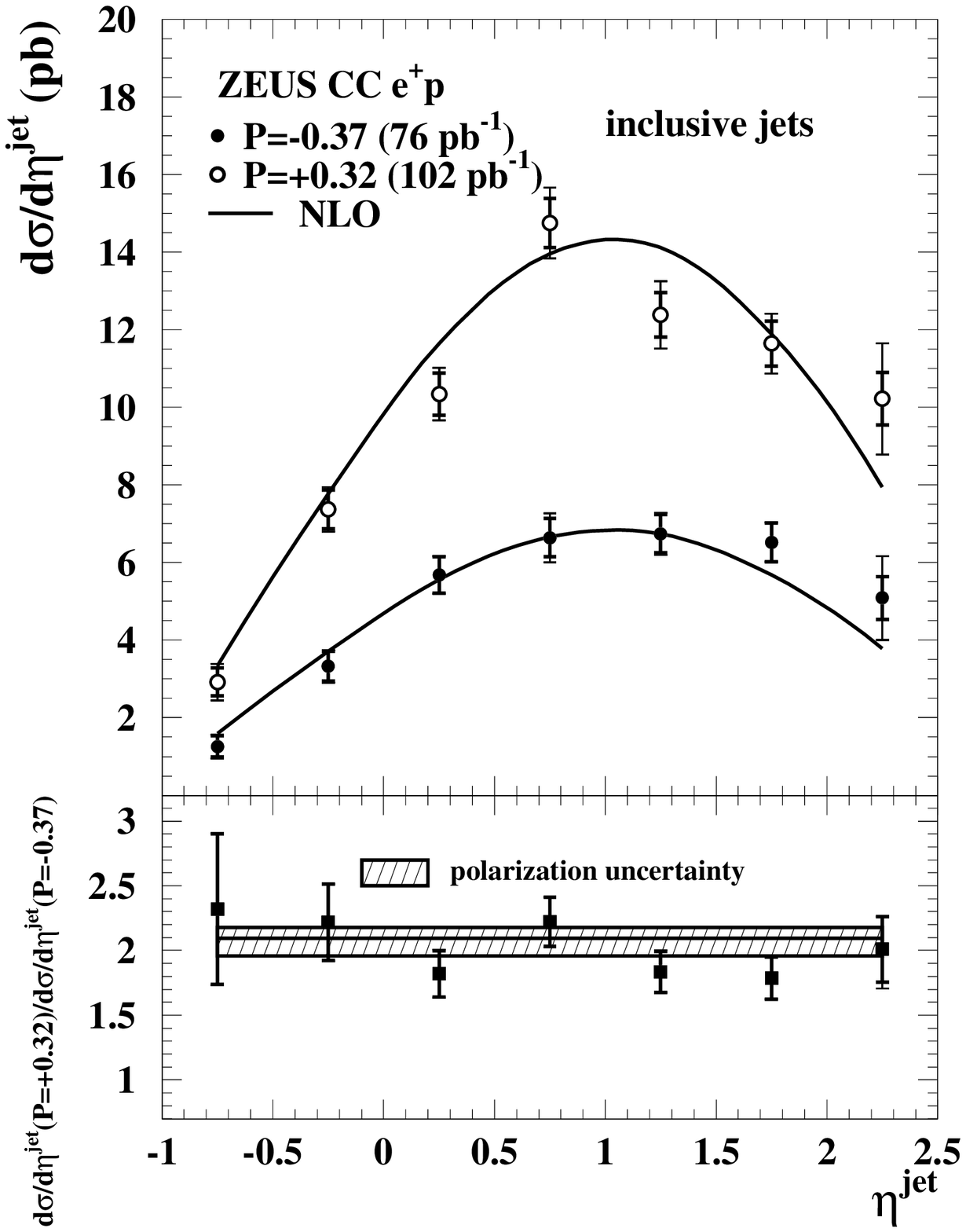,width=10cm}}
\put (8.0,9.5){\epsfig{figure=\figdir 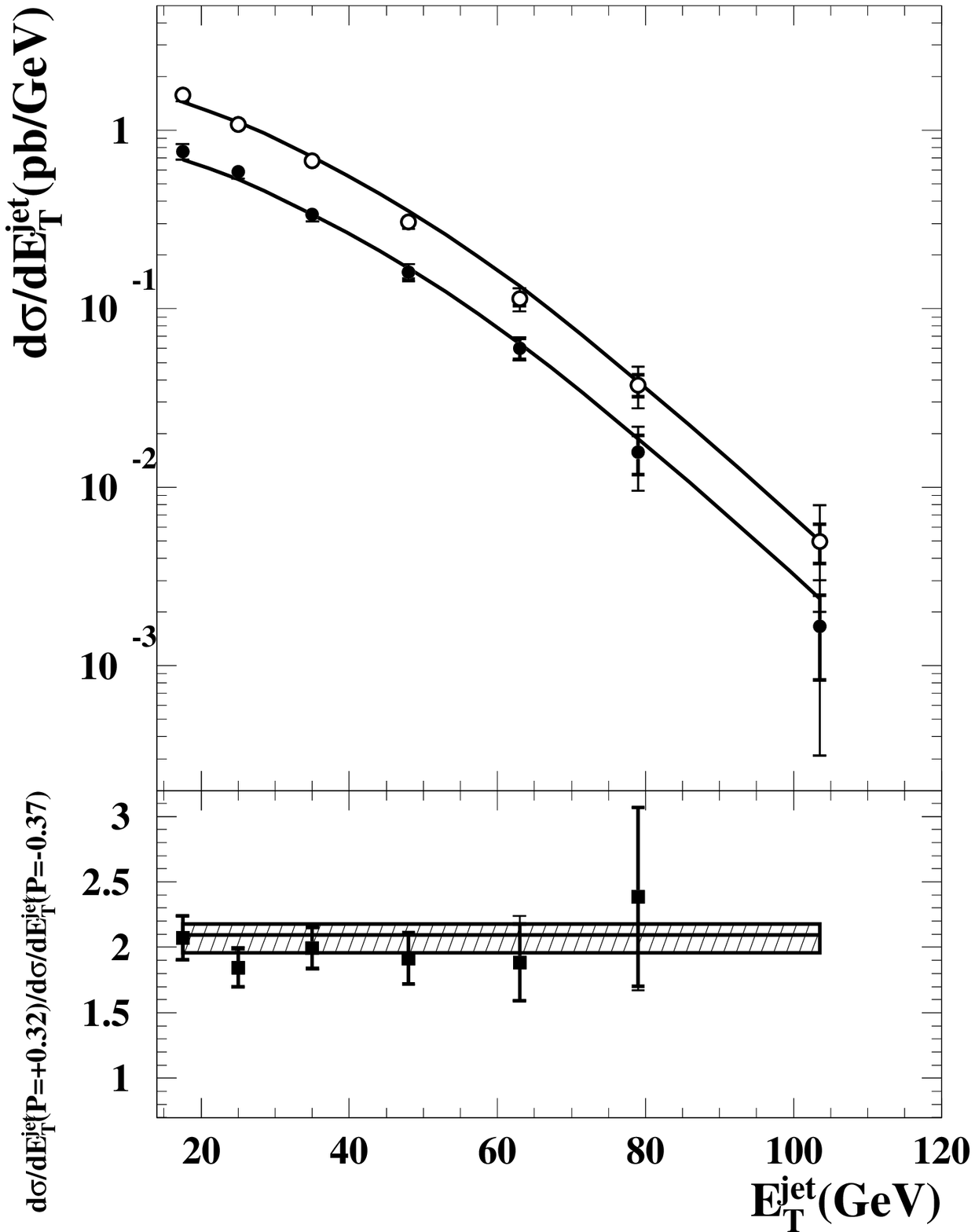,width=10cm}}
\put (-1.0,-0.5){\epsfig{figure=\figdir 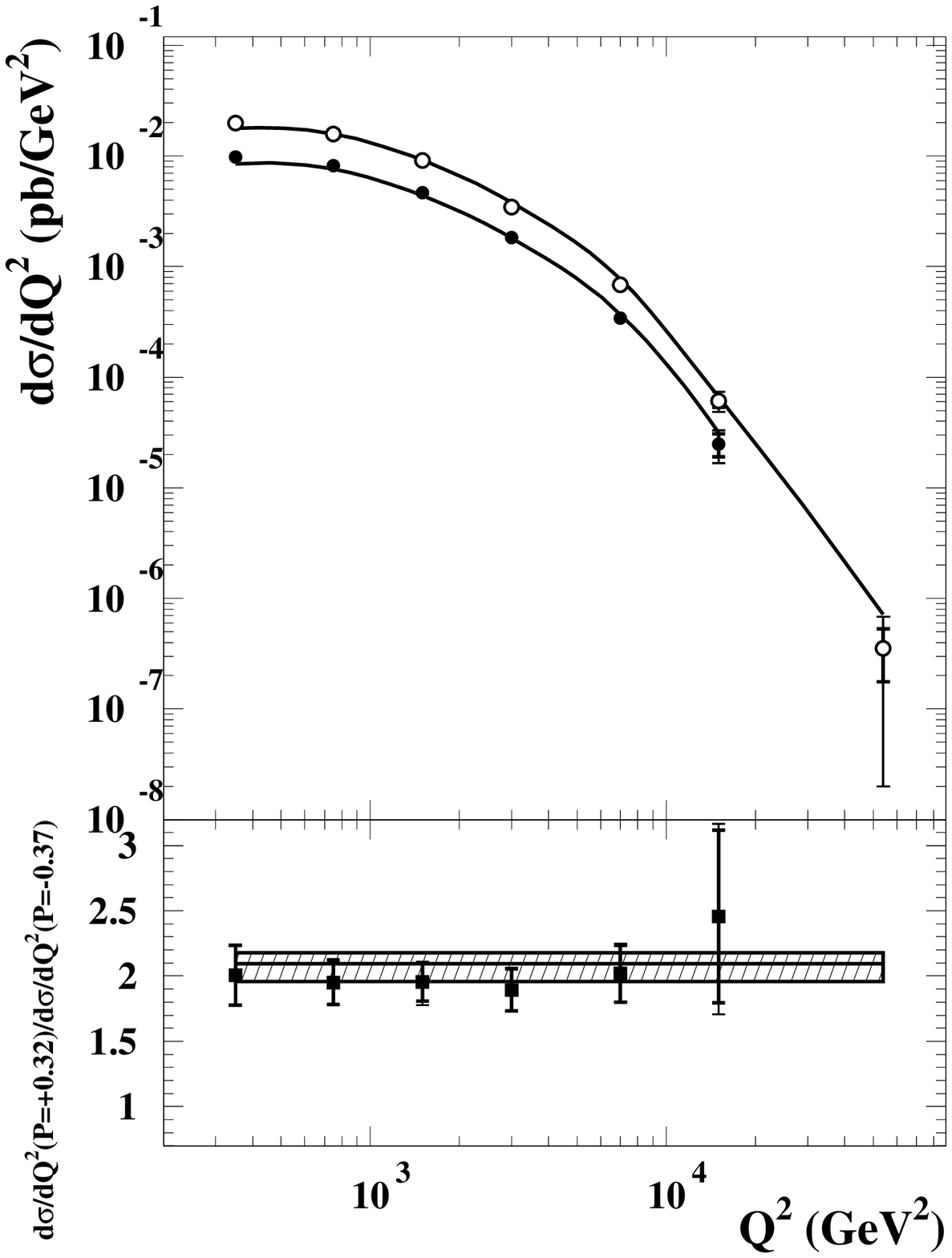,width=10cm}}
\put (8.0,-0.5){\epsfig{figure=\figdir 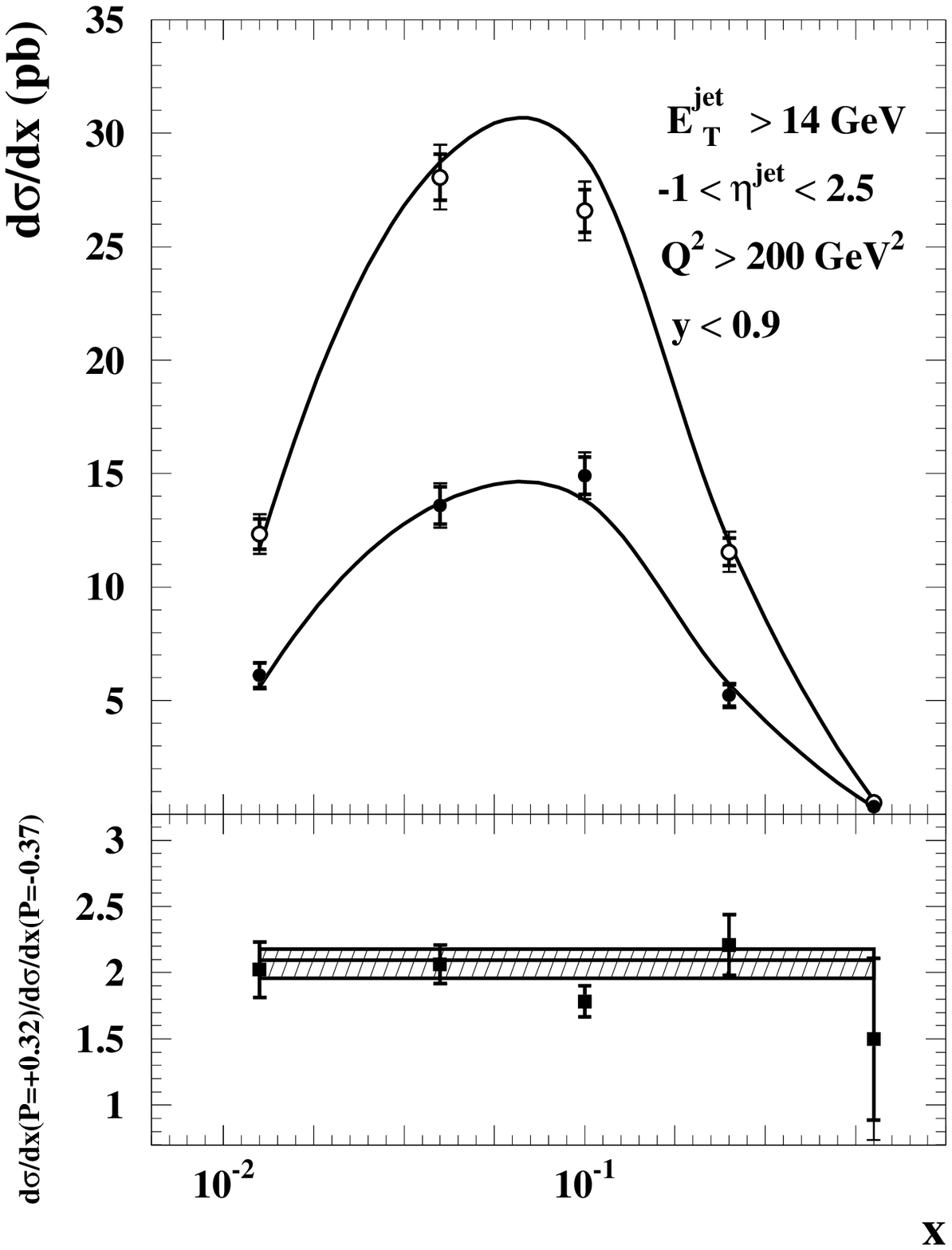,width=10cm}}
\put (6.3,18.4){\bf\small (a)}
\put (10.3,18.4){\bf\small (b)}
\put (6.3,8.4){\bf\small (c)}
\put (10.3,8.4){\bf\small (d)}
\end{picture}
\caption
{\it 
Measured inclusive-jet cross sections in CC $e^+p$ DIS for jets with
$\etjet>14$ GeV and $-1<\etajet<2.5$ in the kinematic regime given by
$\q2>200$~\g2\ and $y<0.9$ as functions of (a) $\etajet$, (b)
$\etjet$, (c) $\q2$ and (d) Bjorken $x$ for negative (dots) and
positive (open circles) longitudinally polarized electron beams.
Other details as in the caption to Fig.~\ref{fig5}.
}
\label{fig6}
\vfill
\end{figure}

The inclusive-jet differential cross sections as functions of $\etajet$,
$\etjet$, $\q2$ and $x$ for negatively- and
positively-polarized $e^-p$ ($e^+p$) collisions are shown in
Fig.~\ref{fig5} (\ref{fig6}) and Tables~\ref{tabthree} to
\ref{tabsix}. The predictions of the NLO calculations are compared to
the data in the figures. The lower parts of the figures show the ratio
of the cross sections for negatively- and positively-polarized lepton
beams, which is in agreement with the measured polarization ratio, 
\mbox{$(1-P_{e^-}^{\rm neg})/(1-P_{e^-}^{\rm pos})=1.79\pm 0.05$} for
$e^-p$ and \mbox{$(1+P_{e^+}^{\rm pos})/(1+P_{e^+}^{\rm
    neg})=2.10^{+0.08}_{-0.14}$} for $e^+p$. 
The integrated polarized inclusive-jet cross sections, $\sigma_{\rm
  jets}$, are shown in Table~\ref{tabone}. The measured
cross sections are in good agreement with the predictions of the SM as
given by the NLO QCD calculations, also shown in Table~\ref{tabone},
in the kinematic range studied.

\subsection{Unpolarized inclusive-jet cross sections}

\begin{figure}[p]
\vfill
\setlength{\unitlength}{1.0cm}
\begin{picture} (18.0,17.0)
\put (0.0,11.0){\centerline{\epsfig{figure=\figdir zeus.eps,width=10cm}}}
\put (-2.0,8.5){\epsfig{figure=\figdir 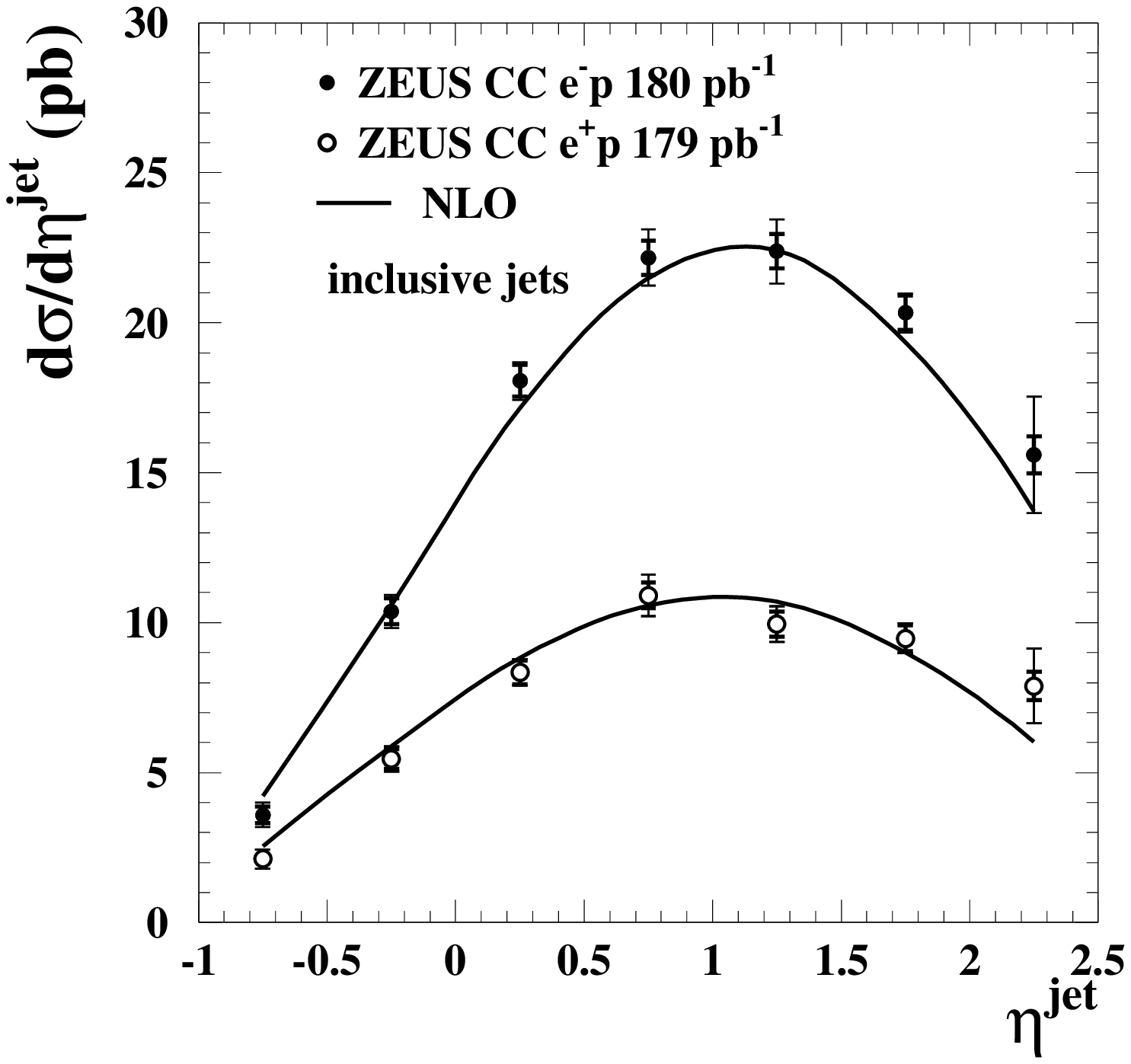,width=12cm}}
\put (7.0,8.5){\epsfig{figure=\figdir 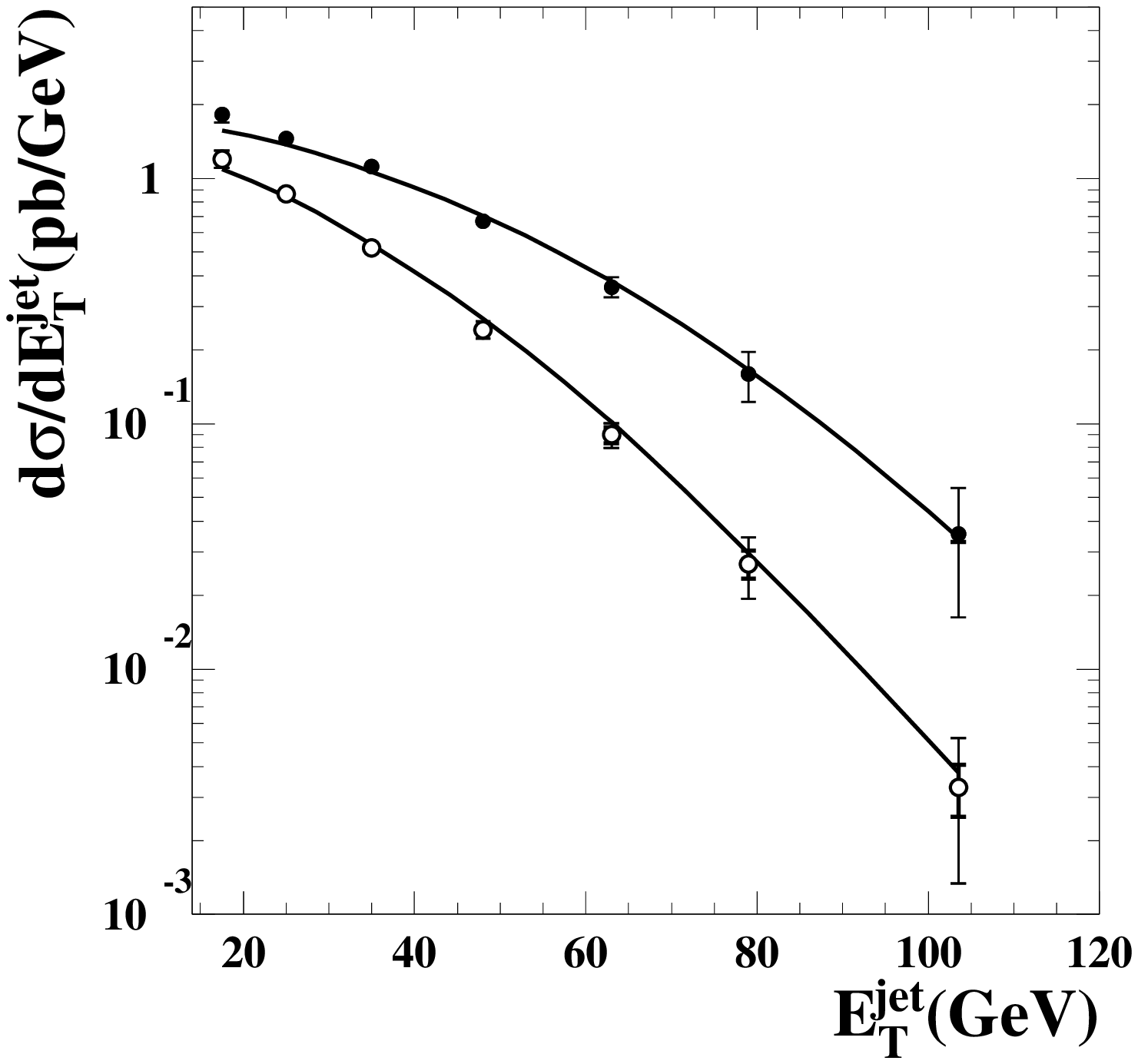,width=12cm}}
\put (-2.0,-1.5){\epsfig{figure=\figdir 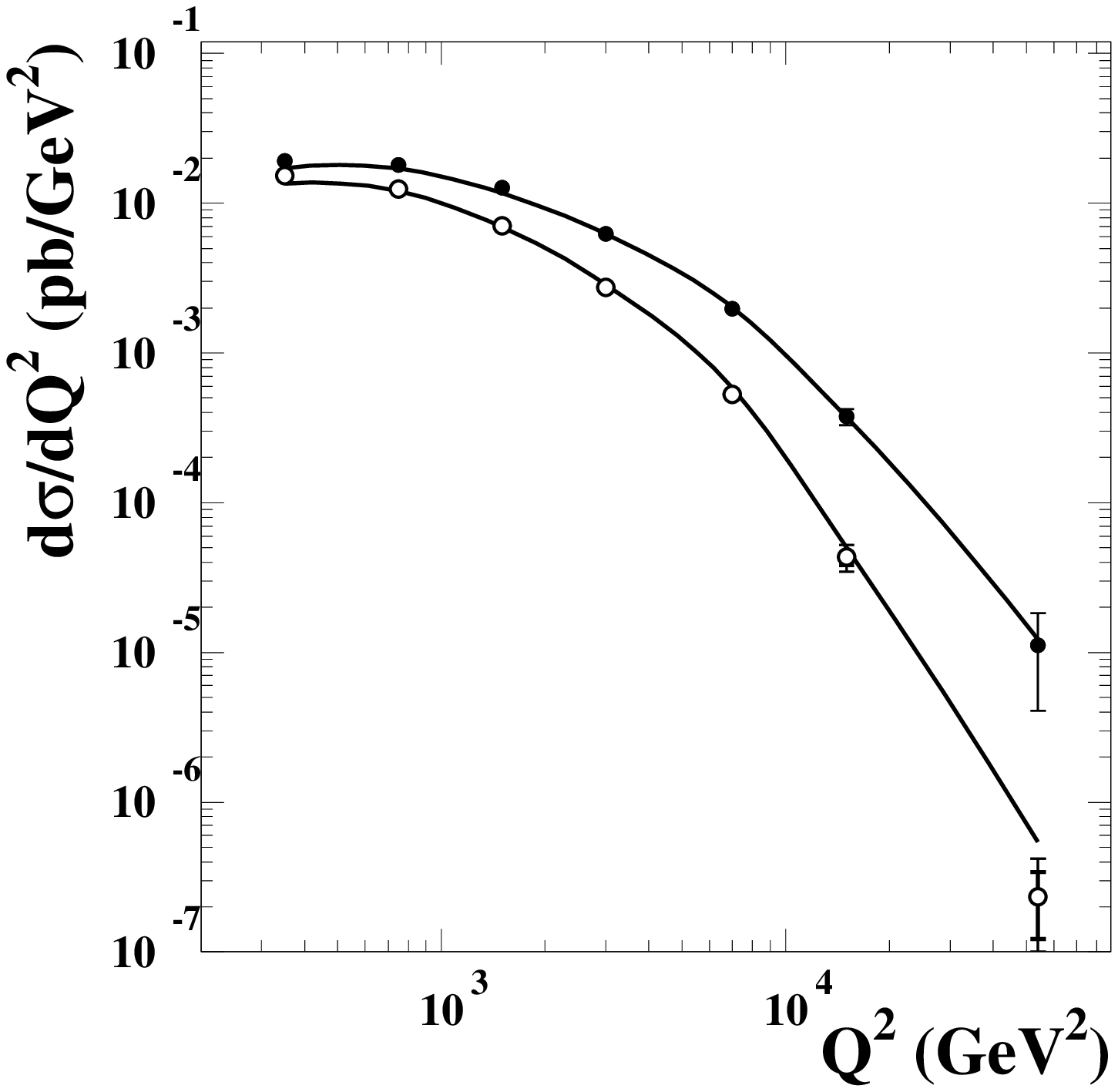,width=12cm}}
\put (7.0,-1.5){\epsfig{figure=\figdir 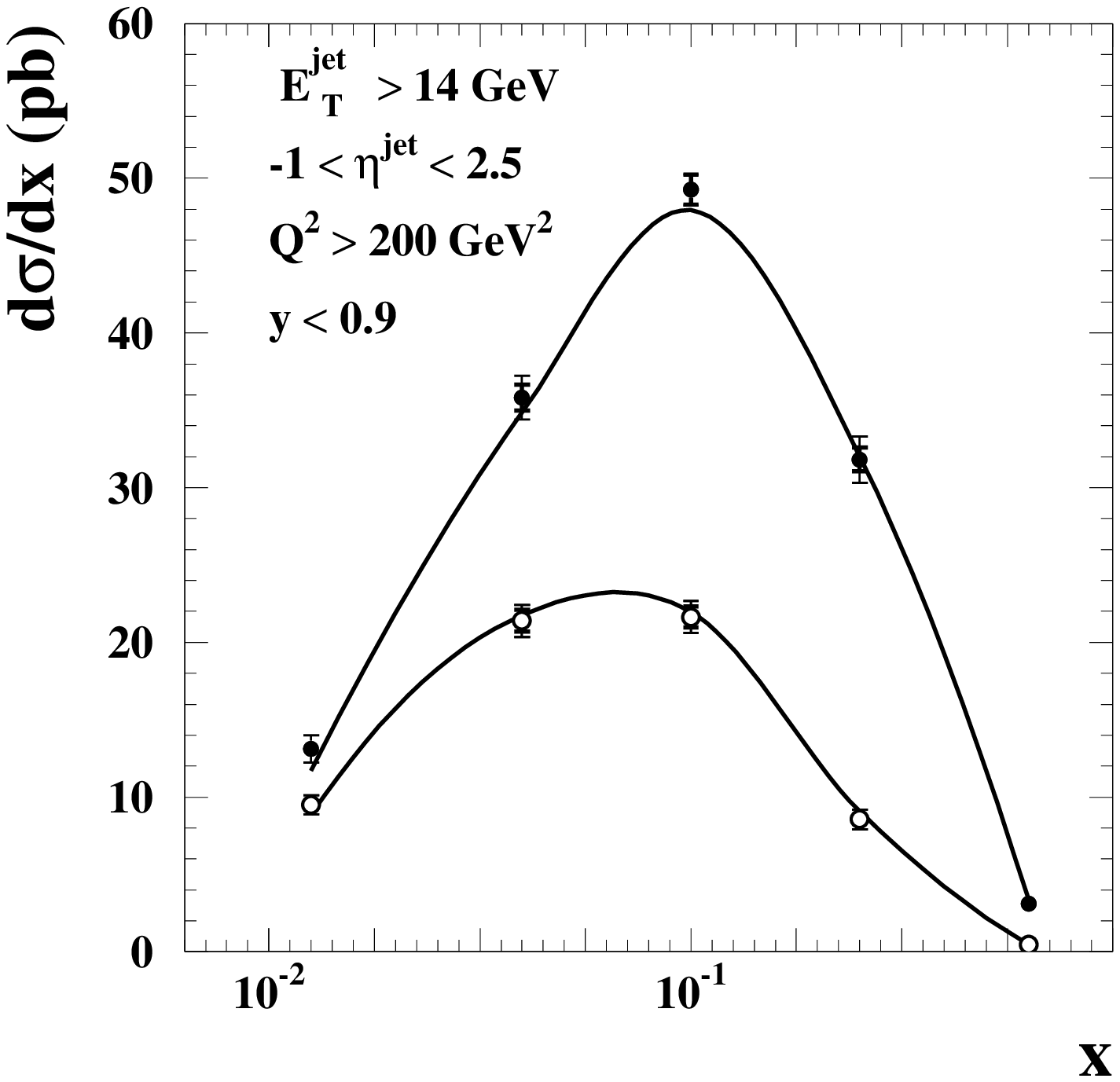,width=12cm}}
\put (6.7,17.2){\bf\small (a)}
\put (15.7,17.2){\bf\small (b)}
\put (6.7,7.2){\bf\small (c)}
\put (15.7,7.2){\bf\small (d)}
\end{picture}
\caption
{\it 
Measured unpolarized inclusive-jet cross sections in CC DIS
for jets with $\etjet>14$ GeV and $-1<\etajet<2.5$ in the kinematic
regime given by $\q2>200$~\g2\ and $y<0.9$ as functions of (a)
$\etajet$, (b) $\etjet$, (c) $\q2$ and (d) Bjorken $x$ in $e^-p$
(dots) and $e^+p$ (open circles) collisions. For comparison, the
predictions of NLO QCD based on the {\sc Mepjet} calculations using
the ZEUS PDF sets (solid lines) are also shown.
Other details as in the caption to Fig.~\ref{fig5}.
}
\label{fig7}
\vfill
\end{figure}

Figure~\ref{fig7} and Tables~\ref{tabseven} to \ref{tabfourteen} show
the unpolarized inclusive-jet differential cross sections as functions
of $\etajet$, $\etjet$, $\q2$ and $x$ in CC $e^{\pm}p$ DIS. The
unpolarized cross section for an observable $A$ was obtained via

$$\frac{d\sigma}{dA}(e^{\pm})=\frac{N_{\rm data}^{\rm unpol}(e^{\pm})}{{\cal L}_{e^{\pm}}\cdot \Delta A}\cdot\frac{N^{\rm had}_{\rm MC}}{N^{\rm det}_{\rm MC}},$$

where 

$$N_{\rm data}^{\rm unpol}(e^{\pm})=\frac{N_{\rm data}^{\rm neg}(e^{\pm})}{1\pm P_{e^{\pm}}^{\rm neg}}+\frac{N_{\rm data}^{\rm pos}(e^{\pm})}{1\pm P_{e^{\pm}}^{\rm pos}},$$

${\cal L}_{e^{\pm}}$ is the total integrated luminosity for $e^{\pm}p$
collisions and $\Delta A$ is the bin width.

\begin{figure}[p]
\vfill
\setlength{\unitlength}{1.0cm}
\begin{picture} (18.0,17.0)
\put (0.0,11.0){\centerline{\epsfig{figure=\figdir zeus.eps,width=10cm}}}
\put (-2.0,8.5){\epsfig{figure=\figdir 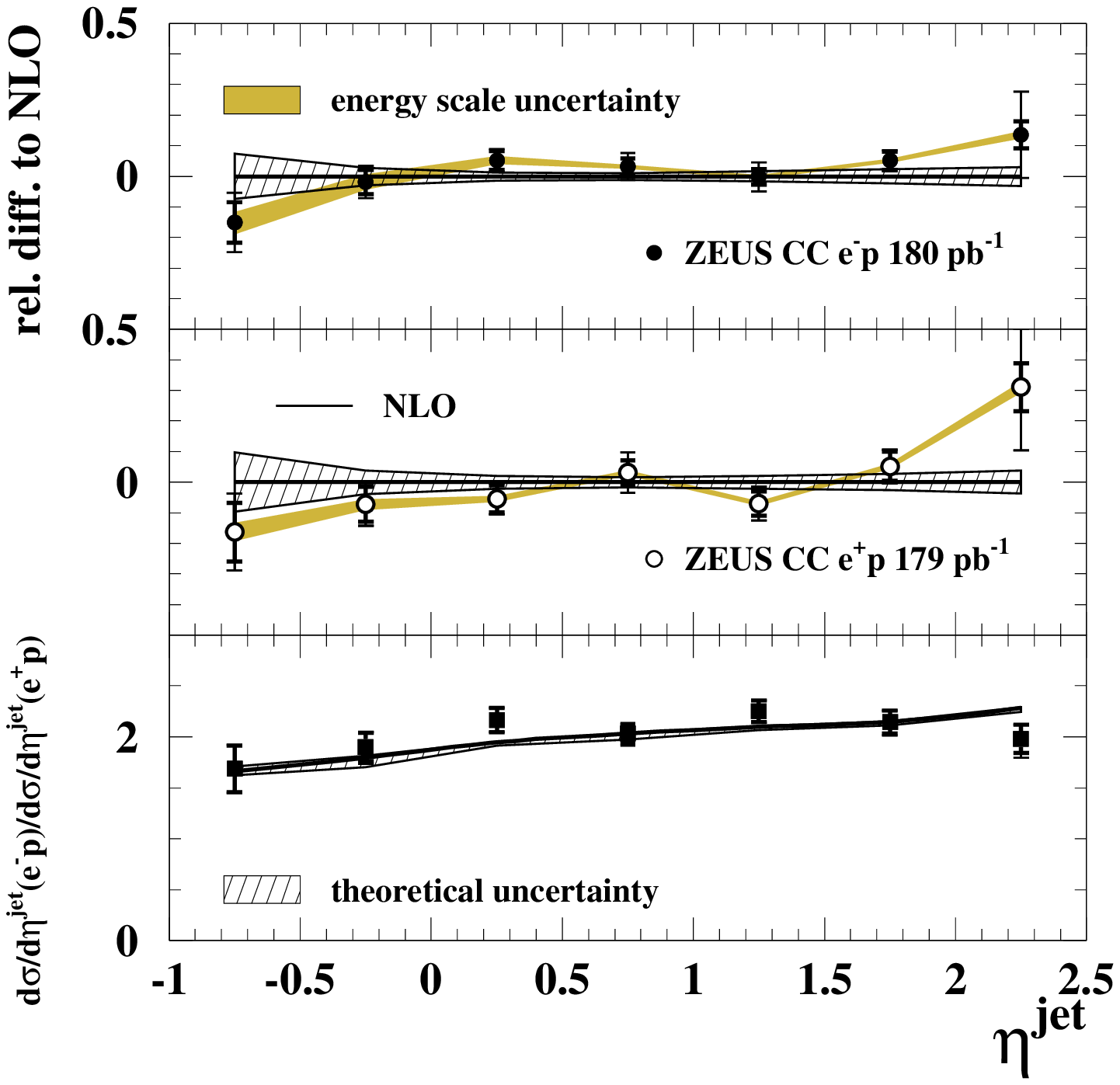,width=12cm}}
\put (7.0,8.5){\epsfig{figure=\figdir 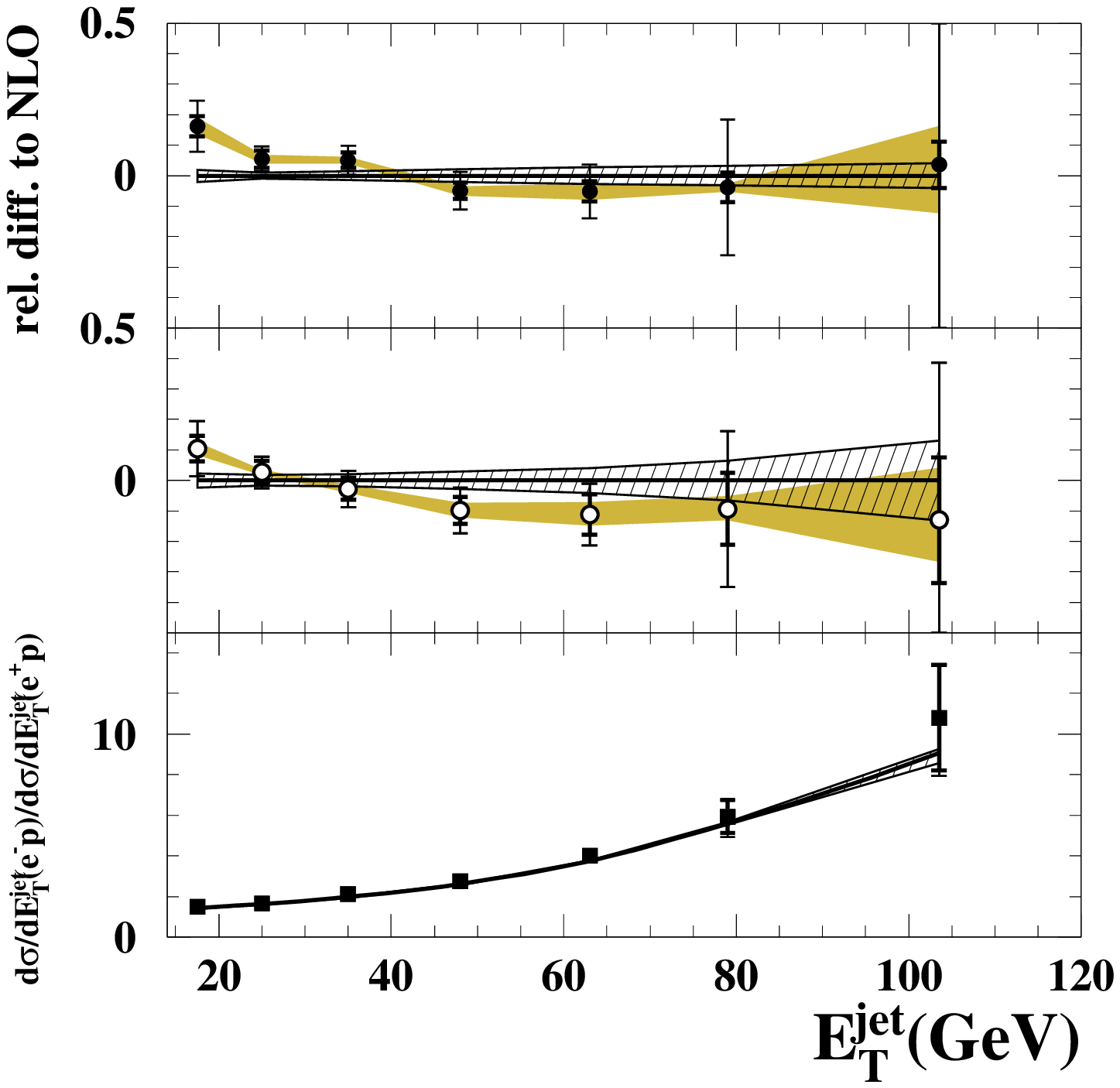,width=12cm}}
\put (-2.0,-1.5){\epsfig{figure=\figdir 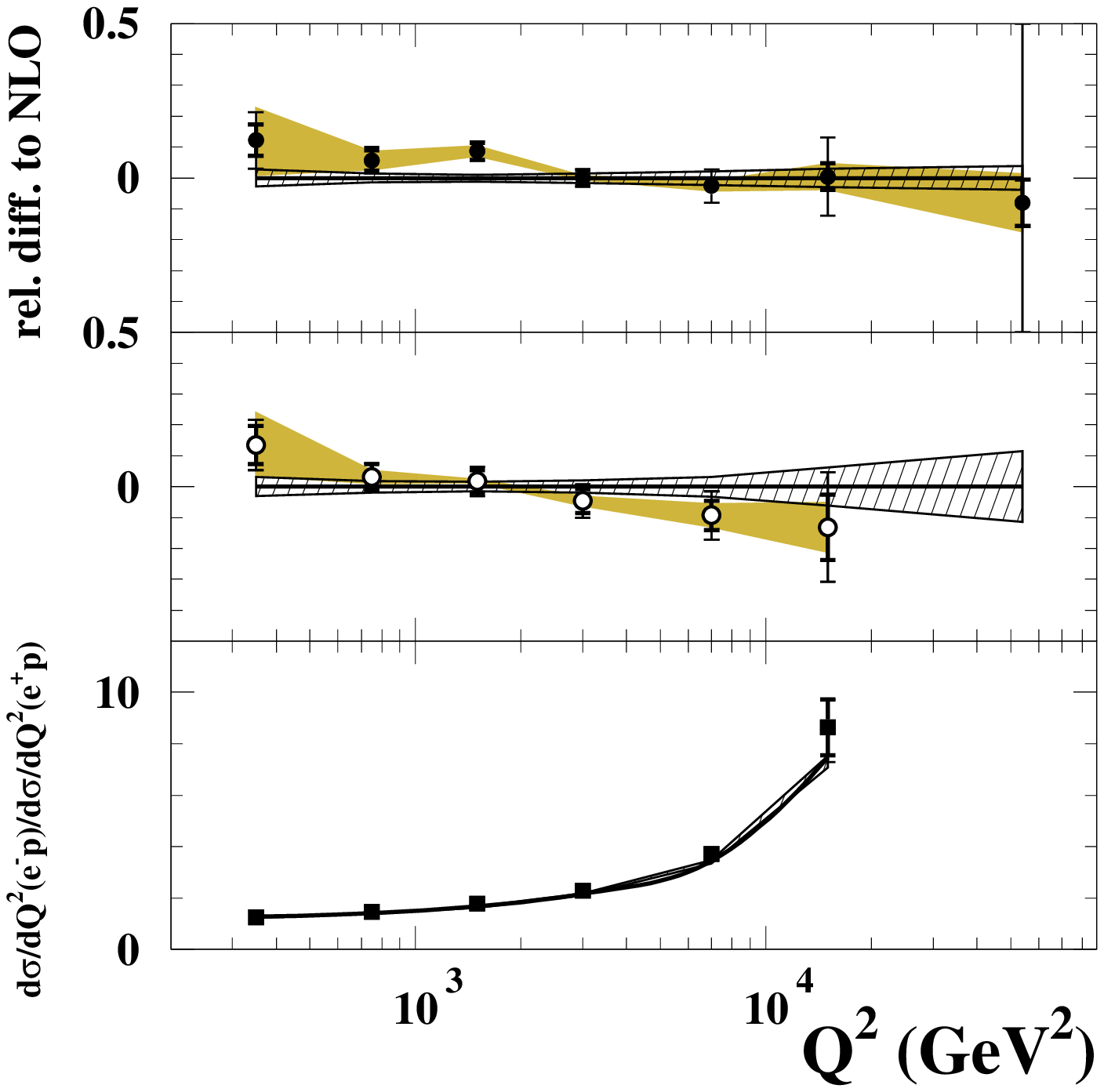,width=12cm}}
\put (7.0,-1.5){\epsfig{figure=\figdir 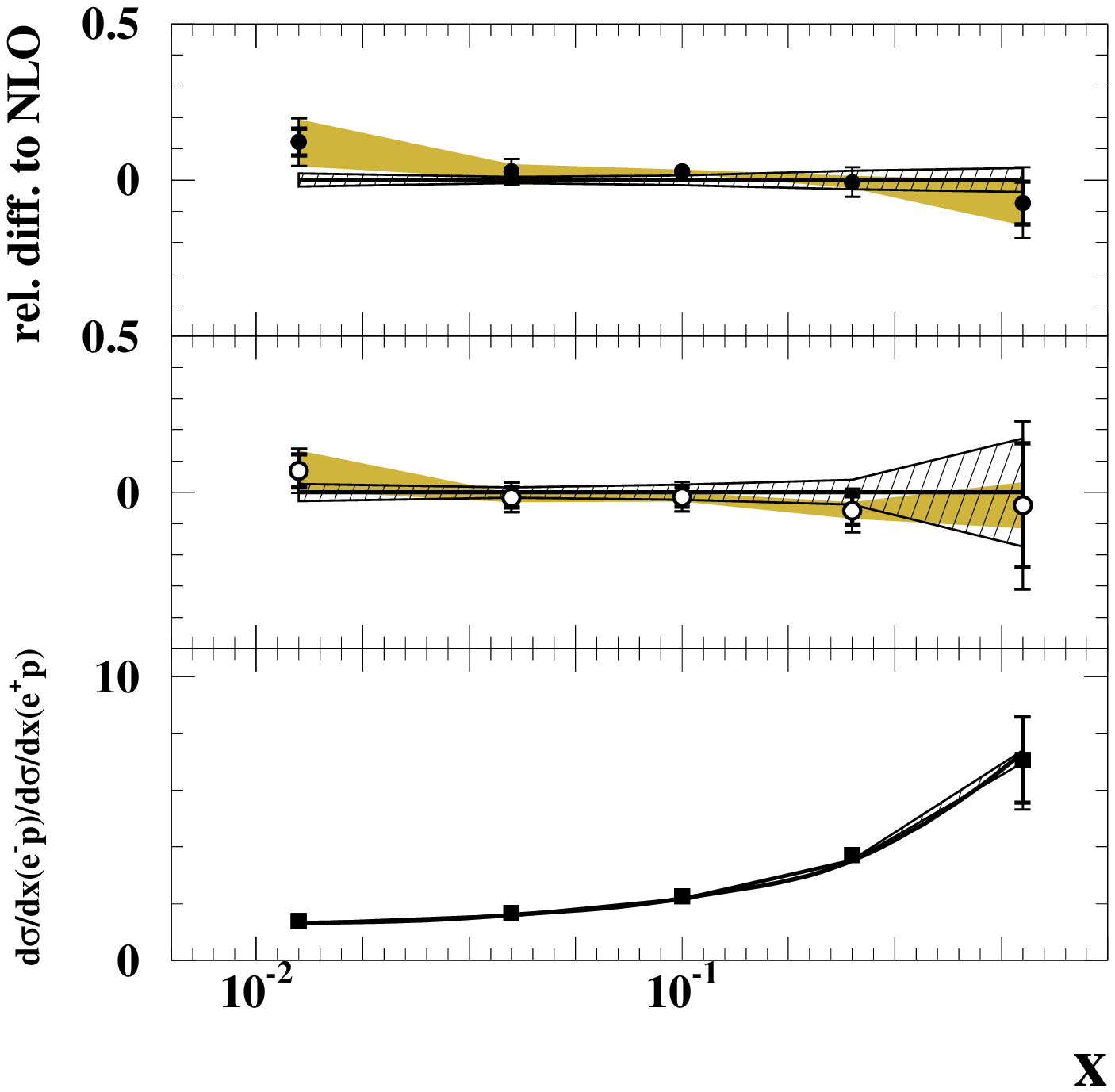,width=12cm}}
\put (6.4,17.6){\bf\small (a)}
\put (15.6,17.6){\bf\small (b)}
\put (6.4,7.6){\bf\small (c)}
\put (15.7,7.6){\bf\small (d)}
\end{picture}
\caption
{\it 
Relative difference between the measured cross sections of
Fig.~\ref{fig7} and the corresponding NLO QCD predictions in $e^-p$
(dots) and $e^+p$ (open circles) collisions as functions of (a)
$\etajet$, (b) $\etjet$, (c) $\q2$ and (d) Bjorken $x$. The lower parts
of the figures display the ratio of the cross sections for $e^-p$ and
$e^+p$ collisions (squares). The hatched areas display the theoretical
uncertainty and the shaded areas display the uncertainty due to the 
absolute energy scale. Other details as in the caption to
Fig.~\ref{fig5}.
}
\label{fig8}
\vfill
\end{figure}

The measured $\seta$ has a maximum at $\etajet\approx 1$. The measured
$\set$ exhibits a fall-off of two (three) orders of magnitude in the
$e^-p$ ($e^+p$) sample. Values of $\etjet$ of more than $100$ GeV are
accessible with the present statistics. For $200<\q2\lesssim 2000$ \g2, 
the distributions display a weak dependence on $\q2$. The cross
sections as functions of $\etjet$ and $\q2$ show a less rapid
fall-off than what is observed in NC DIS processes due to the presence
of the massive $W$ propagator. Furthermore, the measured cross
sections for the $e^+p$ sample decrease more rapidly as a function of
$\etjet$ and $\q2$ than for the $e^-p$ sample (see below). The values
in $x$ accessible by the data are within the range $0.013<x<0.63$, as
shown in Fig.~\ref{fig7}d. The NLO QCD predictions using the ZEUS-S
PDF sets are compared to the data in Fig.~\ref{fig7}. Figure~\ref{fig8}
shows the relative difference between the data and the
predictions. The NLO QCD predictions give a reasonable description of
the shape and normalization of the data.

Figure~\ref{fig8} also displays the ratio of the $e^-p$ and $e^+p$
differential cross sections. The measured ratio as a function of
$\etajet$ is constant and $\approx 2$ as predicted by QCD, since the
two reactions probe a different parton content of the proton. The ratio
as a function of $\etjet$ ($\q2$) increases as $\etjet$ ($\q2$)
increases, in good agreement with the prediction. The increase at high
values of $\etjet$ and $\q2$ is expected due to the increasing
contribution from the valence-quark densities in the proton at high
$x$ and the fact that both reactions are sensitive to different quark
flavors. The behavior observed in the ratio of the measured cross
sections as a function of $\q2$ is similar to the ratio of $u$ and $d$
parton densities. The same behavior is observed as a function of $x$.

\begin{figure}[p]
\vfill
\setlength{\unitlength}{1.0cm}
\begin{picture} (18.0,17.0)
\put (-2.0,8.5){\epsfig{figure=\figdir 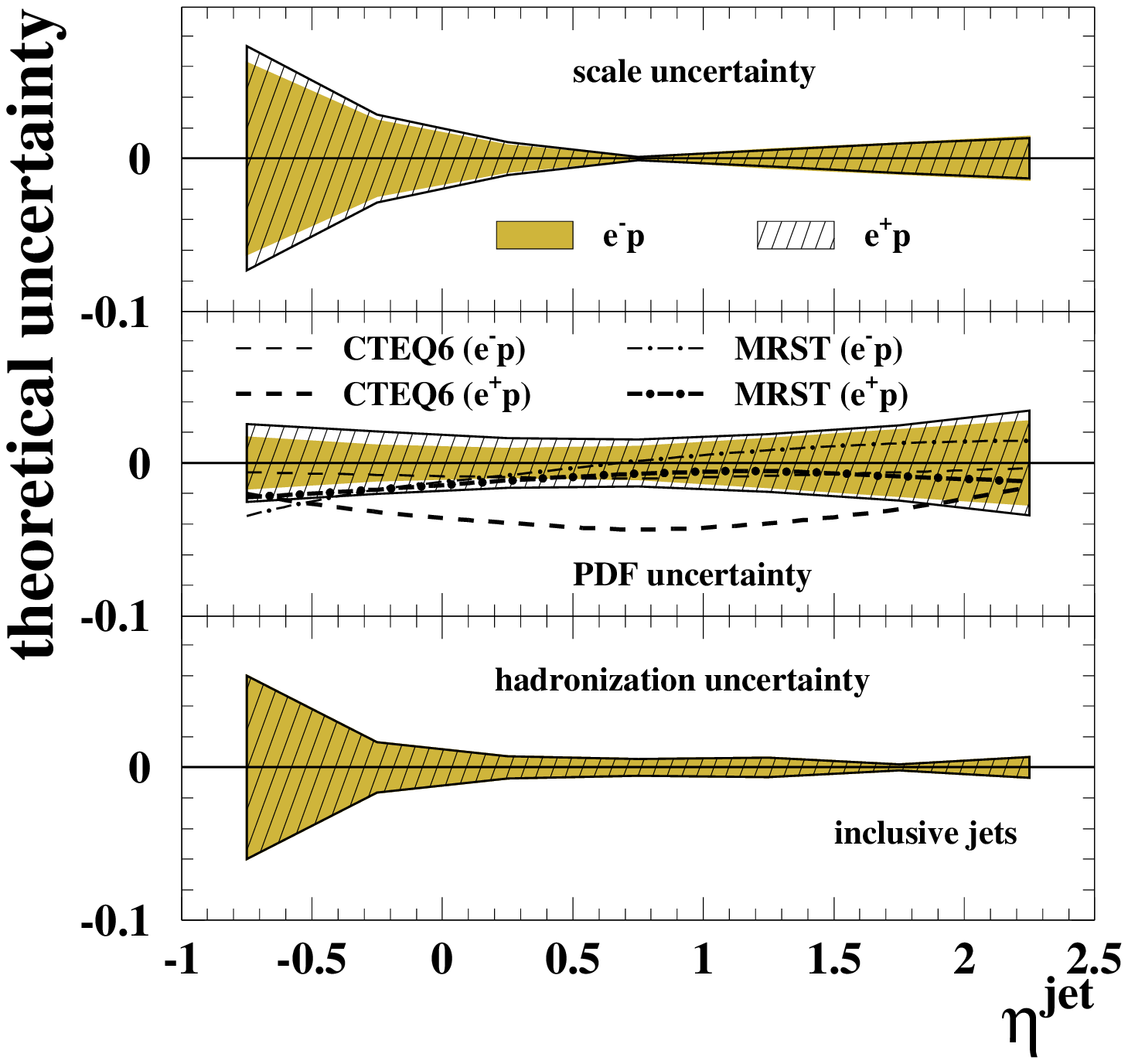,width=12cm}}
\put (7.0,8.5){\epsfig{figure=\figdir 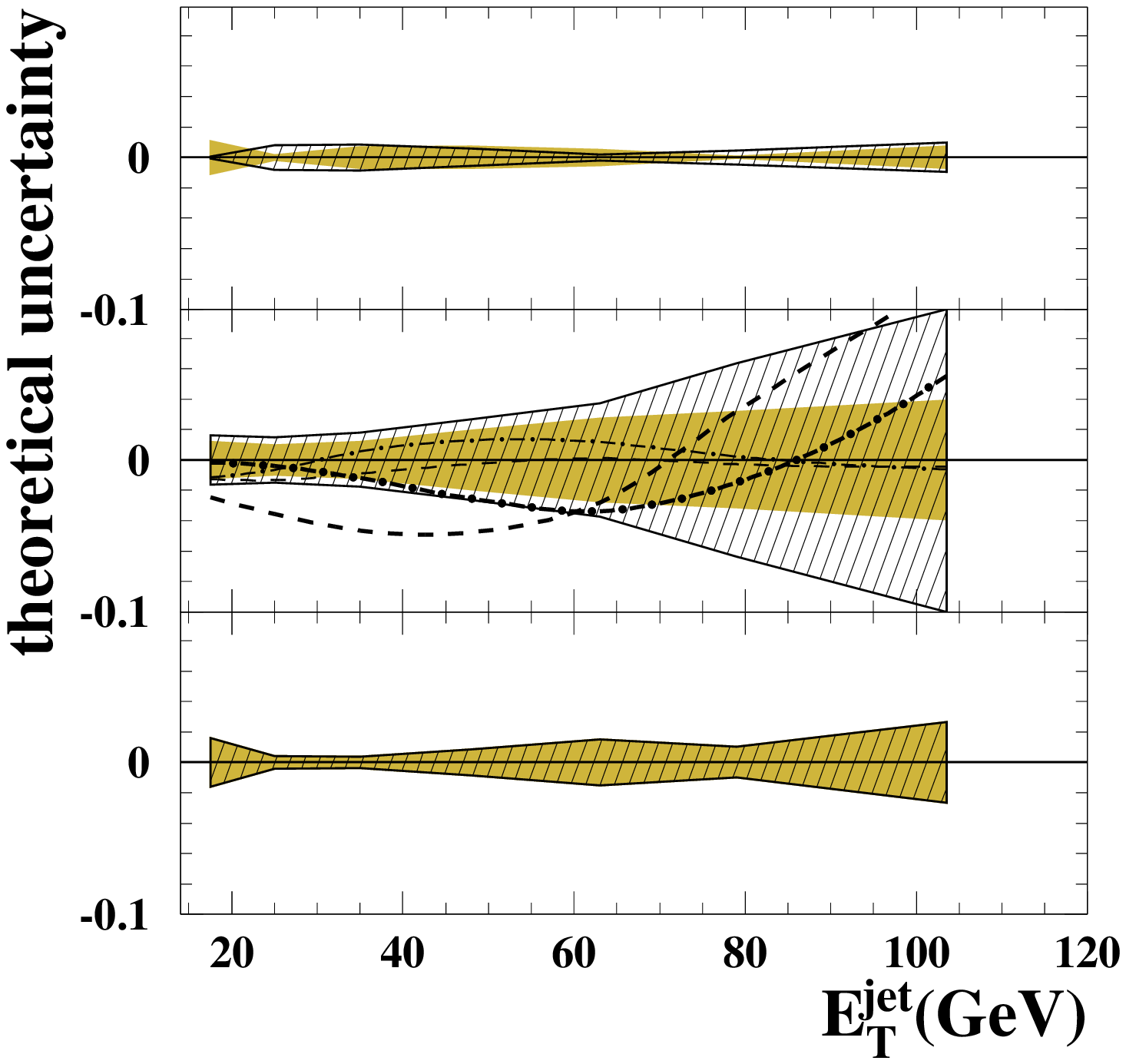,width=12cm}}
\put (-2.0,-1.5){\epsfig{figure=\figdir 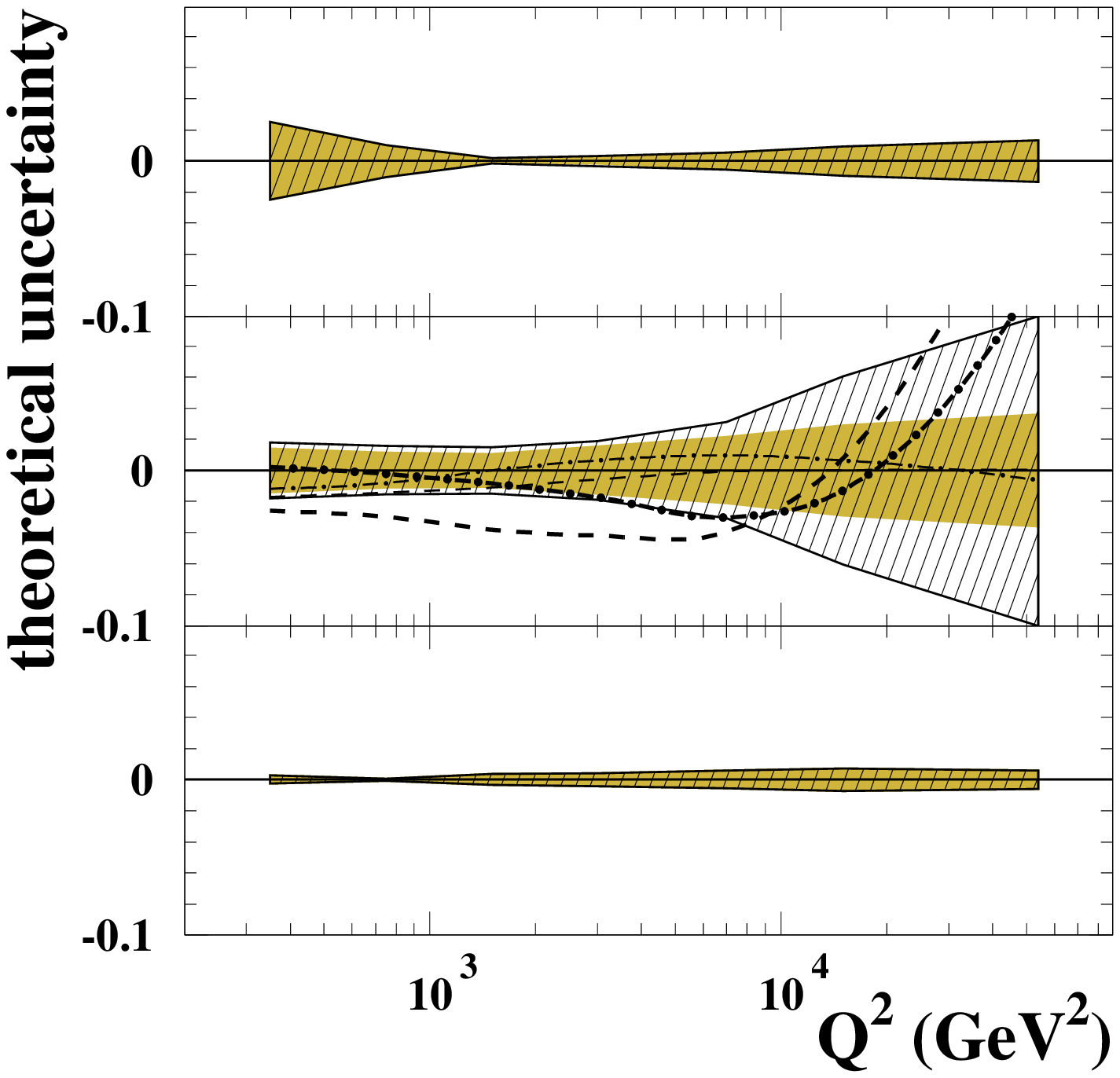,width=12cm}}
\put (7.0,-1.5){\epsfig{figure=\figdir 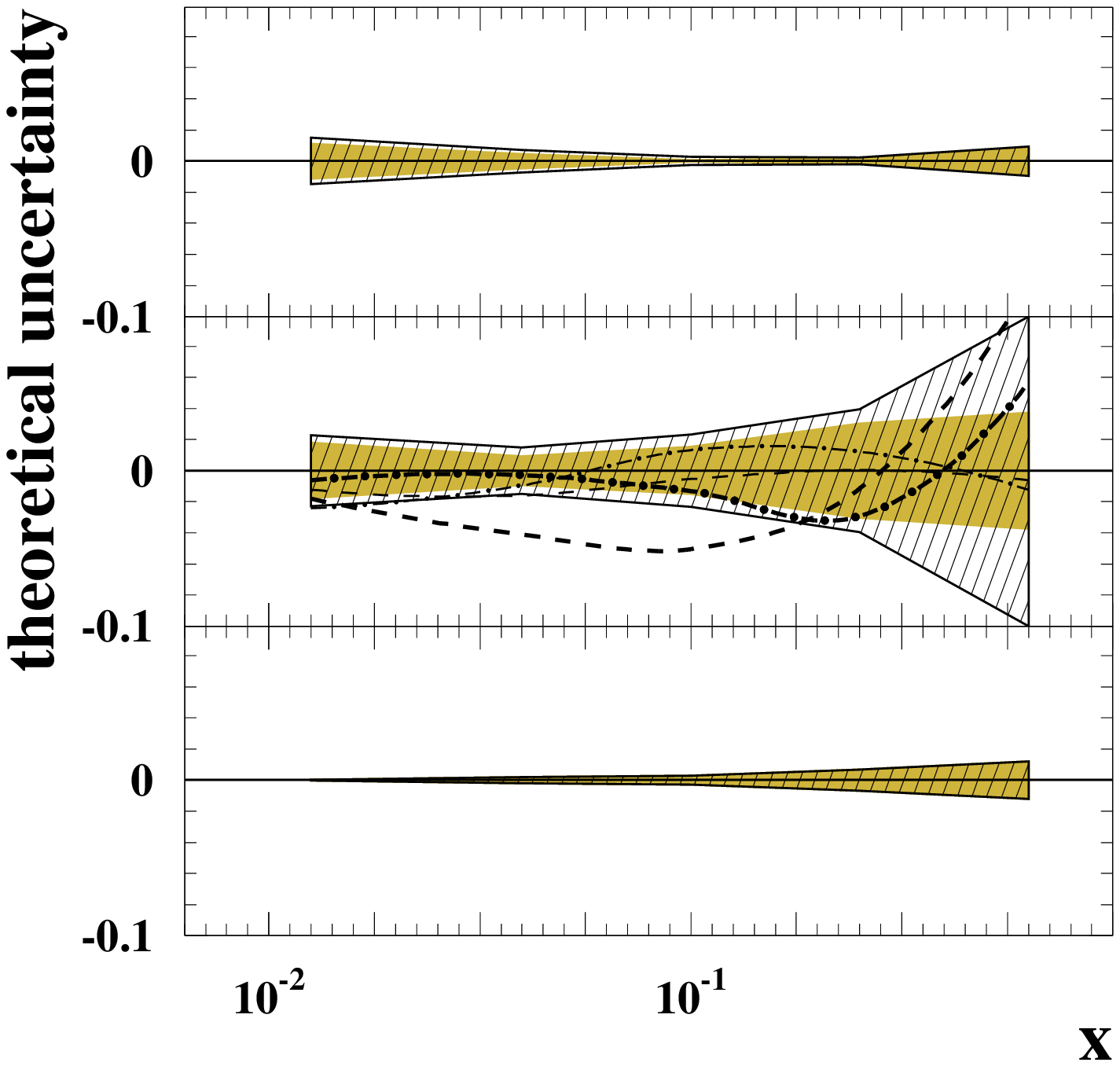,width=12cm}}
\put (6.7,17.6){\bf\small (a)}
\put (15.7,17.6){\bf\small (b)}
\put (6.7,7.7){\bf\small (c)}
\put (15.7,7.7){\bf\small (d)}
\end{picture}
\caption
{\it 
Overview of theoretical uncertainties for the inclusive-jet cross
sections in CC DIS for jets with $\etjet>14$ GeV and $-1<\etajet<2.5$
in the kinematic regime given by $\q2>200$~\g2\ and $y<0.9$ as
functions of (a) $\etajet$, (b) $\etjet$, (c) $\q2$ and (d) Bjorken
$x$ in $e^-p$ (shaded areas) and $e^+p$ (hatched areas)
collisions. Shown are the relative uncertainties induced by the
variation of the renormalization scale $\mu_R$, the uncertainties on
the proton PDFs and hadronisation model. Also shown are the relative
differences between the NLO QCD calculations using the CTEQ6 (dashed
lines) or MRST (dot-dashed lines) PDF sets to the calculations based
on the ZEUS sets.
}
\label{fig9}
\vfill
\end{figure}

Figure~\ref{fig9} shows the contributions to the theoretical
uncertainty from the terms beyond NLO, the parton-shower model and
that coming from the uncertainty in the PDFs separately for $e^-p$ and
$e^+p$ collisions. Also shown are calculations using other PDF
sets. For inclusive-jet $e^{\pm}p$ CC cross sections, the uncertainty
coming from that on the PDFs is dominant. At high $\etjet$, $\q2$ and
$x$, the uncertainty in the predicted cross sections for positron
beams is larger than those for electron beams. This difference in the
uncertainty due to the PDFs in the calculations for $e^-$ and $e^+$
beams can be attributed to the different flavor content probed: in
$e^-p$ ($e^+p$) at high $x$ the $W^-$ ($W^+$) will couple
predominantly to the $u$ ($d$) valence quark in the proton; at
present, the uncertainty in the $d$ parton density is larger than that
for the $u$ quark. Furthermore, the comparison with the calculations
using other PDF sets shows a wide spread in the predictions,
especially for positron beams. Therefore, these measurements, in a
phase-space region where the other theoretical uncertainties are well
under control, have the potential to constrain the flavor content of
the proton if used together with other data in global fits. A fast and
accurate method to perform fits to extract the proton PDFs on data
sets that included jet cross sections in NC DIS and photoproduction
was recently developed by the ZEUS
Collaboration~\cite{epj:c42:1}; the result was a sizable reduction of
the uncertainty on the gluon density at medium and high $x$. Using the
data presented here and extending such a method to jet cross sections
in CC DIS may help to constrain the $u$ and $d$ valence quark
distributions at high $x$.

The integrated unpolarized inclusive-jet cross sections,
$\sigma_{\rm jets}$, are shown in Table~\ref{tabtwo}. The measured
cross sections are in good agreement with the predictions of NLO QCD,
also shown in the table using different PDF sets.

\subsection{Dijet cross sections}

\begin{figure}[p]
\vfill
\setlength{\unitlength}{1.0cm}
\begin{picture} (18.0,17.0)
\put (0.0,11.0){\centerline{\epsfig{figure=\figdir zeus.eps,width=10cm}}}
\put (-2.0,8.5){\epsfig{figure=\figdir 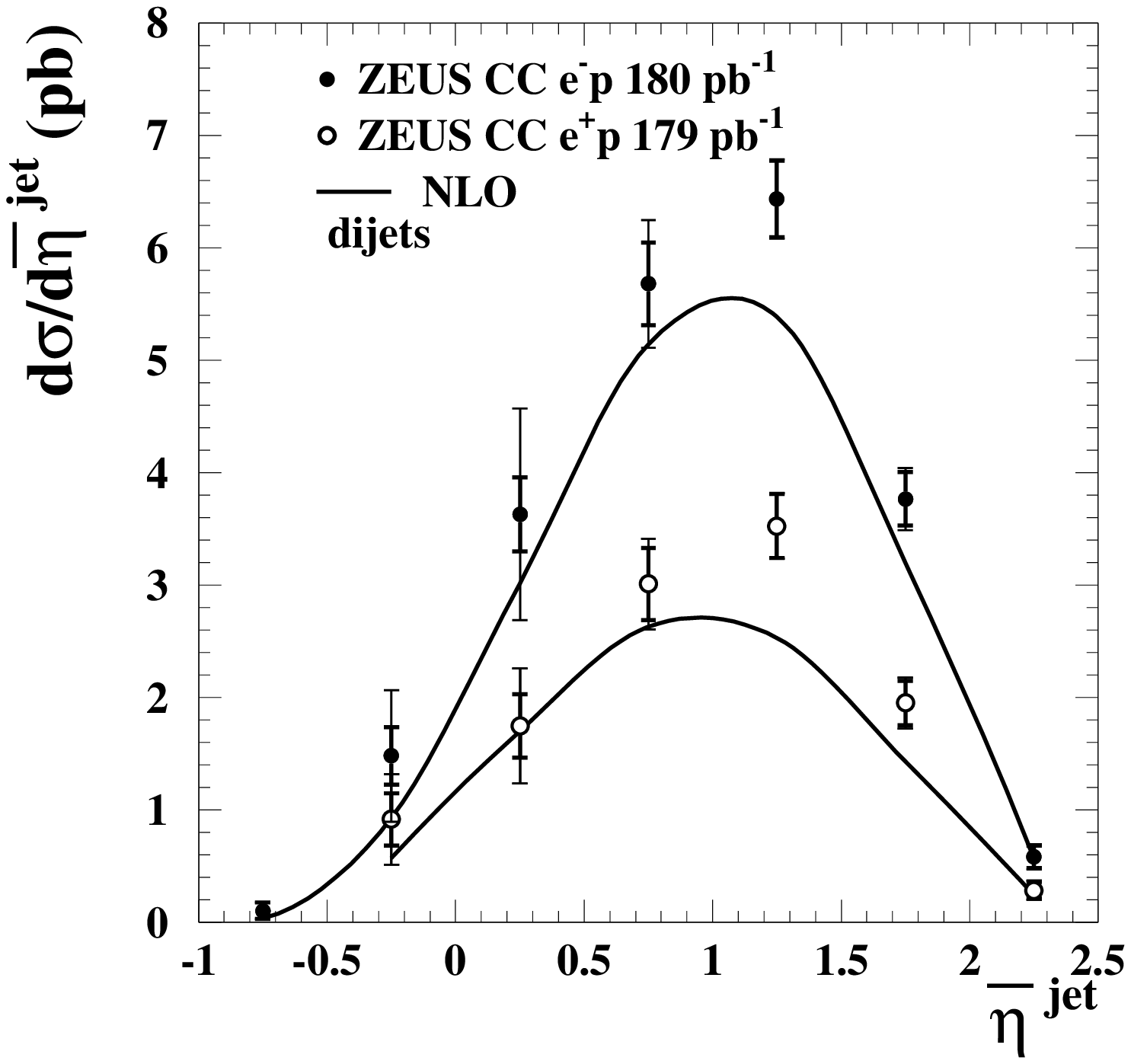,width=12cm}}
\put (7.0,8.5){\epsfig{figure=\figdir 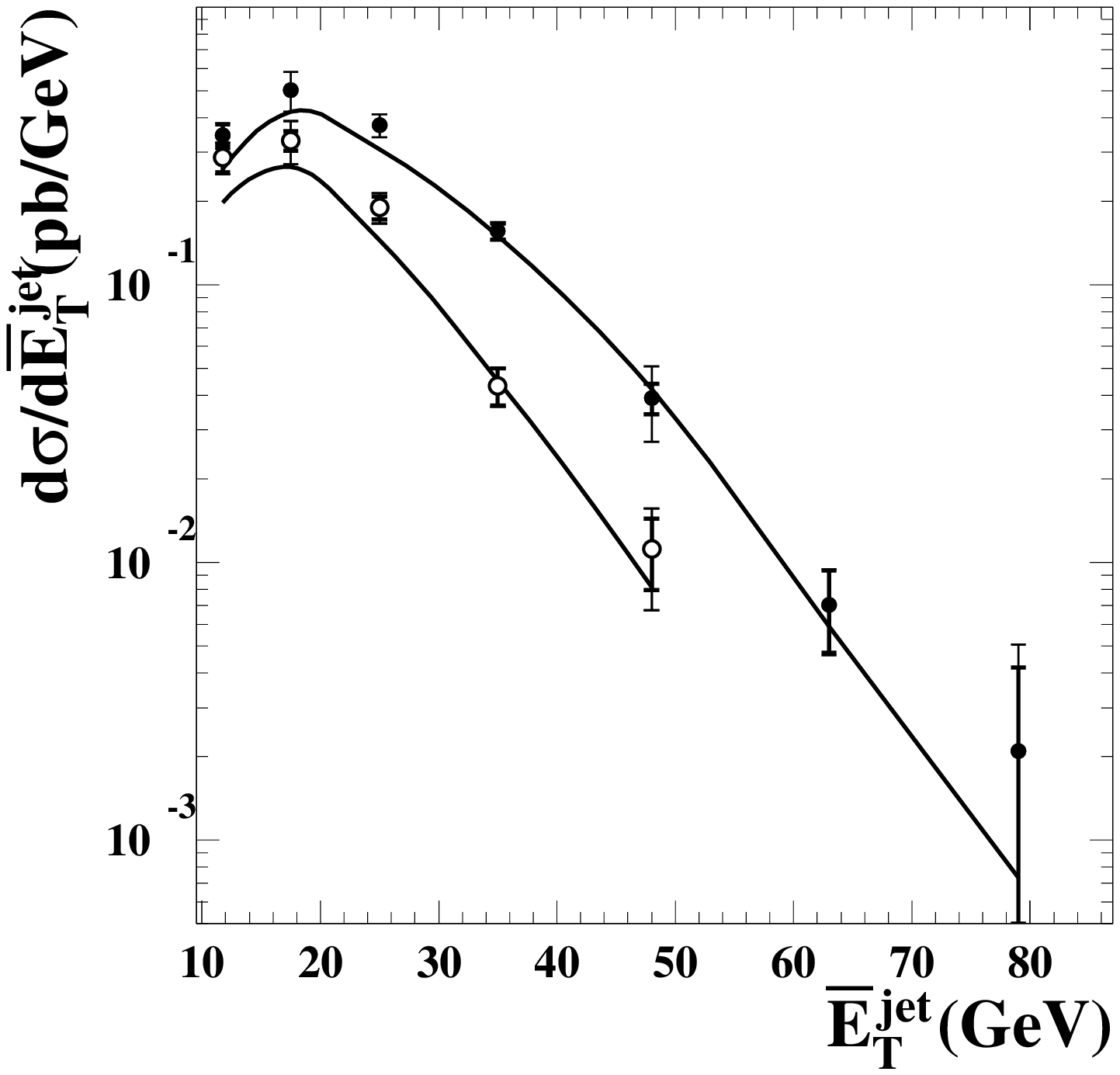,width=12cm}}
\put (-2.0,-1.5){\epsfig{figure=\figdir 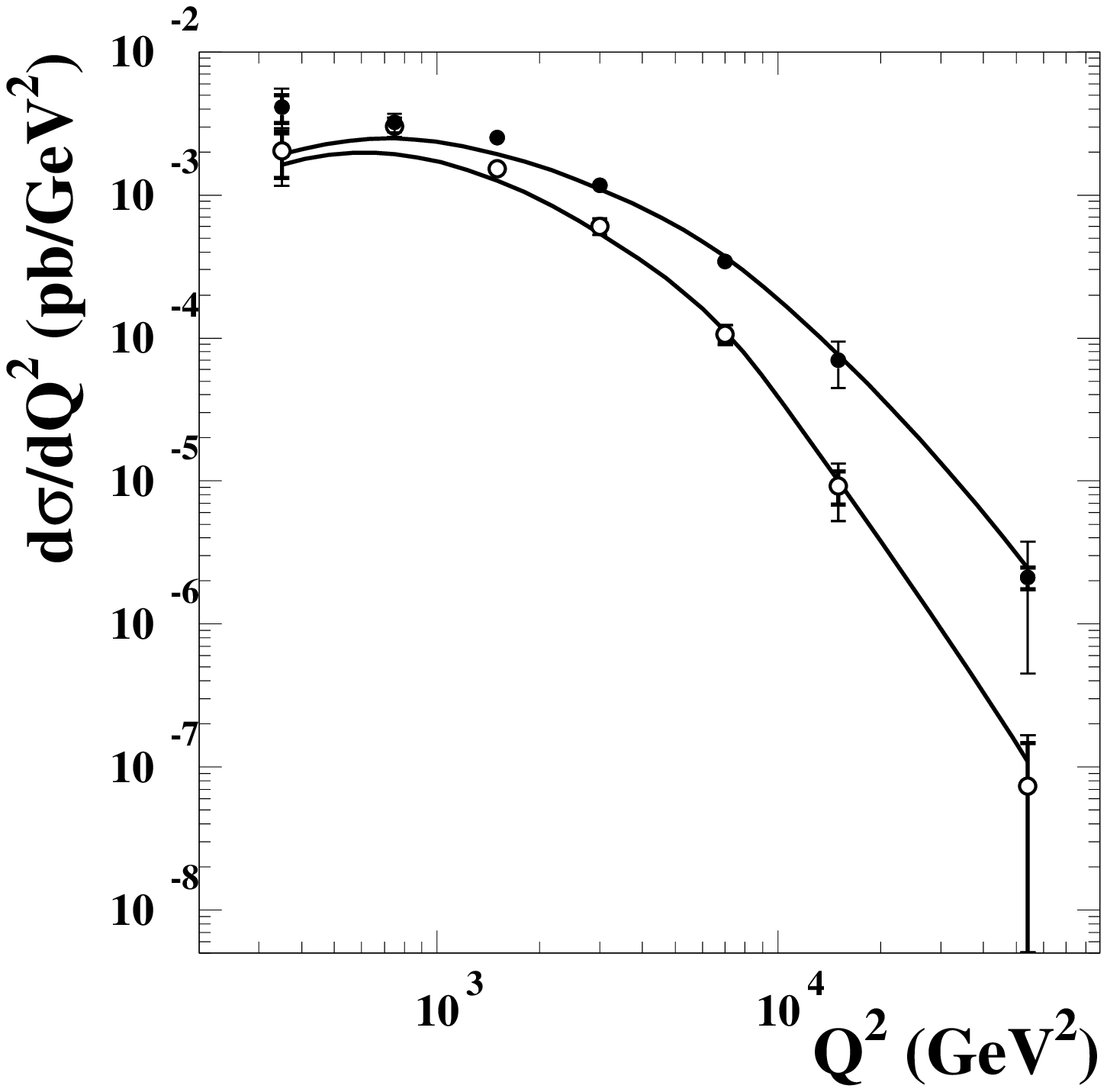,width=12cm}}
\put (7.0,-1.5){\epsfig{figure=\figdir 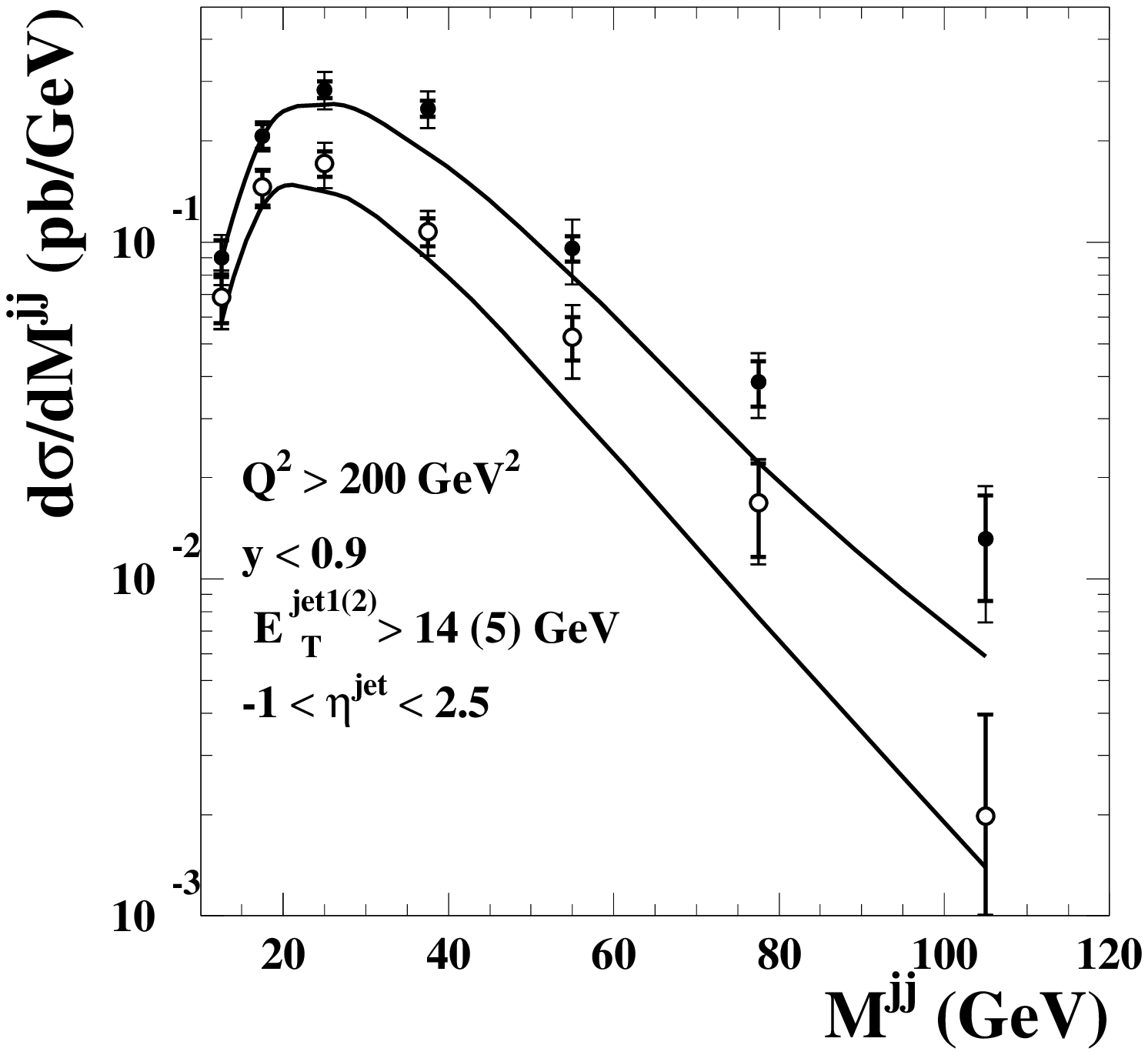,width=12cm}}
\put (6.7,17.2){\bf\small (a)}
\put (15.7,17.2){\bf\small (b)}
\put (6.7,7.2){\bf\small (c)}
\put (15.7,7.2){\bf\small (d)}
\end{picture}
\caption
{\it 
Measured unpolarized dijet cross sections in CC DIS for jets with
$\etj>14$ GeV, $\etjj>5$ GeV and $-1<\etajet<2.5$ in the kinematic
regime given by $\q2>200$~\g2\ and $y<0.9$ as functions of (a)
$\etabar$, (b) $\etbar$, (c) $\q2$ and (d) $\mj$
in $e^-p$ (dots) and $e^+p$ (open circles) collisions. Other details
as in the caption to Fig.~\ref{fig7}.
}
\label{fig10}
\vfill
\end{figure}

Unpolarized dijet differential cross sections were measured in the
kinematic regime $\q2>200$ \g2\ and $y<0.9$. The cross sections were
determined for jets with $\etj>14$ GeV, $\etjj>5$ GeV and
$-1<\etajet<2.5$. Figure~\ref{fig10} and Tables~\ref{tabseven} to 
\ref{tabtwelve}, \ref{tabfifteen} and \ref{tabsixteen} show the
unpolarized dijet differential cross sections as functions of
$\etabar$, $\etbar$, $\q2$ and the dijet invariant mass, $\mj$, where
$\etabar=(\etaj+\etajj)/2$ and $\etbar=(\etj+\etjj)/2$ in CC
$e^{\pm}p$ DIS. The measured $\etabar$ cross section has a maximum at
$\etabar\sim 1.25$. The measured cross section as a function of
$\etbar$ exhibits a fall-off of two orders of magnitude for
$\etbar\gtrsim 20$~GeV. For $200<\q2\lesssim 2000$ \g2, the
distribution displays a weak dependence on $\q2$. Values of $\mj$
above $100$ GeV are accessible with the present statistics.

\begin{figure}[p]
\vfill
\setlength{\unitlength}{1.0cm}
\begin{picture} (18.0,17.0)
\put (0.0,11.0){\centerline{\epsfig{figure=\figdir zeus.eps,width=10cm}}}
\put (-2.0,8.5){\epsfig{figure=\figdir 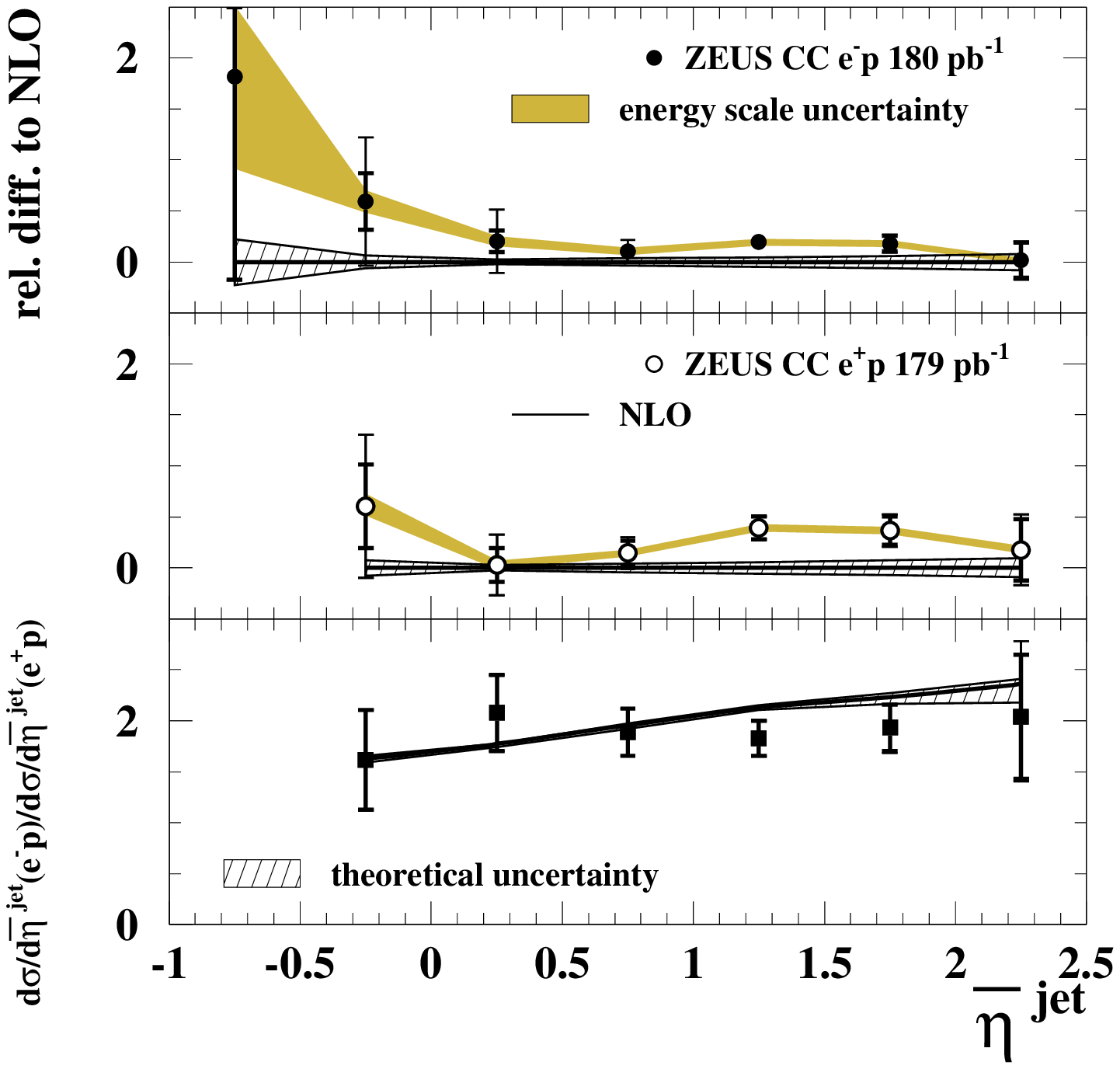,width=12cm}}
\put (7.0,8.5){\epsfig{figure=\figdir 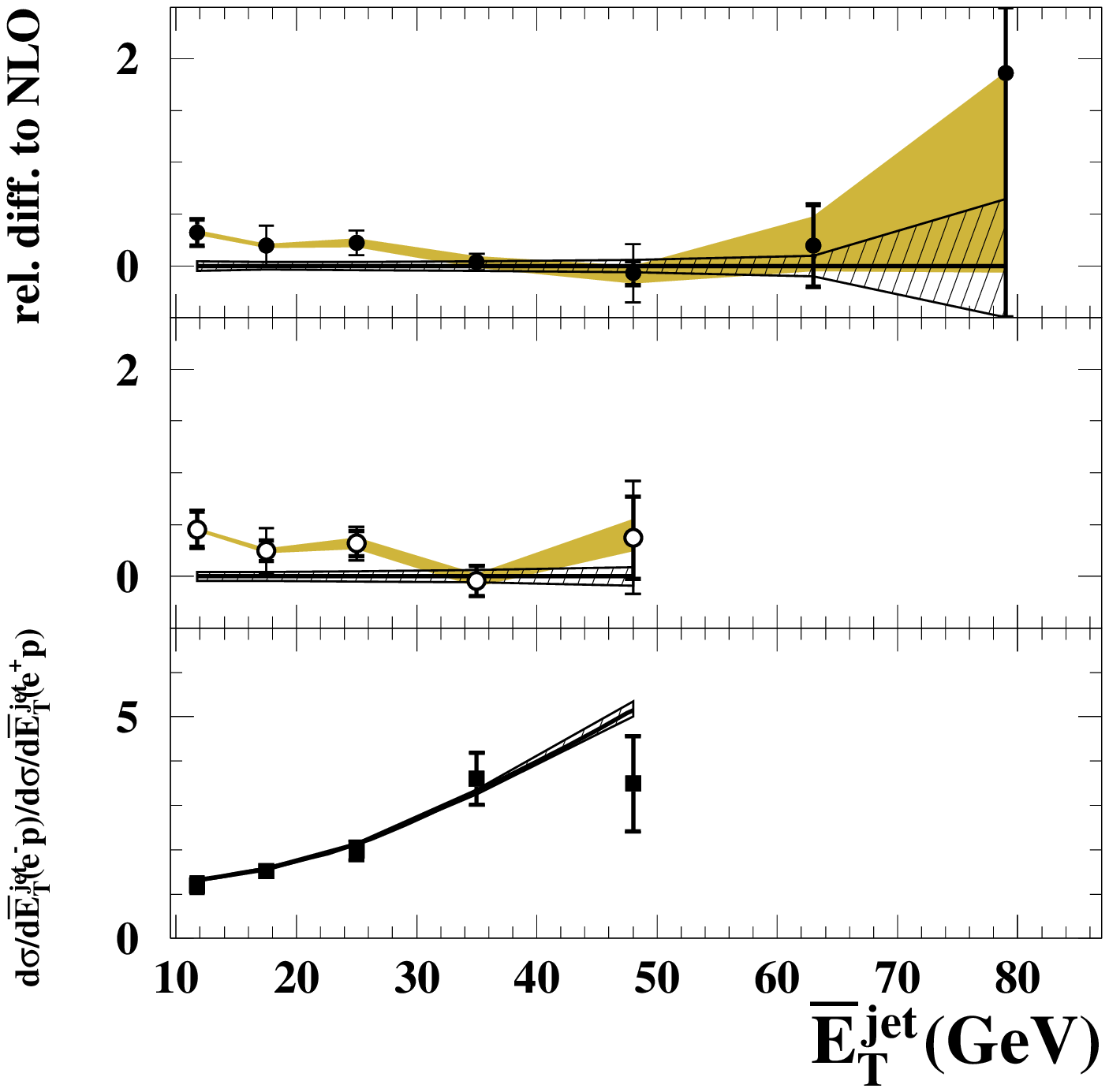,width=12cm}}
\put (-2.0,-1.5){\epsfig{figure=\figdir 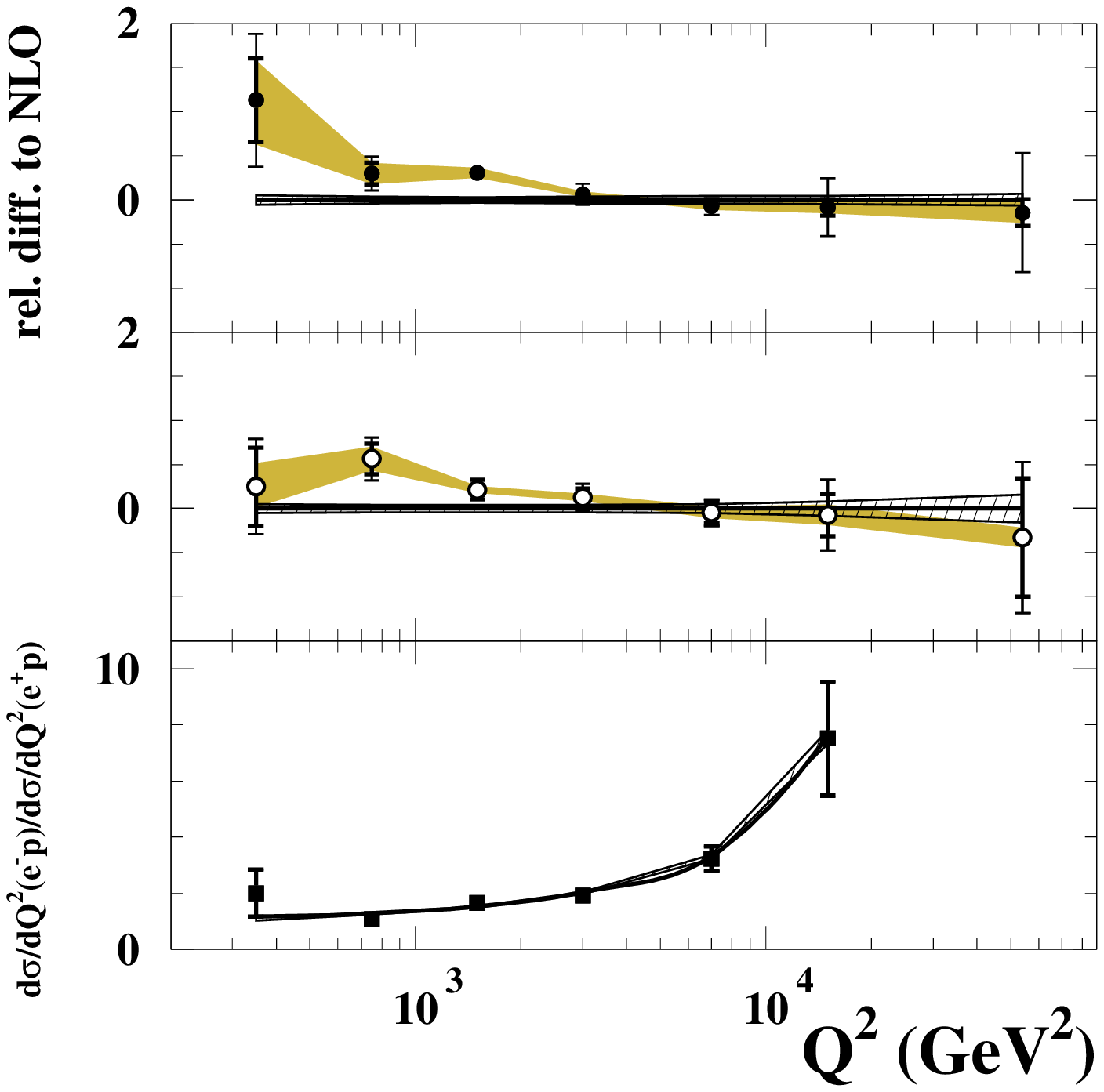,width=12cm}}
\put (7.0,-1.5){\epsfig{figure=\figdir 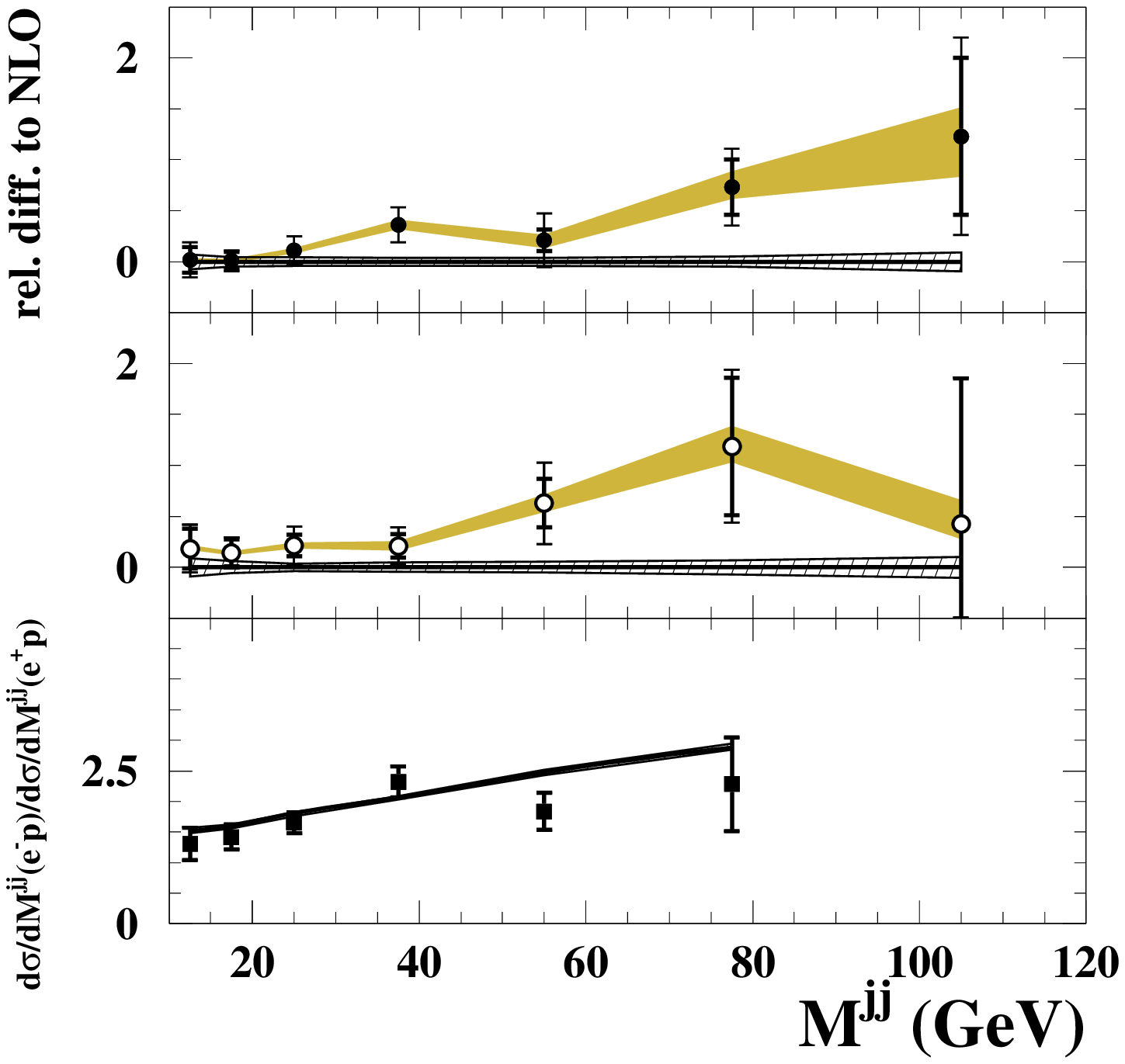,width=12cm}}
\put (6.8,17.2){\bf\small (a)}
\put (9.7,17.2){\bf\small (b)}
\put (6.7,7.4){\bf\small (c)}
\put (15.7,7.4){\bf\small (d)}
\end{picture}
\caption
{\it 
Relative difference between the measured cross sections of
Fig.~\ref{fig10} and the corresponding NLO QCD predictions in $e^-p$
(dots) and $e^+p$ (open circles) collisions as functions of (a)
$\etabar$, (b) $\etbar$, (c) $\q2$ and (d)
$\mj$. The lower parts of the figures display the ratio of the cross
sections for $e^-p$ and $e^+p$ collisions (squares). Other details as
in the caption to Fig.~\ref{fig8}.
}
\label{fig11}
\vfill
\end{figure}

The NLO QCD predictions are compared to the data in
Fig.~\ref{fig10}. Figure~\ref{fig11} shows the relative difference to
the predictions. The NLO predictions do not give an adequate
description in shape and normalization of the measured differential
cross sections over the entire phase space considered. In particular,
for $\mj$, the data tend to be above the predictions for $\mj\gtrsim
70$ GeV. It was reported~\cite{hep-ph:9910448} that calculations
of jet cross sections in NC DIS computed using the {\sc Mepjet}
program differ by $5-8\%$ from the results from other NLO
programs. Comparisons of inclusive-jet calculations for NC DIS in the
kinematic range of the measurements presented here performed using
{\sc Mepjet} and {\sc Disent}~\cite{np:b485:291} showed an agreement
better than $1\%$. However, similar comparisons for dijet cross
sections showed relative differences above $\sim 5\%$. For CC DIS, it
is not possible to quantify the degree of accuracy of the
calculations of {\sc Mepjet} since no alternative program exists.
The NLO predictions give a reasonable description of the ratios of the
cross sections for $e^-p$ and $e^+p$ interactions (see
Fig.~\ref{fig11}). New implementations of the theory are crucially
needed to use the differential dijet cross sections presented here in
global fits to extract the proton PDFs.

The integrated unpolarized dijet cross sections are shown in
Table~\ref{tabtwo}. The measured cross sections are larger than the
predictions of NLO QCD.

\subsection{Measurements of three-jet cross sections and observation
  of four-jet events}

Differential three-jet cross sections were measured in the kinematic
regime $\q2>200$ \g2\ and $y<0.9$. The cross sections were determined
for jets with $\etj>14$ GeV, $\etjj>5$ GeV, $\etjjj>5$ GeV and
$-1<\etajet<2.5$. Three-jet cross sections in CC DIS were measured for
the first time in $e^{\pm}p$ collisions.

\begin{figure}[p]
\vfill
\setlength{\unitlength}{1.0cm}
\begin{picture} (18.0,12.0)
\put (0.0,5.0){\centerline{\epsfig{figure=\figdir zeus.eps,width=10cm}}}
\put (0.0,6.5){\epsfig{figure=\figdir 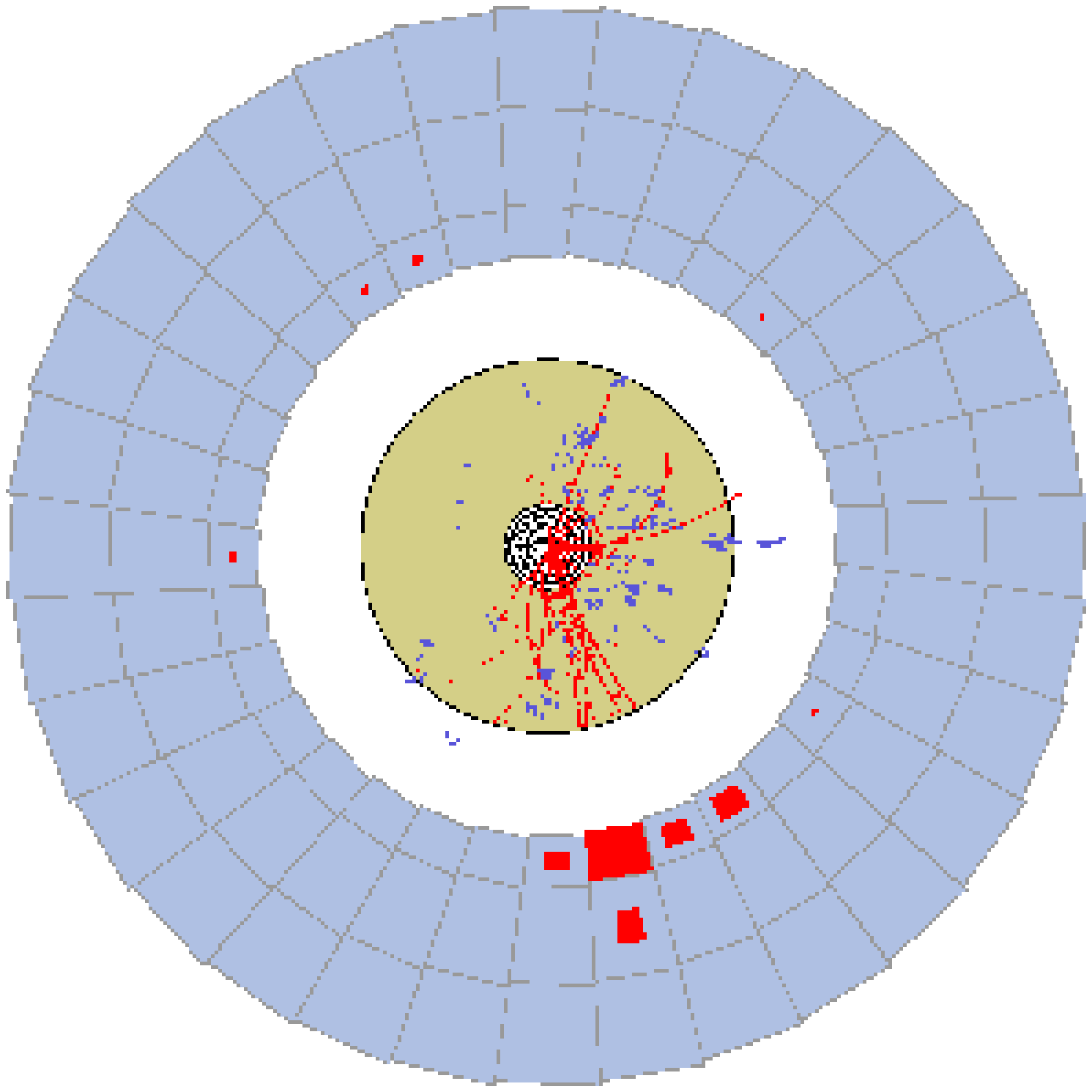,width=5.7cm}}
\put (6.0,6.5){\epsfig{figure=\figdir 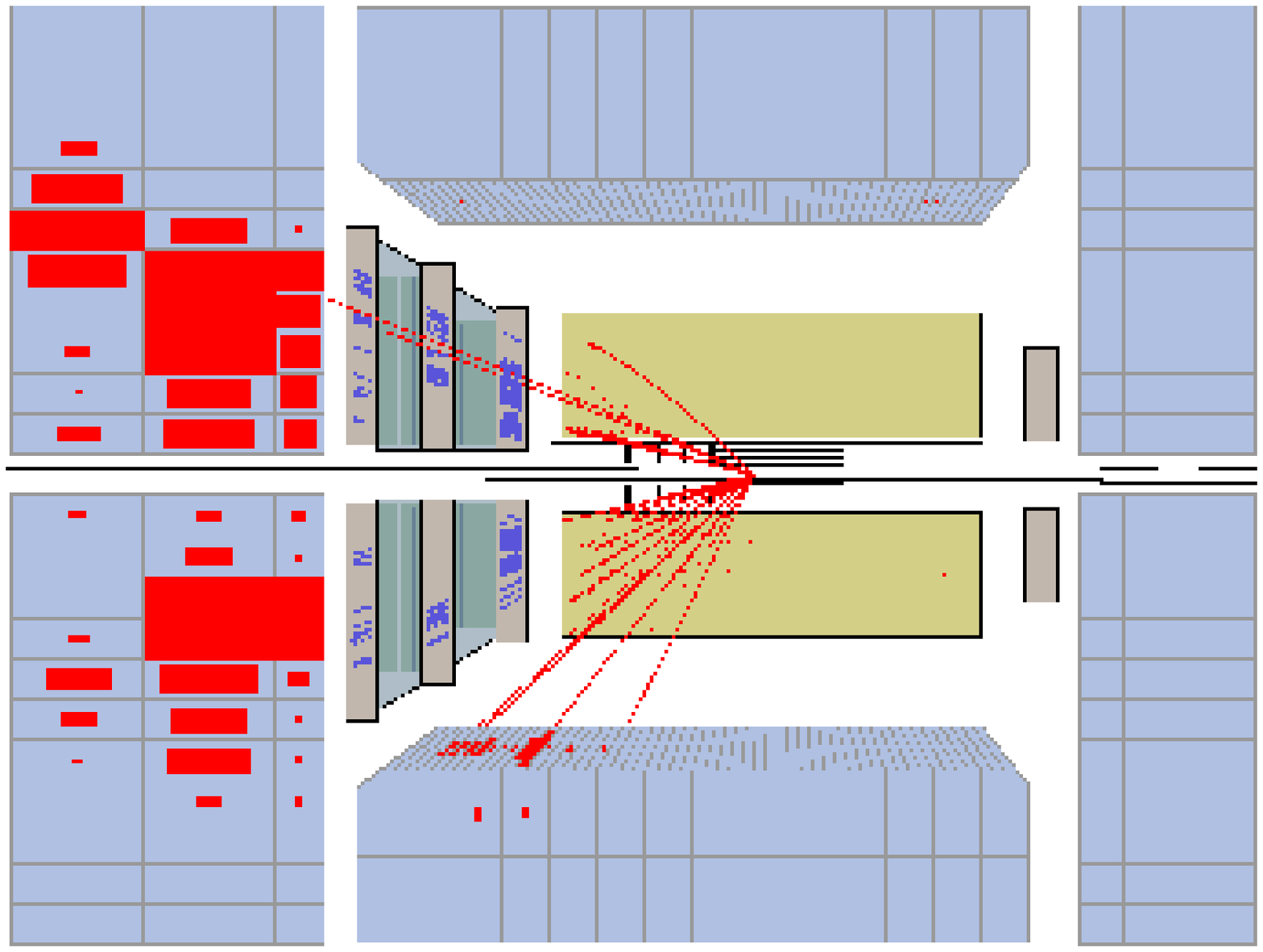,width=9cm,height=5.7cm}}
\put (0.0,0.0){\centerline{\epsfig{figure=\figdir 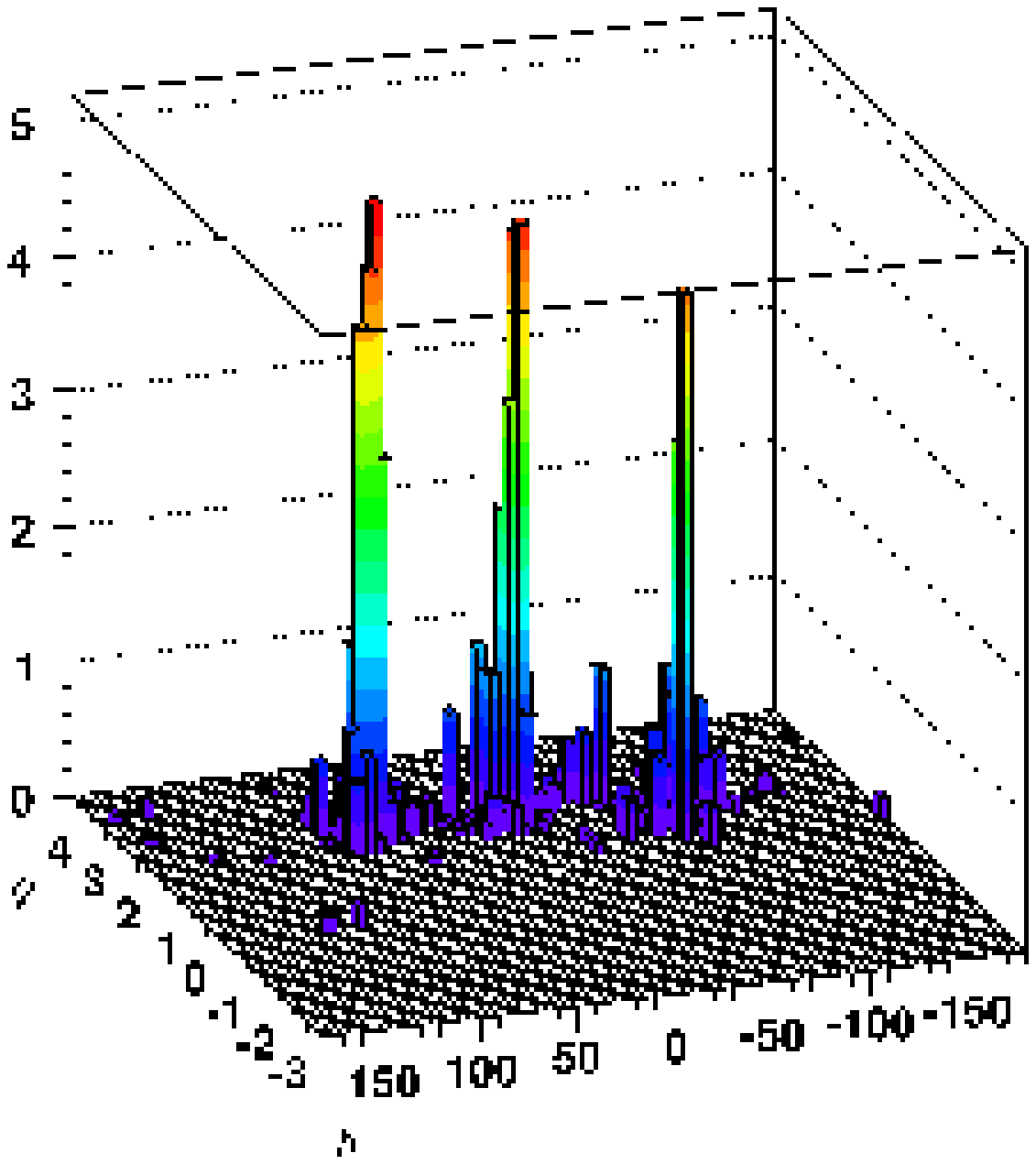,width=9cm,height=5.7cm}}}
\put (6.5,4.5){\bf\small jet 1}
\put (7.5,4.3){\bf\small jet 2}
\put (8.8,4.0){\bf\small jet 3}
\end{picture}
\caption
{\it 
Three-jet candidate event in CC DIS in the ZEUS detector. The energy
deposition in the CAL is proportional to the size and density of
shading in the CAL cells. The lego plots show the CAL transverse energy
deposition projected in the $\etaphi$ plane. In the $X-Y$ view, only
the energy deposition in the barrel calorimeter is shown.
}
\label{fig12}
\vfill
\end{figure}

\begin{figure}[p]
\vfill
\setlength{\unitlength}{1.0cm}
\begin{picture} (18.0,12.0)
\put (0.0,5.0){\centerline{\epsfig{figure=\figdir zeus.eps,width=10cm}}}
\put (0.0,6.5){\epsfig{figure=\figdir 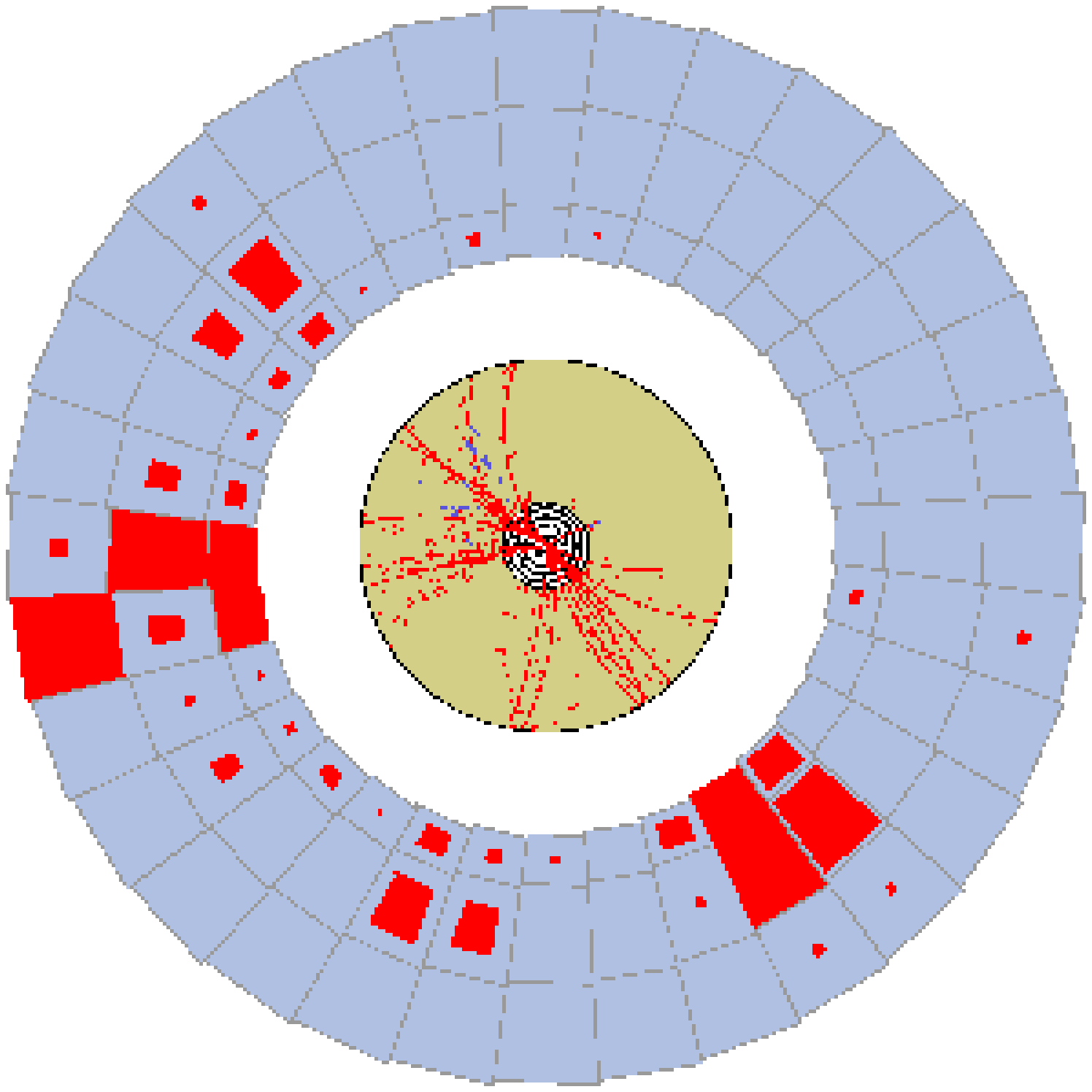,width=5.7cm}}
\put (6.0,6.5){\epsfig{figure=\figdir 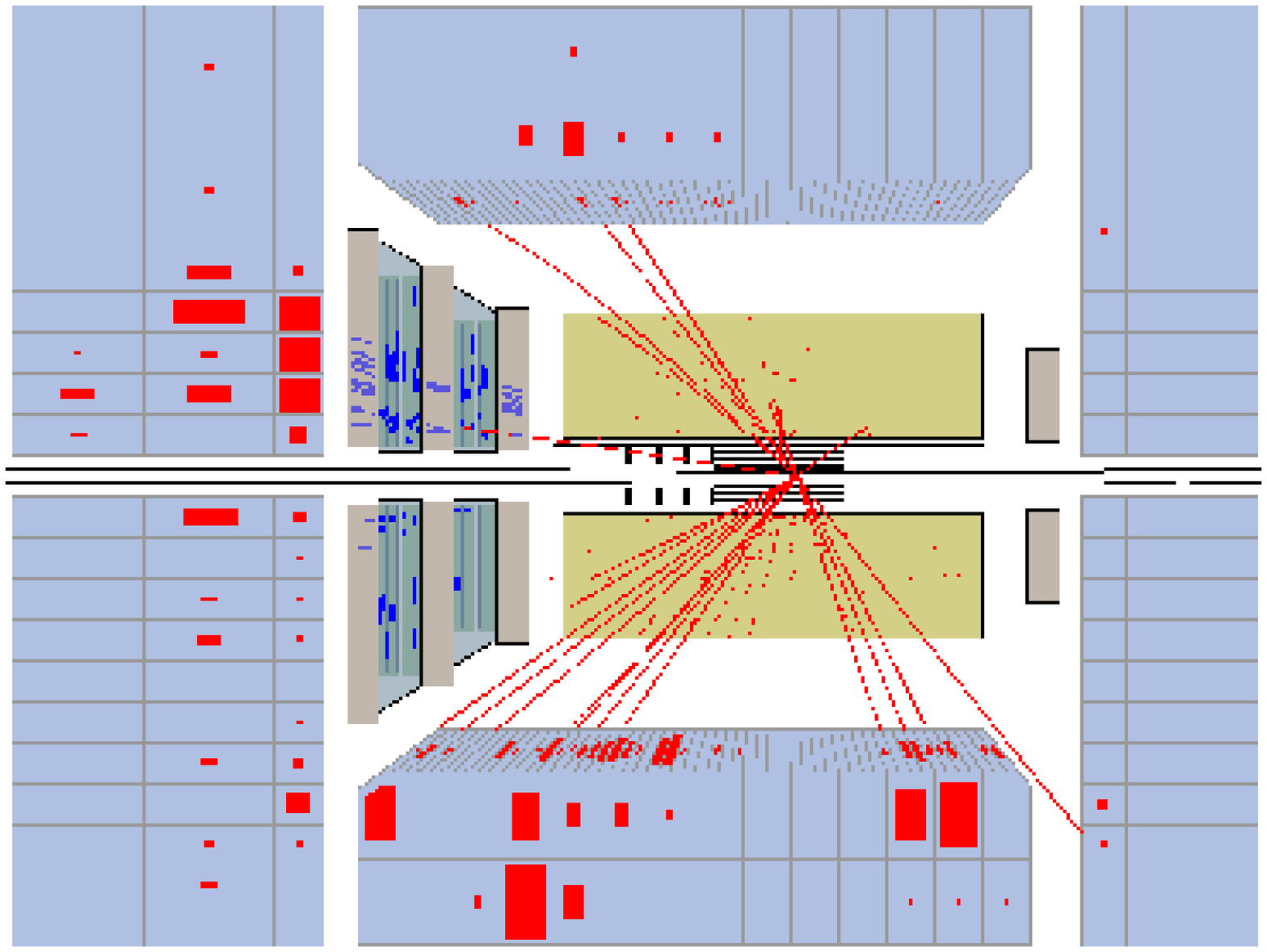,width=9cm,height=5.7cm}}
\put (0.0,0.0){\centerline{\epsfig{figure=\figdir 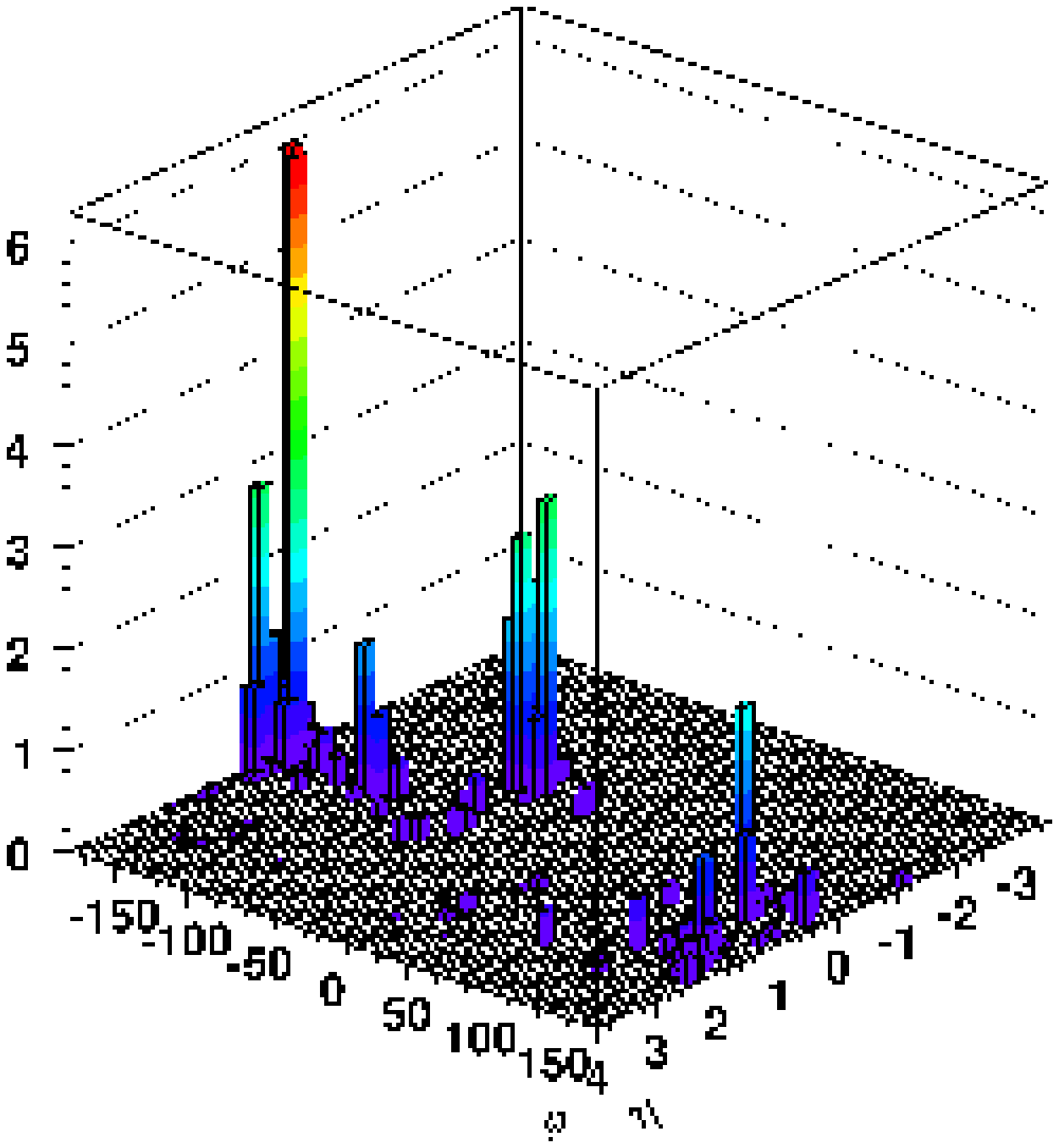,width=9cm,height=5.7cm}}}
\put (5.7,4.8){\bf\small jet 1}
\put (7.5,3.1){\bf\small jet 2}
\put (9.0,2.2){\bf\small jet 3}
\put (6.4,2.5){\bf\small jet 4}
\end{picture}
\caption
{\it 
Four-jet candidate event in CC DIS in the ZEUS detector. The energy
deposition in the CAL is proportional to the size and density of
shading in the CAL cells. The lego plots show the CAL transverse energy
deposition projected in the $\etaphi$ plane. In the $X-Y$ view, only
the energy deposition in the barrel calorimeter is shown.
}
\label{fig13}
\vfill
\end{figure}

Figure~\ref{fig12} shows a three-jet candidate event in the ZEUS
detector: a clear three-jet topology and large transverse momentum are
observed. The three-jet selected sample also contains 9 $e^-p$ and
2 $e^+p$ candidates with a fourth jet of transverse energy above 5 GeV
in the $\etajet$ range considered. One of these candidates is
displayed in Fig.~\ref{fig13}: the fourth jet is clearly observed in
the ZEUS detector.

\begin{figure}[p]
\vfill
\setlength{\unitlength}{1.0cm}
\begin{picture} (18.0,17.0)
\put (0.0,11.0){\centerline{\epsfig{figure=\figdir zeus.eps,width=10cm}}}
\put (-2.0,8.5){\epsfig{figure=\figdir 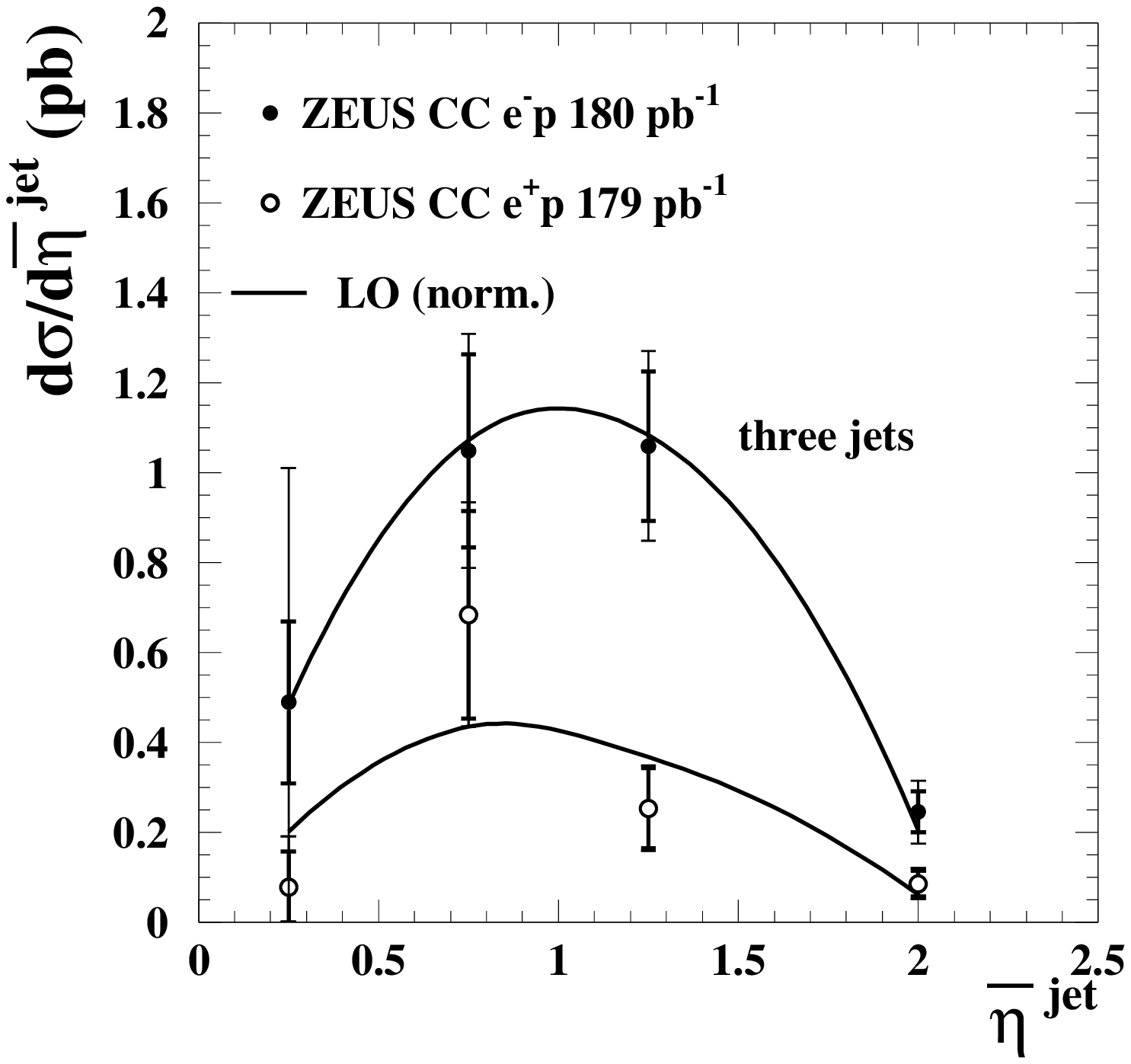,width=12cm}}
\put (7.0,8.5){\epsfig{figure=\figdir 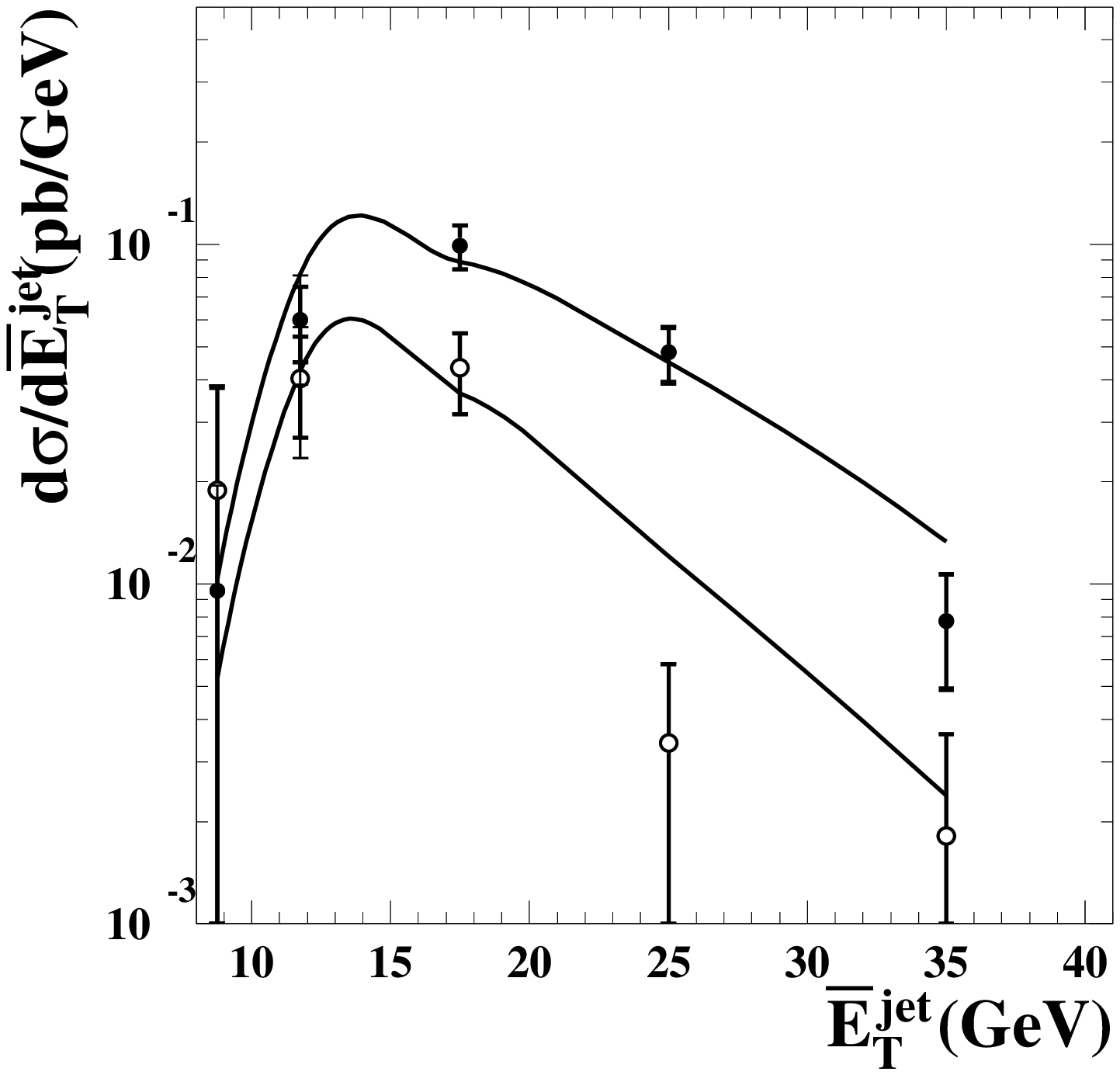,width=12cm}}
\put (-2.0,-1.5){\epsfig{figure=\figdir 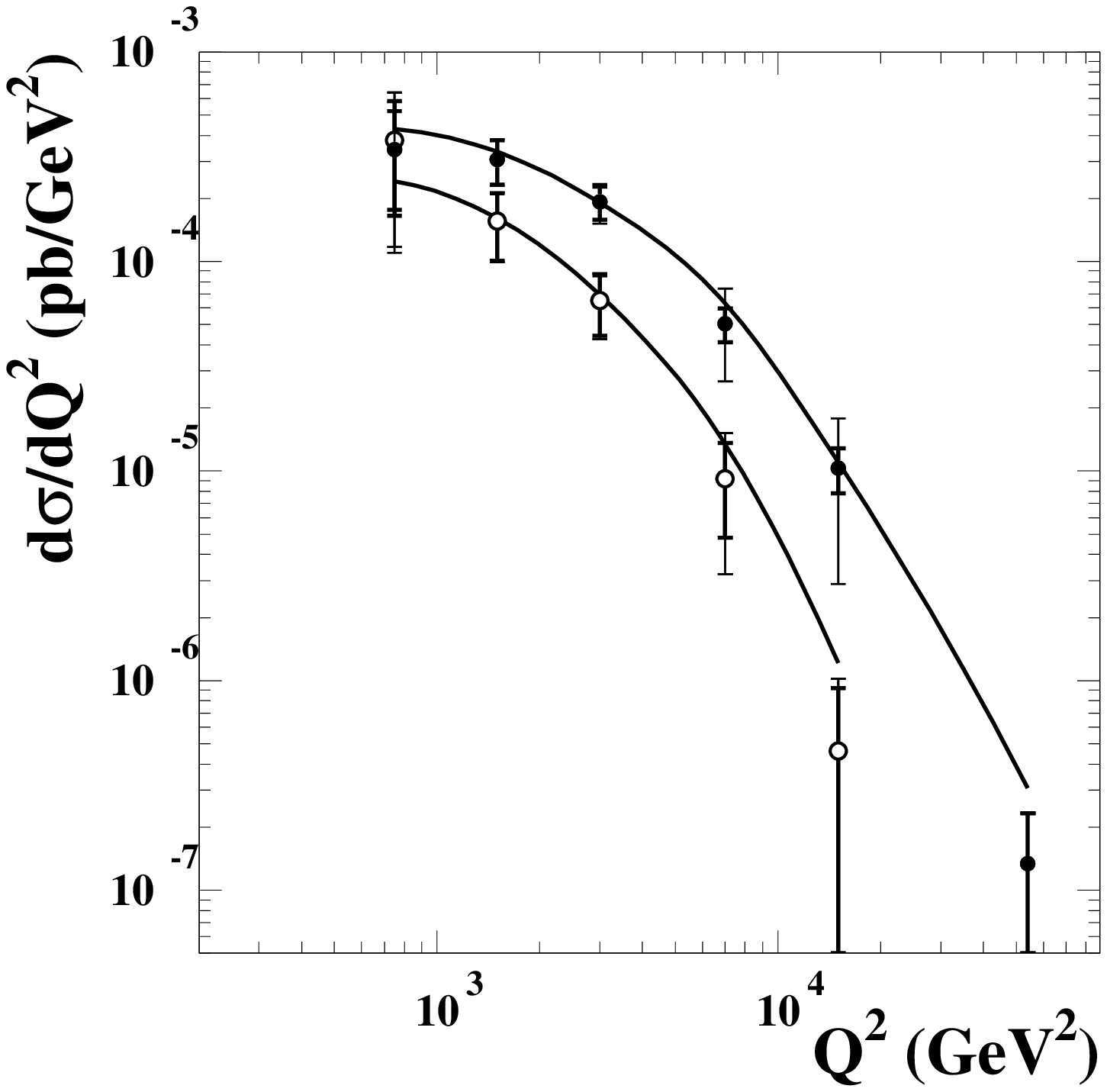,width=12cm}}
\put (7.0,-1.5){\epsfig{figure=\figdir 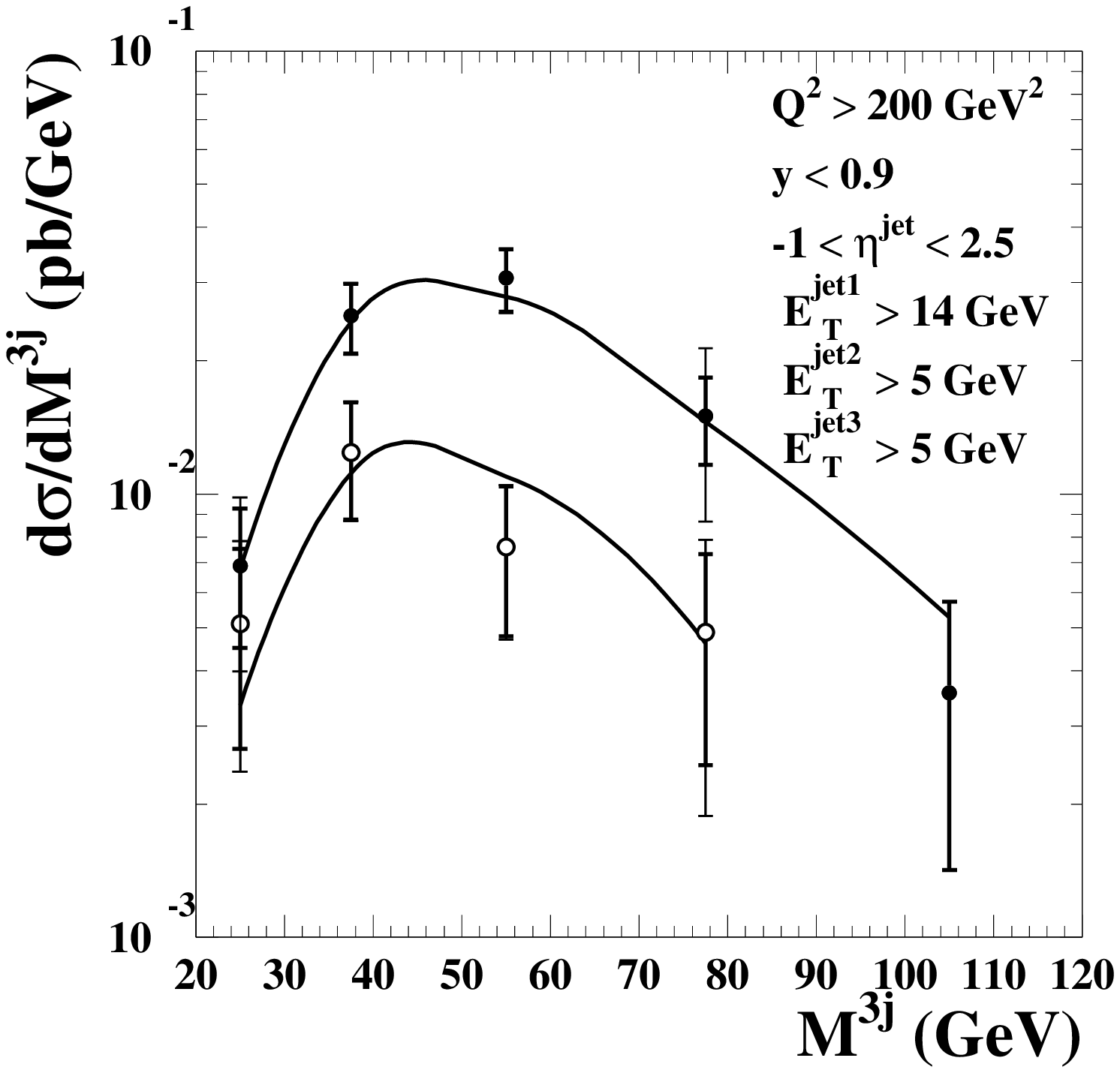,width=12cm}}
\put (6.7,17.1){\bf\small (a)}
\put (15.7,17.2){\bf\small (b)}
\put (6.7,7.2){\bf\small (c)}
\put (15.7,7.0){\bf\small (d)}
\end{picture}
\caption
{\it 
Measured unpolarized three-jet cross sections in CC DIS for
jets with $\etj>14$ GeV, $\etjj>5$ GeV, $\etjjj>5$ GeV and
$-1<\etajet<2.5$ in the kinematic regime given by $\q2>200$~\g2\ and
$y<0.9$ as functions of (a) $\etabar$, (b)
$\etbar$, (c) $\q2$ and (d) $\m3j$ in $e^-p$
(dots) and $e^+p$ (open circles) collisions. For comparison, the
predictions of LO QCD based on the {\sc Mepjet} calculations using
the ZEUS PDF sets (solid lines) are also shown. The predicted cross
sections are normalized to the measured three-jet cross sections
($\times 1.92$ for $e^-p$ and $\times 1.42$ for $e^+p$).
Other details as in the caption to Fig.~\ref{fig7}.
}
\label{fig14}
\vfill
\end{figure}

Figure~\ref{fig14} and Tables~\ref{tabseven} to \ref{tabtwelve},
\ref{tabfifteen} and \ref{tabsixteen} show the unpolarized three-jet 
differential cross sections as functions of $\etabar$,
$\etbar$, $\q2$ and the three-jet invariant mass, $\m3j$,
where $\etabar=(\etaj+\etajj+\etajjj)/3$ and
$\etbar=(\etj+\etjj+\etjjj)/3$ in CC $e^{\pm}p$ DIS. Values
of $\m3j$ above $100$ GeV are accessible with the present statistics.

\begin{figure}[p]
\vfill
\setlength{\unitlength}{1.0cm}
\begin{picture} (18.0,17.0)
\put (0.0,11.0){\centerline{\epsfig{figure=\figdir zeus.eps,width=10cm}}}
\put (-2.0,8.5){\epsfig{figure=\figdir 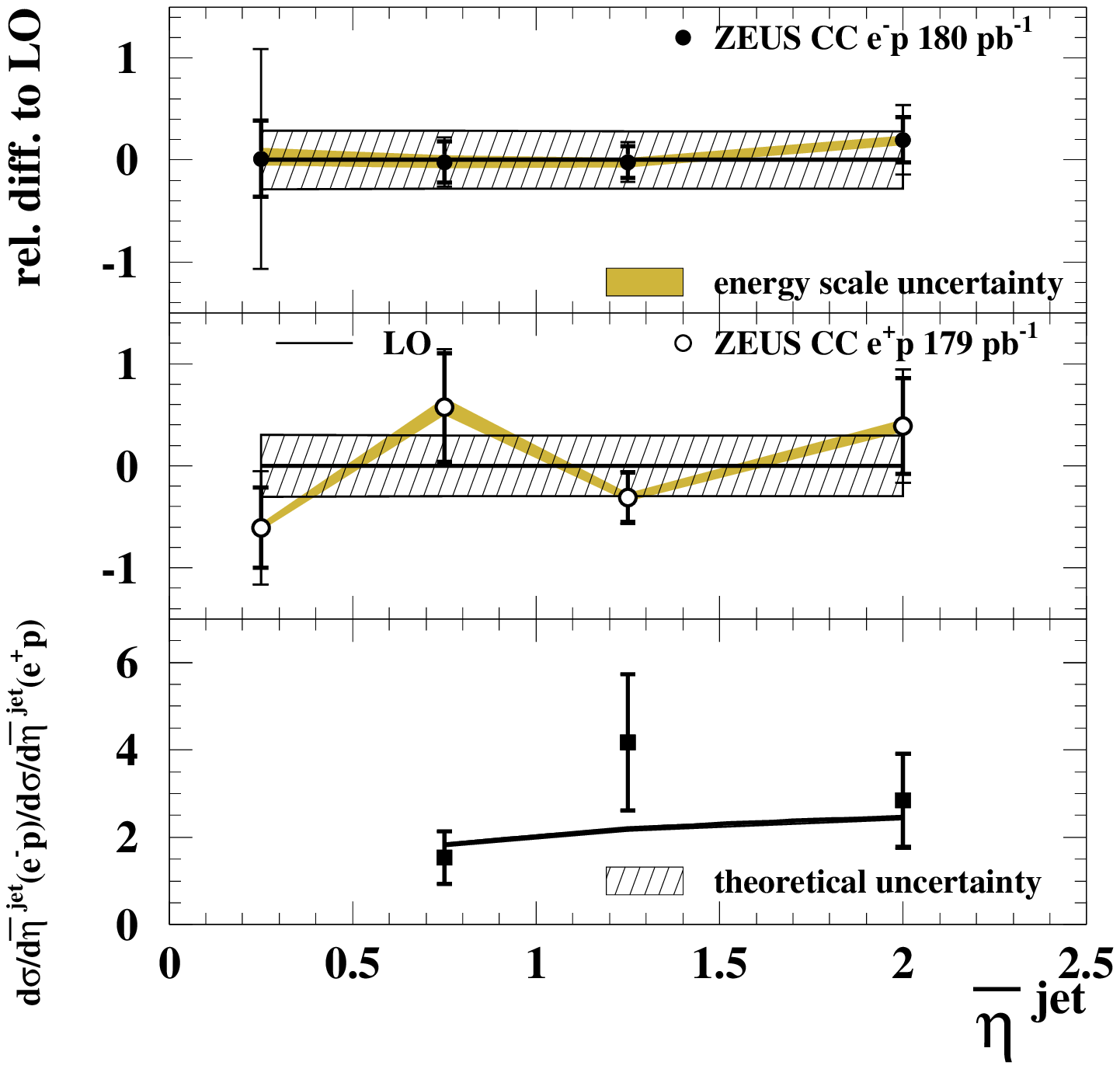,width=12cm}}
\put (7.0,8.5){\epsfig{figure=\figdir 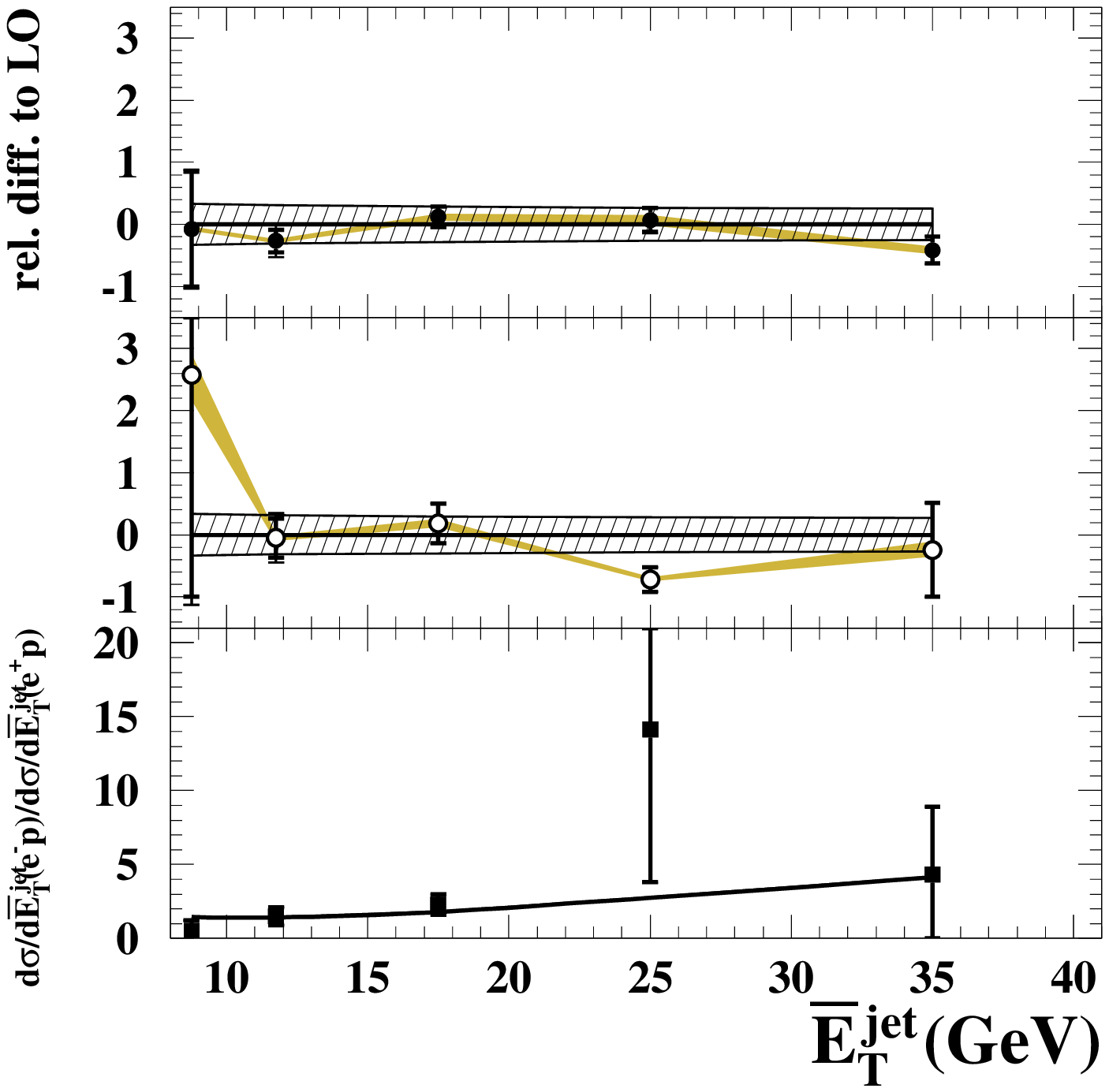,width=12cm}}
\put (-2.0,-1.5){\epsfig{figure=\figdir 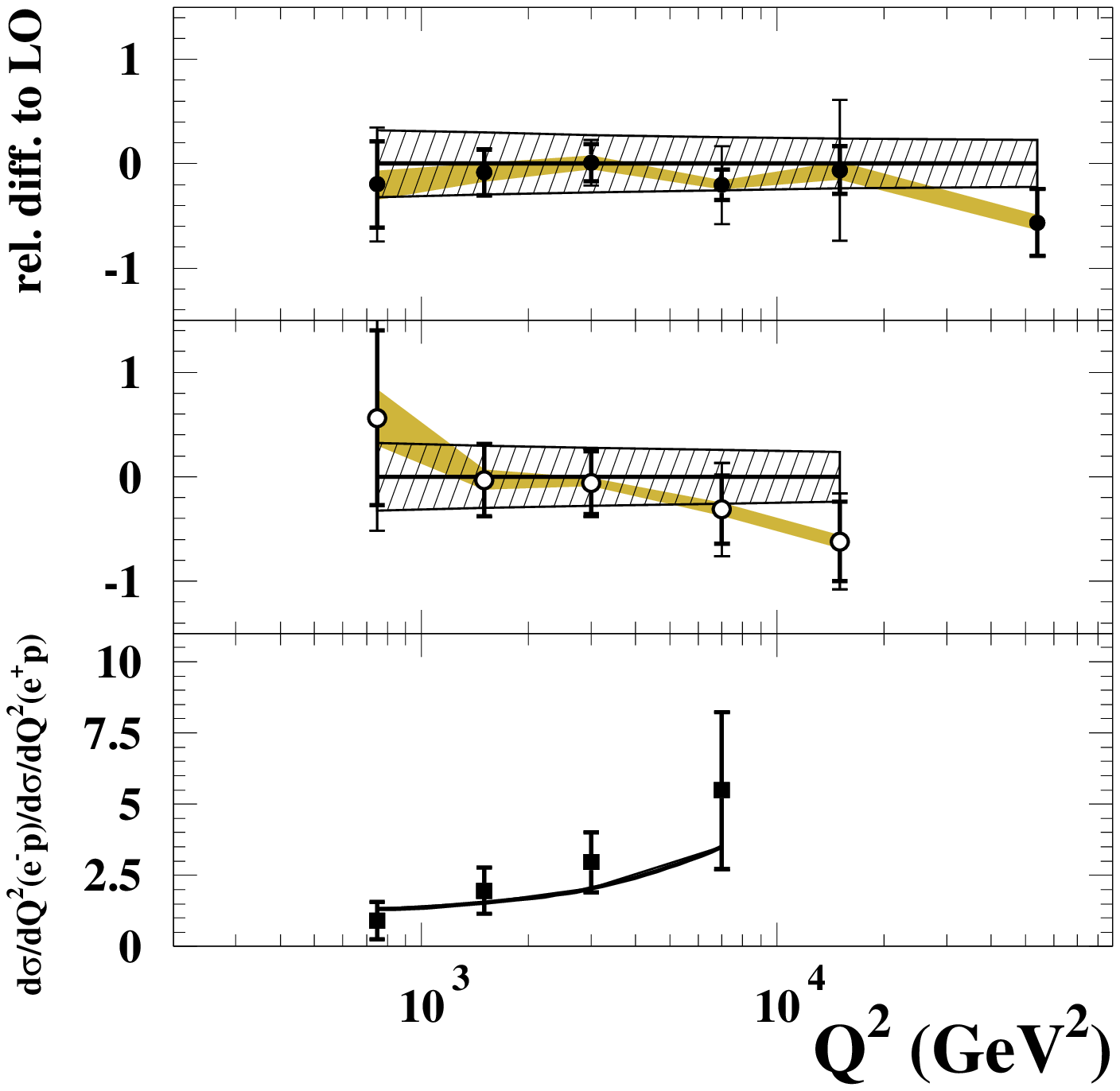,width=12cm}}
\put (7.0,-1.5){\epsfig{figure=\figdir 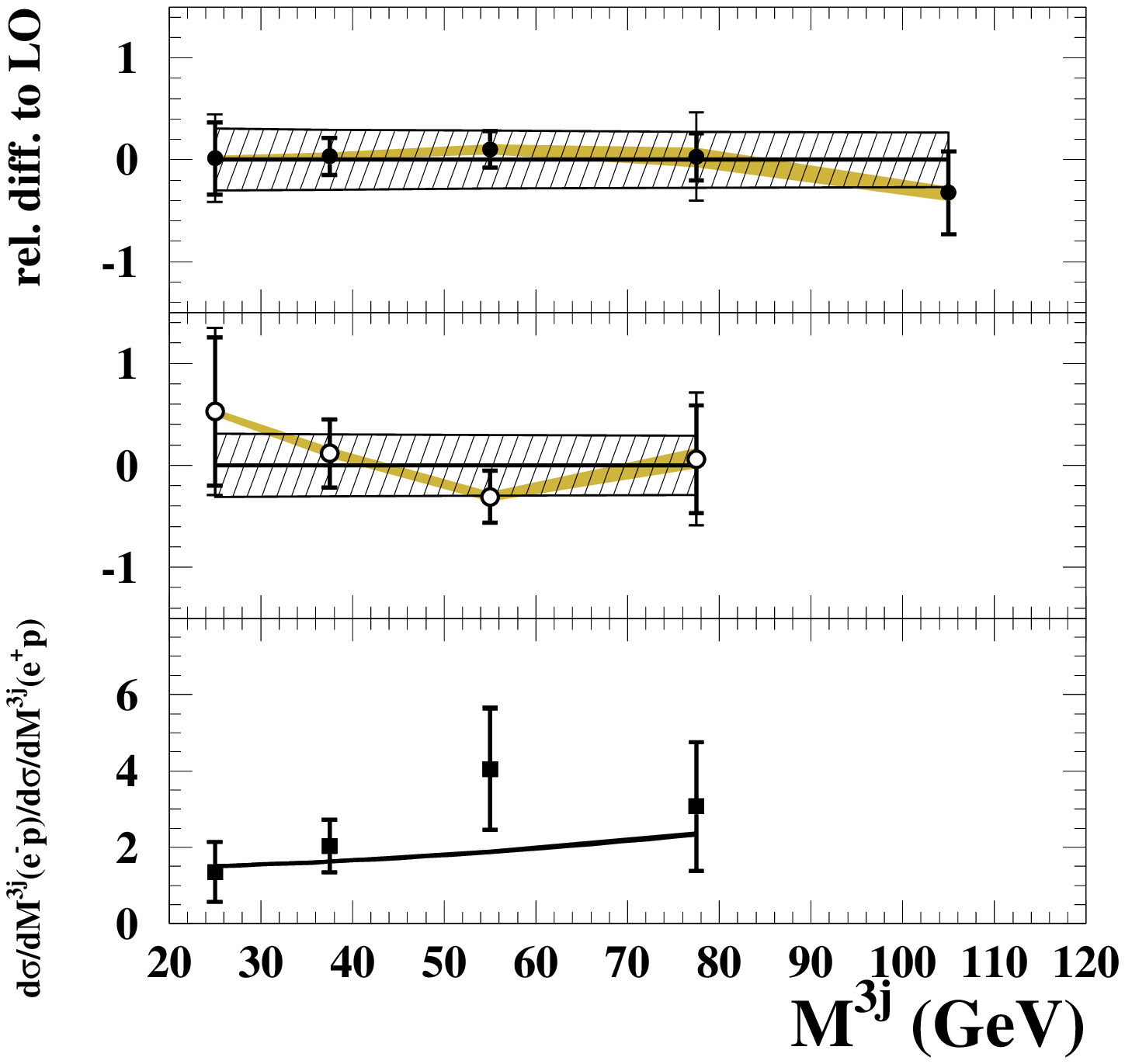,width=12cm}}
\put (6.7,17.2){\bf\small (a)}
\put (15.7,17.2){\bf\small (b)}
\put (0.7,7.2){\bf\small (c)}
\put (15.7,7.2){\bf\small (d)}
\end{picture}
\caption
{\it 
Relative difference between the measured cross sections of
Fig.~\ref{fig14} and the corresponding LO QCD predictions in $e^-p$
(dots) and $e^+p$ (open circles) collisions as functions of (a)
$\etabar$, (b) $\etbar$, (c) $\q2$ and (d)
$\m3j$. The lower parts of the figures display the ratio of the cross
sections for $e^-p$ and $e^+p$ collisions (squares). Other details as
in the caption to Fig.~\ref{fig8}.
}
\label{fig15}
\vfill
\end{figure}

The predictions of LO QCD are compared to the data in
Fig.~\ref{fig14}. The currently available QCD calculations are only
lowest order and cannot predict the normalization of the
data. Therefore, they were scaled by 1.92 and 1.42 for $e^-p$ and
$e^+p$ collisions, respectively, so as to reproduce the measured
integrated three-jet cross section. The scaled LO calculations give a
good description of the shape of the data. Figure~\ref{fig15} shows the 
relative difference between the data and the scaled predictions. The
lower part of Fig.~\ref{fig15} shows the ratio of the differential
cross sections for the $e^-p$ and $e^+p$ samples.

The integrated unpolarized three-jet cross sections are shown in
Table~\ref{tabtwo}. The predictions of LO QCD are also shown in the
table.

\section{Summary}

Measurements of polarized and unpolarized integrated and
differential multi-jet cross sections in CC $e^{\pm}p$ DIS
were made using $0.36$~\fb1\ of data collected with the ZEUS
detector at HERA II. The measurements were made in
the kinematic region defined by $\q2>200$~\g2\ and $y<0.9$.
Jets were identified in the laboratory frame using the $\kt$ cluster
algorithm in the longitudinally invariant inclusive mode. 

Polarized inclusive-jet cross sections were measured integrated over
the phase-space region considered and differentially as functions of
$\etajet$, $\etjet$, $\q2$ and $x$ for jets with $\etjet>14$
GeV and $-1<\etajet<2.5$. The measured cross sections are in good
agreement with the SM predictions. The ratios of the differential
cross sections for negative and positive longitudinally-polarized
lepton beams are also well described by the predictions.

Unpolarized differential inclusive-jet cross sections were
measured as functions of $\etajet$, $\etjet$, $\q2$ and $x$. The
ratio of the differential cross sections for $e^-p$ and $e^+p$
collisions as a function of $\etajet$ is $\approx 2$ in the $\etajet$
range measured, as predicted by the theory. The ratio as a function of
$\etjet$ ($\q2$) increases as $\etjet$ ($\q2$) increases, in agreement
with the expected increased contribution from the valence-quark
densities in the proton at high $x$ and the fact that both reactions
are sensitive to different quark flavors. Dijet differential cross
sections were measured for jets with $\etj>14$ GeV, $\etjj>5$ GeV and
$-1<\etajet<2.5$.

Next-to-leading-order QCD predictions computed using the program {\sc
  Mepjet} were compared to the data. The NLO QCD predictions give
a good description of the measured inclusive-jet cross sections. A
detailed study of the theoretical uncertainties was performed:
they are dominated by the contribution from the PDFs. Furthermore, the
uncertainties due to the PDFs are larger for $e^+p$ than for $e^-p$
collisions. Therefore, these measurements, if used together with other
data in global fits, have the potential to constrain the flavor
content of the proton at high $x$.

The comparison of the predictions with the measured dijet differential
cross sections shows a poor agreement in shape and
normalization. Improved implementations of the theory are crucially
needed to use these dijet measurements in a global fit to constrain the
proton PDFs.

Three-jet differential cross sections were measured for the first time
in $e^{\pm}p$ collisions for jets with $\etj>14$ GeV, $\etjj>5$ GeV,
$\etjjj>5$ GeV and $-1<\etajet<2.5$. The leading-order QCD predictions
give a good description of the shape of the data. The three-jet sample
also contains eleven candidates with a fourth jet of $\etjet>5$ GeV in
the $\etajet$ range considered.

\vspace{0.5cm}
\noindent {\Large\bf Acknowledgements}
\vspace{0.3cm}

We are grateful to the DESY directorate for their strong support and
encouragement. We thank the HERA machine group whose outstanding
efforts resulted in a successful upgrade of the HERA accelerator which
made this work possible. We also thank the HERA polarimeter group for
providing the measurements of the lepton-beam polarization. The
design, construction and installation of the ZEUS detector were
made possible by the efforts of many people not listed as authors.

\vfill\eject

\providecommand{\etal}{et al.\xspace}
\providecommand{\coll}{Collaboration}
\catcode`\@=11
\def\@bibitem#1{%
\ifmc@bstsupport
  \mc@iftail{#1}%
    {;\newline\ignorespaces}%
    {\ifmc@first\else.\fi\orig@bibitem{#1}}
  \mc@firstfalse
\else
  \mc@iftail{#1}%
    {\ignorespaces}%
    {\orig@bibitem{#1}}%
\fi}%
\catcode`\@=12

\clearpage
\newpage
\begin{table}
\begin{center}
    \begin{tabular}{||c|cccc||}
\hline
  $\etajet$ bin
& $d\sigma/d\etajet$ (pb)
& $\delta_{\rm stat}$
& $\delta_{\rm syst}$
& $\delta_{\rm ES}$\\
\hline
\multicolumn{5}{||c||}{$P_{e^{-}}=-0.27$ - inclusive jets} \\
\hline
$-1.0 , -0.5$ & $ 4.22$ & $\pm 0.38$ & $\pm 0.32$ & $_{-0.16}^{+0.17}$ \\
$-0.5 ,  0.0$ & $13.35$ & $\pm 0.62$ & $\pm 0.51$ & $_{-0.29}^{+0.30}$ \\
$ 0.0 ,  0.5$ & $23.40$ & $\pm 0.77$ & $\pm 0.29$ & $_{-0.22}^{+0.23}$ \\
$ 0.5 ,  1.0$ & $27.77$ & $\pm 0.81$ & $\pm 0.82$ & $\pm 0.09$ \\
$ 1.0 ,  1.5$ & $28.30$ & $\pm 0.82$ & $\pm 0.92$ & $_{-0.09}^{+0.08}$ \\
$ 1.5 ,  2.0$ & $24.71$ & $\pm 0.79$ & $\pm 0.40$ & $_{-0.09}^{+0.10}$ \\
$ 2.0 ,  2.5$ & $19.41$ & $\pm 0.87$ & $\pm 2.36$ & $\pm 0.15$ \\
\hline
\multicolumn{5}{||c||}{$P_{e^{-}}=+0.30$ - inclusive jets} \\
\hline
$-1.0 , -0.5$ & $ 2.81$ & $\pm 0.37$ & $\pm 0.21$ & $_{-0.10}^{+0.11}$ \\
$-0.5 ,  0.0$ & $ 7.22$ & $\pm 0.55$ & $\pm 0.28$ & $\pm 0.16$ \\
$ 0.0 ,  0.5$ & $12.45$ & $\pm 0.67$ & $\pm 0.47$ & $_{-0.11}^{+0.12}$ \\
$ 0.5 ,  1.0$ & $16.05$ & $\pm 0.74$ & $\pm 0.48$ & $\pm 0.05$ \\
$ 1.0 ,  1.5$ & $15.99$ & $\pm 0.74$ & $\pm 0.50$ & $\pm 0.05$ \\
$ 1.5 ,  2.0$ & $15.34$ & $\pm 0.75$ & $\pm 0.42$ & $_{-0.05}^{+0.06}$ \\
$ 2.0 ,  2.5$ & $11.39$ & $\pm 0.80$ & $\pm 1.37$ & $\pm 0.09$ \\
\hline
\multicolumn{5}{||c||}{$P_{e^{+}}=-0.37$ - inclusive jets} \\
\hline
$-1.0 , -0.5$ & $ 1.26$ & $\pm 0.28$ & $\pm 0.14$ & $\pm 0.04$ \\
$-0.5 ,  0.0$ & $ 3.33$ & $\pm 0.38$ & $\pm 0.16$ & $\pm 0.05$ \\
$ 0.0 ,  0.5$ & $ 5.68$ & $\pm 0.47$ & $\pm 0.14$ & $\pm 0.04$ \\
$ 0.5 ,  1.0$ & $ 6.64$ & $\pm 0.49$ & $\pm 0.40$ & $\pm 0.04$ \\
$ 1.0 ,  1.5$ & $ 6.74$ & $\pm 0.49$ & $\pm 0.24$ & $\pm 0.03$ \\
$ 1.5 ,  2.0$ & $ 6.52$ & $\pm 0.50$ & $\pm 0.07$ & $\pm 0.04$ \\
$ 2.0 ,  2.5$ & $ 5.09$ & $\pm 0.55$ & $\pm 0.93$ & $\pm 0.06$ \\
\hline
\multicolumn{5}{||c||}{$P_{e^{+}}=+0.32$ - inclusive jets} \\
\hline
$-1.0 , -0.5$ & $ 2.92$ & $\pm 0.36$ & $\pm 0.30$ & $\pm 0.09$ \\
$-0.5 ,  0.0$ & $ 7.37$ & $\pm 0.49$ & $\pm 0.29$ & $\pm 0.11$ \\
$ 0.0 ,  0.5$ & $10.34$ & $\pm 0.54$ & $\pm 0.41$ & $\pm 0.08$ \\
$ 0.5 ,  1.0$ & $14.75$ & $\pm 0.63$ & $\pm 0.65$ & $\pm 0.08$ \\
$ 1.0 ,  1.5$ & $12.38$ & $\pm 0.58$ & $\pm 0.65$ & $\pm 0.06$ \\
$ 1.5 ,  2.0$ & $11.64$ & $\pm 0.58$ & $\pm 0.50$ & $_{-0.06}^{+0.07}$ \\
$ 2.0 ,  2.5$ & $10.22$ & $\pm 0.68$ & $\pm 1.26$ & $\pm 0.12$ \\
\hline
    \end{tabular}
 \caption{
   Differential polarized inclusive-jet cross-sections
   $d\sigma/d\etajet$ for jets of hadrons in the laboratory frame
   selected with the longitudinally invariant $\kt$ cluster
   algorithm. The statistical, uncorrelated systematic and
   energy-scale ({\rm ES}) uncertainties are shown separately.}
 \label{tabthree}
\end{center}
\end{table}

\clearpage
\newpage
\begin{table}
\begin{center}
    \begin{tabular}{||c|cccc||}
\hline
  $\etjet$ bin (GeV)
& $d\sigma/d\etjet$ (pb/GeV)
& $\delta_{\rm stat}$
& $\delta_{\rm syst}$
& $\delta_{\rm ES}$\\
\hline
\multicolumn{5}{||c||}{$P_{e^{-}}=-0.27$ - inclusive jets} \\
\hline
$ 14, 21$ & $2.318$ & $\pm 0.076$ & $\pm 0.140$ & $_{-0.051}^{+0.045}$ \\
$ 21, 29$ & $1.776$ & $\pm 0.056$ & $\pm 0.079$ & $_{-0.018}^{+0.019}$ \\
$ 29, 41$ & $1.419$ & $\pm 0.038$ & $\pm 0.070$ & $_{-0.010}^{+0.009}$ \\
$ 41, 55$ & $0.834$ & $\pm 0.027$ & $\pm 0.051$ & $_{-0.010}^{+0.011}$ \\
$ 55, 71$ & $0.450$ & $\pm 0.018$ & $\pm 0.022$ & $\pm 0.011$ \\
$ 71, 87$ & $0.212$ & $\pm 0.012$ & $\pm 0.057$ & $_{-0.002}^{+0.003}$ \\
$ 87, 120$ & $0.0421$ & $\pm 0.0035$ & $\pm 0.0225$ & $_{-0.0050}^{+0.0063}$ \\
\hline
\multicolumn{5}{||c||}{$P_{e^{-}}=+0.30$ - inclusive jets} \\
\hline
$ 14, 21$ & $1.284$ & $\pm 0.068$ & $\pm 0.089$ & $_{-0.028}^{+0.025}$ \\
$ 21, 29$ & $1.081$ & $\pm 0.052$ & $\pm 0.049$ & $_{-0.011}^{+0.012}$ \\
$ 29, 41$ & $0.795$ & $\pm 0.035$ & $\pm 0.041$ & $_{-0.006}^{+0.005}$ \\
$ 41, 55$ & $0.486$ & $\pm 0.024$ & $\pm 0.030$ & $\pm 0.006$ \\
$ 55, 71$ & $0.263$ & $\pm 0.017$ & $\pm 0.016$ & $_{-0.006}^{+0.007}$ \\
$ 71, 87$ & $0.106$ & $\pm 0.010$ & $\pm 0.029$ & $\pm 0.001$ \\
$ 87, 120$ & $0.0276$ & $\pm 0.0035$ & $\pm 0.0149$ & $_{-0.0033}^{+0.0041}$ \\
\hline
\multicolumn{5}{||c||}{$P_{e^{+}}=-0.37$ - inclusive jets} \\
\hline
$ 14,  21$ & $0.761$ & $\pm 0.053$ & $\pm 0.058$ & $\pm 0.012$ \\
$ 21,  29$ & $0.584$ & $\pm 0.039$ & $\pm 0.031$ & $_{-0.004}^{+0.003}$ \\
$ 29,  41$ & $0.338$ & $\pm 0.023$ & $\pm 0.019$ & $_{-0.003}^{+0.004}$ \\
$ 41,  55$ & $0.160$ & $\pm 0.014$ & $\pm 0.011$ & $\pm 0.004$ \\
$ 55,  71$ & $0.0602$ & $\pm 0.0079$ & $\pm 0.0037$ & $_{-0.0023}^{+0.0025}$ \\
$ 71,  87$ & $0.0157$ & $\pm 0.0039$ & $\pm 0.0047$ & $_{-0.0006}^{+0.0007}$ \\
$ 87, 120$ & $0.00166$ & $\pm 0.00083$ & $\pm 0.00106$ & $_{-0.00025}^{+0.00032}$ \\
\hline
\multicolumn{5}{||c||}{$P_{e^{+}}=+0.32$ - inclusive jets} \\
\hline
$ 14,  21$ & $1.576$ & $\pm 0.066$ & $\pm 0.112$ & $_{-0.024}^{+0.026}$ \\
$ 21,  29$ & $1.077$ & $\pm 0.045$ & $\pm 0.020$ & $_{-0.007}^{+0.006}$ \\
$ 29,  41$ & $0.675$ & $\pm 0.028$ & $\pm 0.037$ & $_{-0.006}^{+0.007}$ \\
$ 41,  55$ & $0.307$ & $\pm 0.017$ & $\pm 0.020$ & $\pm 0.007$ \\
$ 55,  71$ & $0.1135$ & $\pm 0.0094$ & $\pm 0.0143$ & $_{-0.0043}^{+0.0047}$ \\
$ 71,  87$ & $0.0374$ & $\pm 0.0052$ & $\pm 0.0083$ & $_{-0.0014}^{+0.0016}$ \\
$ 87, 120$ & $0.0050$ & $\pm 0.0012$ & $\pm 0.0027$ & $_{-0.0008}^{+0.0010}$ \\
\hline
    \end{tabular}
 \caption{
   Differential polarized inclusive-jet cross-sections
   $d\sigma/d\etjet$. Other details as in the caption to
   Table~\ref{tabthree}.}
 \label{tabfour}
\end{center}
\end{table}

\clearpage
\newpage
\begin{table}
\begin{center}
    \begin{tabular}{||c|cccc||}
\hline
  $\q2$ bin (GeV$^2$)
& $d\sigma/d\q2$ (pb/GeV$^2$)
& $\delta_{\rm stat}$
& $\delta_{\rm syst}$
& $\delta_{\rm ES}$\\
\hline
\multicolumn{5}{||c||}{$P_{e^{-}}=-0.27$ - inclusive jets} \\
\hline
$  200,   500$ & $0.0246$ & $\pm 0.0013$ & $\pm 0.0015$ & $_{-0.0023}^{+0.0024}$ \\
$  500,  1000$ & $0.02226$ & $\pm 0.00079$ & $\pm 0.00090$ & $_{-0.00059}^{+0.00061}$ \\
$ 1000,  2000$ & $0.01578$ & $\pm 0.00046$ & $\pm 0.00053$ & $\pm 0.00022$ \\
$ 2000,  4000$ & $0.00797$ & $\pm 0.00023$ & $\pm 0.00007$ & $\pm 0.00004$ \\
$ 4000, 10000$ & $0.002532$ & $\pm 0.000072$ & $\pm 0.000134$ & $\pm 0.000038$ \\
$10000, 20000$ & $0.000429$ & $\pm 0.000022$ & $\pm 0.000029$ & $\pm 0.000018$ \\
$20000, 88000$ & $1.4\cdot 10^{-5}$ & $\pm 1.3\cdot 10^{-6}$ & $\pm
5.7\cdot 10^{-6}$ & $\pm 1.4\cdot 10^{-6}$ \\
\hline
\multicolumn{5}{||c||}{$P_{e^{-}}=+0.30$ - inclusive jets} \\
\hline
$  200,   500$ & $0.0135$ & $\pm 0.0011$ & $\pm 0.0011$ & $\pm 0.0013$ \\
$  500,  1000$ & $0.01332$ & $\pm 0.00074$ & $\pm 0.00051$ & $_{-0.00035}^{+0.00036}$ \\
$ 1000,  2000$ & $0.00926$ & $\pm 0.00042$ & $\pm 0.00030$ & $\pm 0.00013$ \\
$ 2000,  4000$ & $0.00437$ & $\pm 0.00020$ & $\pm 0.00004$ & $\pm 0.00002$ \\
$ 4000, 10000$ & $0.001364$ & $\pm 0.000064$ & $\pm 0.000075$ & $\pm 0.000020$ \\
$10000, 20000$ & $30.6\cdot 10^{-5}$ & $\pm 2.2\cdot 10^{-5}$ & $\pm 2.0\cdot 10^{-5}$ & $\pm 1.3\cdot 10^{-5}$ \\
$20000, 88000$ & $8.1\cdot 10^{-6}$ & $\pm 1.2\cdot 10^{-6}$ & $\pm
3.3\cdot 10^{-6}$ & $\pm 0.8\cdot 10^{-6}$ \\
\hline
\multicolumn{5}{||c||}{$P_{e^{+}}=-0.37$ - inclusive jets} \\
\hline
$  200,   500$ & $0.00984$ & $\pm 0.00096$ & $\pm 0.00049$ & $_{-0.00087}^{+0.00091}$ \\
$  500,  1000$ & $0.00814$ & $\pm 0.00058$ & $\pm 0.00026$ & $\pm 0.00016$ \\
$ 1000,  2000$ & $0.00463$ & $\pm 0.00030$ & $\pm 0.00022$ & $\pm 0.00002$ \\
$ 2000,  4000$ & $0.00182$ & $\pm 0.00013$ & $\pm 0.00007$ & $\pm 0.00003$ \\
$ 4000, 10000$ & $34.1\cdot 10^{-5}$ & $\pm 3.1\cdot 10^{-5}$ & $\pm
2.0\cdot 10^{-5}$ & $\pm 1.4\cdot 10^{-5}$ \\
$10000, 20000$ & $25.0\cdot 10^{-6}$ & $\pm 5.9\cdot 10^{-6}$ & $\pm
5.7\cdot 10^{-6}$ & $\pm 2.3\cdot 10^{-6}$ \\
\hline
\multicolumn{5}{||c||}{$P_{e^{+}}=+0.32$ - inclusive jets} \\
\hline
$  200,   500$ & $0.0198$ & $\pm 0.0012$ & $\pm 0.0010$ & $_{-0.0017}^{+0.0018}$ \\
$  500,  1000$ & $0.01586$ & $\pm 0.00070$ & $\pm 0.00023$ & $_{-0.00031}^{+0.00030}$ \\
$ 1000,  2000$ & $0.00905$ & $\pm 0.00036$ & $\pm 0.00027$ & $_{-0.00004}^{+0.00004}$ \\
$ 2000,  4000$ & $0.00345$ & $\pm 0.00016$ & $\pm 0.00014$ & $_{-0.00005}^{+0.00005}$ \\
$ 4000, 10000$ & $68.8\cdot 10^{-5}$ & $\pm 3.9\cdot 10^{-5}$ & $\pm
5.6\cdot 10^{-5}$ & $\pm 2.7\cdot 10^{-5}$ \\
$10000, 20000$ & $61.3\cdot 10^{-6}$ & $\pm 8.0\cdot 10^{-6}$ & $\pm
10.0\cdot 10^{-6}$ & $\pm 5.6\cdot 10^{-6}$ \\
$20000, 88000$ & $3.5\cdot 10^{-7}$ & $\pm 1.8\cdot 10^{-7}$ & $\pm
2.8\cdot 10^{-7}$ & $\pm 0.6\cdot 10^{-7}$ \\
\hline
    \end{tabular}
 \caption{
   Differential polarized inclusive-jet cross-sections
   $d\sigma/d\q2$. Other details as in the caption to
   Table~\ref{tabthree}.}
 \label{tabfive}
\end{center}
\end{table}

\clearpage
\newpage
\begin{table}
\begin{center}
    \begin{tabular}{||c|cccc||}
\hline
  $x$ bin
& $d\sigma/dx$ (pb/GeV)
& $\delta_{\rm stat}$
& $\delta_{\rm syst}$
& $\delta_{\rm ES}$\\
\hline
\multicolumn{5}{||c||}{$P_{e^{-}}=-0.27$ - inclusive jets} \\
\hline
$0.006 , 0.025$ & $16.17$ & $\pm 0.70$ & $\pm 0.83$ & $_{-1.01}^{+1.06}$ \\
$0.025 , 0.063$ & $46.9$ & $\pm 1.3$ & $\pm 1.6$ & $\pm 0.9$ \\
$0.063 , 0.16$ & $60.6$ & $\pm 1.4$ & $\pm 0.9$ & $\pm 0.2$ \\
$0.16 , 0.40$ & $39.7$ & $\pm 1.1$ & $\pm 1.7$ & $\pm 0.7$ \\
$0.40 , 1.0$ & $ 3.94$ & $\pm 0.33$ & $\pm 0.53$ & $\pm 0.29$ \\
\hline
\multicolumn{5}{||c||}{$P_{e^{-}}=+0.30$ - inclusive jets} \\
\hline
$0.006 , 0.025$ & $ 9.71$ & $\pm 0.65$ & $\pm 0.76$ & $_{-0.61}^{+0.64}$ \\
$0.025 , 0.063$ & $24.3$ & $\pm 1.1$ & $\pm 0.9$ & $\pm 0.5$ \\
$0.063 , 0.16$ & $36.6$ & $\pm 1.3$ & $\pm 0.6$ & $\pm 0.1$ \\
$0.16 , 0.40$ & $23.2$ & $\pm 1.0$ & $\pm 1.2$ & $\pm 0.4$ \\
$0.40 , 1.0$ & $ 2.21$ & $\pm 0.30$ & $\pm 0.38$ & $\pm 0.16$ \\
\hline
\multicolumn{5}{||c||}{$P_{e^{+}}=-0.37$ - inclusive jets} \\
\hline
$0.006 , 0.025$ & $ 6.10$ & $\pm 0.54$ & $\pm 0.28$ & $_{-0.34}^{+0.35}$ \\
$0.025 , 0.063$ & $13.59$ & $\pm 0.81$ & $\pm 0.53$ & $\pm 0.15$ \\
$0.063 , 0.16$ & $14.90$ & $\pm 0.82$ & $\pm 0.62$ & $\pm 0.16$ \\
$0.16 , 0.40$ & $ 5.22$ & $\pm 0.47$ & $\pm 0.30$ & $\pm 0.13$ \\
$0.40 , 1.0$ & $ 0.33$ & $\pm 0.11$ & $\pm 0.12$ & $\pm 0.02$ \\
\hline
\multicolumn{5}{||c||}{$P_{e^{+}}=+0.32$ - inclusive jets} \\
\hline
$0.006 , 0.025$ & $12.34$ & $\pm 0.67$ & $\pm 0.56$ & $_{-0.69}^{+0.71}$ \\
$0.025 , 0.063$ & $28.1$ & $\pm 1.0$ & $\pm 1.0$ & $\pm 0.3$ \\
$0.063 , 0.16$ & $26.57$ & $\pm 0.95$ & $\pm 0.88$ & $\pm 0.29$ \\
$0.16 , 0.40$ & $11.55$ & $\pm 0.60$ & $\pm 0.65$ & $\pm 0.28$ \\
$0.40 , 1.0$ & $ 0.50$ & $\pm 0.12$ & $\pm 0.06$ & $\pm 0.04$ \\
\hline
    \end{tabular}
 \caption{
   Differential polarized inclusive-jet cross-sections
   $d\sigma/dx$. Other details as in the caption to
   Table~\ref{tabthree}.}
 \label{tabsix}
\end{center}
\end{table}

\clearpage
\newpage
\begin{table}
\begin{center}
    \begin{tabular}{|c||c|c|c|c|c|}
\hline
  lepton and
  polarization
& $\sigma_{\rm jets}$ (pb)
& $\delta_{\rm stat}$ (pb)
& $\delta_{\rm syst}$ (pb)
& $\delta_{\rm ES}$ (pb)
& SM prediction (pb)\\
\hline\hline
$P_{e^-}=-0.27\pm 0.01$ & $70.54$ & $0.97$ & $0.58$ & $_{-0.40}^{+0.43}$ & $69.17$\\
$P_{e^-}=+0.29\pm 0.01$ & $40.53$ & $0.88$ & $0.45$ & $_{-0.23}^{+0.24}$ & $38.67$\\
\hline
$P_{e^+}=-0.37^{+0.01}_{-0.02}$ & $17.55$ & $0.60$ & $0.57$ & $\pm 0.11$ & $16.86$\\
$P_{e^+}=+0.32\pm 0.01$ & $34.51$ & $0.72$ & $1.05$ & $_{-0.22}^{+0.23}$ & $35.33$\\
\hline
    \end{tabular}
 \caption{
   Integrated polarized inclusive-jet cross-sections 
   $\sigma_{\rm jets}$ for jets of hadrons in the laboratory frame
   selected with the longitudinally invariant $\kt$ cluster
   algorithm. The statistical, uncorrelated systematic and
   energy-scale ({\rm ES}) uncertainties are shown separately. The
   uncertainty coming from the luminosity measurement is not
   shown. The predictions of the Standard Model as given by the {\sc
     Mepjet} calculation are shown in the last column.}
 \label{tabone}
\end{center}
\end{table}

\clearpage
\newpage
\begin{table}
\begin{center}
    \begin{tabular}{||c|cccc||c||c||}
\hline
  $\etajet$ bin
& $d\sigma/d\etajet$ (pb)
& $\delta_{\rm stat}$
& $\delta_{\rm syst}$
& $\delta_{\rm ES}$
& $C_{\rm QED}$
& $C_{\rm had}$\\
\hline
\multicolumn{7}{||c||}{unpolarized - inclusive jets} \\
\hline
$-1.0 , -0.5$ & $ 3.58$ & $\pm 0.28$ & $\pm 0.30$ & $_{-0.13}^{+0.15}$ & $0.97$ & $0.93$ \\ 
$-0.5 ,  0.0$ & $10.38$ & $\pm 0.43$ & $\pm 0.35$ & $_{-0.22}^{+0.24}$ & $0.97$ & $0.98$ \\ 
$ 0.0 ,  0.5$ & $18.06$ & $\pm 0.53$ & $\pm 0.34$ & $\pm 0.17$ & $0.98$ & $0.99$ \\ 
$ 0.5 ,  1.0$ & $22.17$ & $\pm 0.57$ & $\pm 0.75$ & $\pm 0.07$ & $0.97$ & $1.00$ \\ 
$ 1.0 ,  1.5$ & $22.38$ & $\pm 0.58$ & $\pm 0.90$ & $\pm 0.07$ & $0.97$ & $1.00$ \\ 
$ 1.5 ,  2.0$ & $20.33$ & $\pm 0.57$ & $\pm 0.31$ & $_{-0.07}^{+0.08}$ & $0.96$ & $1.01$ \\ 
$ 2.0 ,  2.5$ & $15.59$ & $\pm 0.61$ & $\pm 1.85$ & $\pm 0.12$ & $0.96$ & $1.01$ \\ 
\hline
\hline
  $\etabar$ bin
& $d\sigma/d\etabar$ (pb)
& $\delta_{\rm stat}$
& $\delta_{\rm syst}$
& $\delta_{\rm ES}$
& $C_{\rm QED}$
& $C_{\rm had}$\\
\hline
\multicolumn{7}{||c||}{unpolarized - dijets} \\
\hline
$-1.0 , -0.5$ & $ 0.103$ & $\pm 0.073$ & $\pm 0.003$ & $_{-0.032}^{+0.033}$ & $0.80$ & $0.60$ \\ 
$-0.5 ,  0.0$ & $ 1.48$ & $\pm 0.26$ & $\pm 0.52$ & $\pm 0.09$ & $0.96$ & $0.81$ \\ 
$ 0.0 ,  0.5$ & $ 3.63$ & $\pm 0.33$ & $\pm 0.88$ & $\pm 0.12$ & $0.97$ & $0.89$ \\ 
$ 0.5 ,  1.0$ & $ 5.68$ & $\pm 0.37$ & $\pm 0.43$ & $_{-0.13}^{+0.14}$ & $0.98$ & $0.91$ \\ 
$ 1.0 ,  1.5$ & $ 6.43$ & $\pm 0.34$ & $\pm 0.07$ & $_{-0.11}^{+0.12}$ & $0.97$ & $0.92$ \\ 
$ 1.5 ,  2.0$ & $ 3.77$ & $\pm 0.24$ & $\pm 0.14$ & $\pm 0.07$ & $0.95$ & $0.92$ \\ 
$ 2.0 ,  2.5$ & $ 0.58$ & $\pm 0.10$ & $\pm 0.05$ & $\pm 0.01$ & $0.93$ & $0.88$ \\ 
\hline
\multicolumn{7}{||c||}{unpolarized - three jets} \\
\hline
$ 0.0 ,  0.5$ & $ 0.49$ & $\pm 0.18$ & $\pm 0.49$ & $_{-0.05}^{+0.03}$ & $0.95$ & $0.75$ \\ 
$ 0.5 ,  1.0$ & $ 1.05$ & $\pm 0.21$ & $\pm 0.15$ & $_{-0.06}^{+0.05}$ & $0.93$ & $0.78$ \\ 
$ 1.0 ,  1.5$ & $ 1.06$ & $\pm 0.17$ & $\pm 0.13$ & $_{-0.03}^{+0.04}$ & $0.99$ & $0.80$ \\ 
$ 1.5 ,  2.5$ & $ 0.246$ & $\pm 0.046$ & $\pm 0.053$ & $_{-0.007}^{+0.009}$ & $0.99$ & $0.80$ \\ 
\hline
    \end{tabular}
 \caption{
   Differential unpolarized inclusive-jet, dijet and three-jet
   cross-sections $d\sigma/d\etajet$ and $d\sigma/d\etabar$ in $e^-p$
   collisions for jets of hadrons in the laboratory frame selected
   with the longitudinally invariant $\kt$ cluster algorithm. The
   statistical, uncorrelated systematic and jet-energy-scale ({\rm
     ES}) uncertainties are shown separately. The multiplicative
   corrections for QED radiative effects,  $C_{\rm QED}$, and the
   corrections for hadronization effects, $C_{\rm had}$, to be applied
   to the parton-level NLO QCD calculations, are shown in the last two
   columns.}
 \label{tabseven}
\end{center}
\end{table}

\clearpage
\newpage
\begin{table}
\begin{center}
    \begin{tabular}{||c|cccc||c||c||}
\hline
  $\etajet$ bin
& $d\sigma/d\etajet$ (pb)
& $\delta_{\rm stat}$
& $\delta_{\rm syst}$
& $\delta_{\rm ES}$
& $C_{\rm QED}$
& $C_{\rm had}$\\
\hline
\multicolumn{7}{||c||}{unpolarized - inclusive jets} \\
\hline
$-1.0 , -0.5$ & $ 2.12$ & $\pm 0.24$ & $\pm 0.21$ & $_{-0.06}^{+0.07}$ & $0.92$ & $0.94$ \\ 
$-0.5 ,  0.0$ & $ 5.45$ & $\pm 0.34$ & $\pm 0.23$ & $\pm 0.08$ & $0.94$ & $0.98$ \\ 
$ 0.0 ,  0.5$ & $ 8.34$ & $\pm 0.39$ & $\pm 0.22$ & $\pm 0.06$ & $0.95$ & $0.99$ \\ 
$ 0.5 ,  1.0$ & $10.90$ & $\pm 0.43$ & $\pm 0.54$ & $\pm 0.06$ & $0.95$ & $1.00$ \\ 
$ 1.0 ,  1.5$ & $ 9.95$ & $\pm 0.42$ & $\pm 0.42$ & $\pm 0.05$ & $0.95$ & $1.00$ \\ 
$ 1.5 ,  2.0$ & $ 9.48$ & $\pm 0.42$ & $\pm 0.26$ & $_{-0.05}^{+0.06}$ & $0.94$ & $1.01$ \\ 
$ 2.0 ,  2.5$ & $ 7.88$ & $\pm 0.48$ & $\pm 1.15$ & $\pm 0.09$ & $0.93$ & $1.02$ \\ 
\hline
\hline
  $\etabar$ bin
& $d\sigma/d\etabar$ (pb)
& $\delta_{\rm stat}$
& $\delta_{\rm syst}$
& $\delta_{\rm ES}$
& $C_{\rm QED}$
& $C_{\rm had}$\\
\hline
\multicolumn{7}{||c||}{unpolarized - dijets} \\
\hline
$-0.5 ,  0.0$ & $ 0.92$ & $\pm 0.23$ & $\pm 0.33$ & $_{-0.05}^{+0.07}$ & $0.94$ & $0.83$ \\ 
$ 0.0 ,  0.5$ & $ 1.75$ & $\pm 0.28$ & $\pm 0.42$ & $_{-0.05}^{+0.06}$ & $0.94$ & $0.90$ \\ 
$ 0.5 ,  1.0$ & $ 3.01$ & $\pm 0.32$ & $\pm 0.24$ & $_{-0.06}^{+0.07}$ & $0.95$ & $0.94$ \\ 
$ 1.0 ,  1.5$ & $ 3.52$ & $\pm 0.28$ & $\pm 0.06$ & $_{-0.06}^{+0.07}$ & $0.95$ & $0.94$ \\ 
$ 1.5 ,  2.0$ & $ 1.95$ & $\pm 0.20$ & $\pm 0.11$ & $\pm 0.04$ & $0.94$ & $0.94$ \\ 
$ 2.0 ,  2.5$ & $ 0.287$ & $\pm 0.073$ & $\pm 0.043$ & $_{-0.005}^{+0.008}$ & $0.93$ & $0.91$ \\ 
\hline
\multicolumn{7}{||c||}{unpolarized - three jets} \\
\hline
$ 0.0 ,  0.5$ & $ 0.079$ & $\pm 0.079$ & $\pm 0.079$ & $_{-0.003}^{+0.006}$ & $0.92$ & $0.72$ \\ 
$ 0.5 ,  1.0$ & $ 0.68$ & $\pm 0.23$ & $\pm 0.09$ & $_{-0.03}^{+0.04}$ & $0.93$ & $0.79$ \\ 
$ 1.0 ,  1.5$ & $ 0.254$ & $\pm 0.086$ & $\pm 0.033$ & $_{-0.010}^{+0.011}$ & $0.94$ & $0.81$ \\ 
$ 1.5 ,  2.5$ & $ 0.087$ & $\pm 0.029$ & $\pm 0.019$ & $\pm 0.003$ & $0.91$ & $0.83$ \\ 
\hline
    \end{tabular}
 \caption{
   Differential unpolarized inclusive-jet, dijet and three-jet
   cross-sections $d\sigma/d\etajet$ and $d\sigma/\etabar$ in $e^+p$
   collisions. Other details as in the caption to Table~\ref{tabseven}.}
 \label{tabeight}
\end{center}
\end{table}

\clearpage
\newpage
\begin{table}
\begin{center}
    \begin{tabular}{||c|cccc||c||c||}
\hline
  $\etjet$ bin (GeV)
& $d\sigma/d\etjet$ (pb/GeV)
& $\delta_{\rm stat}$
& $\delta_{\rm syst}$
& $\delta_{\rm ES}$
& $C_{\rm QED}$
& $C_{\rm had}$\\
\hline
\multicolumn{7}{||c||}{unpolarized - inclusive jets} \\
\hline
$ 14,  21$ & $1.819$ & $\pm 0.053$ & $\pm 0.120$ & $_{-0.040}^{+0.035}$ & $1.03$ & $0.99$ \\ 
$ 21,  29$ & $1.449$ & $\pm 0.040$ & $\pm 0.040$ & $_{-0.014}^{+0.016}$ & $1.01$ & $1.00$ \\ 
$ 29,  41$ & $1.118$ & $\pm 0.027$ & $\pm 0.042$ & $_{-0.008}^{+0.007}$ & $0.97$ & $1.00$ \\ 
$ 41,  55$ & $0.668$ & $\pm 0.019$ & $\pm 0.039$ & $\pm 0.008$ & $0.95$ & $1.00$ \\ 
$ 55,  71$ & $0.361$ & $\pm 0.013$ & $\pm 0.031$ & $\pm 0.009$ & $0.92$ & $0.99$ \\ 
$ 71,  87$ & $0.1596$ & $\pm 0.0082$ & $\pm 0.0360$ & $_{-0.0015}^{+0.0020}$ & $0.88$ & $0.99$ \\ 
$ 87, 120$ & $0.0355$ & $\pm 0.0026$ & $\pm 0.0190$ & $_{-0.0042}^{+0.0053}$ & $0.83$ & $0.98$ \\ 
\hline
\hline
  $\etbar$ bin (GeV)
& $d\sigma/d\etbar$ (pb/GeV)
& $\delta_{\rm stat}$
& $\delta_{\rm syst}$
& $\delta_{\rm ES}$
& $C_{\rm QED}$
& $C_{\rm had}$\\
\hline
\multicolumn{7}{||c||}{unpolarized - dijets} \\
\hline
$ 9.5 ,  14$ & $0.345$ & $\pm 0.032$ & $\pm 0.017$ & $\pm 0.004$ & $1.03$ & $0.88$ \\ 
$ 14,  21$ & $0.502$ & $\pm 0.028$ & $\pm 0.077$ & $_{-0.004}^{+0.007}$ & $0.99$ & $0.91$ \\ 
$ 21,  29$ & $0.375$ & $\pm 0.021$ & $\pm 0.030$ & $\pm 0.010$ & $0.94$ & $0.92$ \\ 
$ 29,  41$ & $0.156$ & $\pm 0.011$ & $\pm 0.005$ & $\pm 0.007$ & $0.92$ & $0.90$ \\ 
$ 41,  55$ & $0.0390$ & $\pm 0.0049$ & $\pm 0.0107$ & $_{-0.0026}^{+0.0037}$ & $0.88$ & $0.90$ \\ 
$ 55,  71$ & $0.0070$ & $\pm 0.0023$ & $\pm 0.0007$ & $_{-0.0016}^{+0.0014}$ & $0.90$ & $0.95$ \\ 
$ 71,  87$ & $0.0021$ & $\pm 0.0021$ & $\pm 0.0021$ & $_{-0.0000}^{+0.0014}$ & $1.00$ & $0.94$ \\ 
\hline
\multicolumn{7}{||c||}{unpolarized - three jets} \\
\hline
$  8,  9.5$ & $0.0096$ & $\pm 0.0096$ & $\pm 0.0025$ & $_{-0.0017}^{+0.0016}$ & $1.02$ & $0.72$ \\ 
$ 9.5 ,  14$ & $0.060$ & $\pm 0.015$ & $\pm 0.015$ & $\pm 0.001$ & $1.02$ & $0.77$ \\ 
$ 14,  21$ & $0.099$ & $\pm 0.015$ & $\pm 0.003$ & $\pm 0.004$ & $0.96$ & $0.79$ \\ 
$ 21,  29$ & $0.0481$ & $\pm 0.0087$ & $\pm 0.0036$ & $_{-0.0027}^{+0.0030}$ & $0.92$ & $0.78$ \\ 
$ 29,  41$ & $0.0078$ & $\pm 0.0029$ & $\pm 0.0008$ & $_{-0.0006}^{+0.0007}$ & $0.86$ & $0.85$ \\ 
\hline
    \end{tabular}
 \caption{
   Differential unpolarized inclusive-jet, dijet and three-jet
   cross-sections $d\sigma/d\etjet$ and $d\sigma/d\etbar$ in $e^-p$
   collisions. Other details as in the caption to Table~\ref{tabseven}.}
 \label{tabnine}
\end{center}
\end{table}

\clearpage
\newpage
\begin{table}
\begin{center}
\scalebox{0.95}{
    \begin{tabular}{||c|cccc||c||c||}
\hline
  $\etjet$ bin (GeV)
& $d\sigma/d\etjet$ (pb/GeV)
& $\delta_{\rm stat}$
& $\delta_{\rm syst}$
& $\delta_{\rm ES}$
& $C_{\rm QED}$
& $C_{\rm had}$\\
\hline
\multicolumn{7}{||c||}{unpolarized - inclusive jets} \\
\hline
$ 14,  21$ & $1.200$ & $\pm 0.046$ & $\pm 0.087$ & $_{-0.019}^{+0.020}$ & $0.99$ & $1.00$ \\ 
$ 21,  29$ & $0.863$ & $\pm 0.033$ & $\pm 0.029$ & $_{-0.006}^{+0.005}$ & $0.96$ & $1.00$ \\ 
$ 29,  41$ & $0.522$ & $\pm 0.019$ & $\pm 0.025$ & $\pm 0.005$ & $0.93$ & $1.00$ \\ 
$ 41,  55$ & $0.242$ & $\pm 0.012$ & $\pm 0.016$ & $\pm 0.006$ & $0.91$ & $0.99$ \\ 
$ 55,  71$ & $0.0901$ & $\pm 0.0067$ & $\pm 0.0078$ & $_{-0.0034}^{+0.0037}$ & $0.87$ & $0.99$ \\ 
$ 71,  87$ & $0.0269$ & $\pm 0.0035$ & $\pm 0.0067$ & $_{-0.0010}^{+0.0011}$ & $0.84$ & $0.98$ \\ 
$ 87, 120$ & $0.00329$ & $\pm 0.00078$ & $\pm 0.00179$ & $_{-0.00050}^{+0.00064}$ & $0.79$ & $0.97$ \\ 
\hline
\hline
  $\etbar$ bin (GeV)
& $d\sigma/d\etbar$ (pb/GeV)
& $\delta_{\rm stat}$
& $\delta_{\rm syst}$
& $\delta_{\rm ES}$
& $C_{\rm QED}$
& $C_{\rm had}$\\
\hline
\multicolumn{7}{||c||}{unpolarized - dijets} \\
\hline
$ 9.5 ,  14$ & $0.288$ & $\pm 0.033$ & $\pm 0.017$ & $_{-0.002}^{+0.001}$ & $0.99$ & $0.90$ \\ 
$ 14,  21$ & $0.330$ & $\pm 0.026$ & $\pm 0.052$ & $\pm 0.004$ & $0.95$ & $0.93$ \\ 
$ 21,  29$ & $0.190$ & $\pm 0.018$ & $\pm 0.015$ & $_{-0.006}^{+0.007}$ & $0.93$ & $0.93$ \\ 
$ 29,  41$ & $0.0433$ & $\pm 0.0066$ & $\pm 0.0023$ & $_{-0.0022}^{+0.0026}$ & $0.88$ & $0.93$ \\ 
$ 41,  55$ & $0.0112$ & $\pm 0.0032$ & $\pm 0.0031$ & $_{-0.0010}^{+0.0014}$ & $0.90$ & $0.95$ \\ 
\hline
\multicolumn{7}{||c||}{unpolarized - three jets} \\
\hline
$  8,  9.5$ & $0.019$ & $\pm 0.019$ & $\pm 0.005$ & $\pm 0.002$ & $0.82$ & $0.73$ \\ 
$ 9.5 ,  14$ & $0.040$ & $\pm 0.013$ & $\pm 0.010$ & $\pm 0.001$ & $0.96$ & $0.76$ \\ 
$ 14,  21$ & $0.043$ & $\pm 0.012$ & $\pm 0.002$ & $\pm 0.002$ & $0.93$ & $0.80$ \\ 
$ 21,  29$ & $0.0034$ & $\pm 0.0024$ & $\pm 0.0003$ & $_{-0.0002}^{+0.0003}$ & $0.86$ & $0.82$ \\ 
$ 29,  41$ & $0.0018$ & $\pm 0.0018$ & $\pm 0.0002$ & $_{-0.0002}^{+0.0003}$ & $0.88$ & $0.86$ \\ 
\hline
    \end{tabular}}
 \caption{
   Differential unpolarized inclusive-jet, dijet and three-jet
   cross-sections $d\sigma/d\etjet$ and $d\sigma/d\etbar$ in $e^+p$
   collisions. Other details as in the caption to Table~\ref{tabseven}.}
 \label{tabten}
\end{center}
\end{table}

\clearpage
\newpage
\begin{table}
\begin{center}
    \begin{tabular}{||c|cccc||c||c||}
\hline
  $\q2$ bin (\g2)
& $d\sigma/d\q2$ (pb/\g2)
& $\delta_{\rm stat}$
& $\delta_{\rm syst}$
& $\delta_{\rm ES}$
& $C_{\rm QED}$
& $C_{\rm had}$\\
\hline
\multicolumn{7}{||c||}{unpolarized - inclusive jets} \\
\hline
$  200,   500$ & $0.01921$ & $\pm 0.00087$ & $\pm 0.00131$ & $_{-0.00180}^{+0.00190}$ & $0.98$ & $0.97$ \\ 
$  500,  1000$ & $0.01803$ & $\pm 0.00056$ & $\pm 0.00047$ & $_{-0.00048}^{+0.00049}$ & $0.99$ & $1.00$ \\ 
$ 1000,  2000$ & $0.01268$ & $\pm 0.00032$ & $\pm 0.00017$ & $\pm 0.00018$ & $0.98$ & $1.00$ \\ 
$ 2000,  4000$ & $0.00623$ & $\pm 0.00016$ & $\pm 0.00006$ & $\pm 0.00003$ & $0.97$ & $1.00$ \\ 
$ 4000, 10000$ & $0.001963$ & $\pm 0.000050$ & $\pm 0.000097$ & $\pm 0.000029$ & $0.95$ & $1.00$ \\ 
$10000, 20000$ & $0.000376$ & $\pm 0.000016$ & $\pm 0.000045$ & $_{-0.000016}^{+0.000015}$ & $0.94$ & $1.00$ \\ 
$20000, 88000$ & $1.121\cdot 10^{-5}$ & $\pm 0.091\cdot 10^{-5}$ & $\pm 0.706\cdot 10^{-5}$ & $_{-0.114}^{+0.113}\cdot 10^{-5}$ & $0.93$ & $1.00$ \\ 
\hline
\multicolumn{7}{||c||}{unpolarized - dijets} \\
\hline
$  200,   500$ & $0.00409$ & $\pm 0.00091$ & $\pm 0.00112$ & $_{-0.00083}^{+0.00094}$ & $1.00$ & $0.91$ \\ 
$  500,  1000$ & $0.00323$ & $\pm 0.00031$ & $\pm 0.00037$ & $_{-0.00026}^{+0.00027}$ & $0.97$ & $0.92$ \\ 
$ 1000,  2000$ & $0.00251$ & $\pm 0.00016$ & $\pm 0.00004$ & $\pm 0.00009$ & $0.97$ & $0.91$ \\ 
$ 2000,  4000$ & $0.001170$ & $\pm 0.000071$ & $\pm 0.000112$ & $_{-0.000023}^{+0.000024}$ & $0.96$ & $0.91$ \\ 
$ 4000, 10000$ & $0.000343$ & $\pm 0.000021$ & $\pm 0.000029$ & $\pm 0.000011$ & $0.95$ & $0.89$ \\ 
$10000, 20000$ & $69.8\cdot 10^{-6}$ & $\pm 6.9\cdot 10^{-6}$ & $\pm 23.9\cdot 10^{-6}$ & $\pm 4.1\cdot 10^{-6}$ & $0.99$ & $0.87$ \\ 
$20000, 88000$ & $2.11\cdot 10^{-6}$ & $\pm 0.37\cdot 10^{-6}$ & $\pm 1.62\cdot 10^{-6}$ & $\pm 0.26\cdot 10^{-5}$ & $0.96$ & $0.86$ \\ 
\hline
\multicolumn{7}{||c||}{unpolarized - three jets} \\
\hline
$  500,  1000$ & $0.00034$ & $\pm 0.00018$ & $\pm 0.00015$ & $_{-0.00005}^{+0.00006}$ & $0.98$ & $0.78$ \\ 
$ 1000,  2000$ & $0.000306$ & $\pm 0.000073$ & $\pm 0.000025$ & $_{-0.000026}^{+0.000027}$ & $0.96$ & $0.78$ \\ 
$ 2000,  4000$ & $0.000193$ & $\pm 0.000034$ & $\pm 0.000024$ & $_{-0.000010}^{+0.000011}$ & $0.95$ & $0.79$ \\ 
$ 4000, 10000$ & $5.04\cdot 10^{-5}$ & $\pm 0.93\cdot 10^{-5}$ & $\pm 2.18\cdot 10^{-5}$ & $\pm 0.20\cdot 10^{-5}$ & $0.95$ & $0.79$ \\ 
$10000, 20000$ & $1.04\cdot 10^{-5}$ & $\pm 0.25\cdot 10^{-5}$ & $\pm 0.70\cdot 10^{-5}$ & $_{-0.08}^{+0.09}\cdot 10^{-5}$ & $0.95$ & $0.75$ \\ 
$20000, 88000$ & $1.34\cdot 10^{-7}$ & $\pm 0.99\cdot 10^{-7}$ & $\pm 0.06\cdot 10^{-7}$ & $\pm 0.19\cdot 10^{-7}$ & $0.92$ & $0.77$ \\ 
\hline
    \end{tabular}
 \caption{
   Differential unpolarized inclusive-jet, dijet and three-jet
   cross-sections $d\sigma/d\q2$ in $e^-p$ collisions. Other
   details as in the caption to Table~\ref{tabseven}.}
 \label{tabeleven}
\end{center}
\end{table}

\clearpage
\newpage
\begin{table}
\begin{center}
\scalebox{0.95}{
    \begin{tabular}{||c|cccc||c||c||}
\hline
  $\q2$ bin (\g2)
& $d\sigma/d\q2$ (pb/\g2)
& $\delta_{\rm stat}$
& $\delta_{\rm syst}$
& $\delta_{\rm ES}$
& $C_{\rm QED}$
& $C_{\rm had}$\\
\hline
\multicolumn{7}{||c||}{unpolarized - inclusive jets} \\
\hline
$  200,   500$ & $0.01525$ & $\pm 0.00082$ & $\pm 0.00074$ & $_{-0.00134}^{+0.00141}$ & $0.96$ & $0.97$ \\ 
$  500,  1000$ & $0.01241$ & $\pm 0.00050$ & $\pm 0.00018$ & $\pm 0.00024$ & $0.96$ & $1.00$ \\ 
$ 1000,  2000$ & $0.00707$ & $\pm 0.00026$ & $\pm 0.00021$ & $\pm 0.00003$ & $0.95$ & $1.00$ \\ 
$ 2000,  4000$ & $0.00273$ & $\pm 0.00011$ & $\pm 0.00010$ & $\pm 0.00004$ & $0.94$ & $1.00$ \\ 
$ 4000, 10000$ & $0.000530$ & $\pm 0.000027$ & $\pm 0.000037$ & $\pm 0.000021$ & $0.91$ & $1.00$ \\ 
$10000, 20000$ & $4.35\cdot 10^{-5}$ & $\pm 0.53\cdot 10^{-5}$ & $\pm 0.71\cdot 10^{-5}$ & $\pm 0.40\cdot 10^{-5}$ & $0.88$ & $1.00$ \\ 
$20000, 88000$ & $2.3\cdot 10^{-7}$ & $\pm 1.1\cdot 10^{-7}$ & $\pm 1.5\cdot 10^{-7}$ & $\pm 0.4\cdot 10^{-7}$ & $0.81$ & $1.00$ \\ 
\hline
\multicolumn{7}{||c||}{unpolarized - dijets} \\
\hline
$  200,   500$ & $0.00204$ & $\pm 0.00072$ & $\pm 0.00051$ & $_{-0.00037}^{+0.00042}$ & $0.98$ & $0.92$ \\ 
$  500,  1000$ & $0.00301$ & $\pm 0.00033$ & $\pm 0.00033$ & $_{-0.00024}^{+0.00025}$ & $0.96$ & $0.93$ \\ 
$ 1000,  2000$ & $0.00152$ & $\pm 0.00014$ & $\pm 0.00003$ & $_{-0.00003}^{+0.00004}$ & $0.95$ & $0.93$ \\ 
$ 2000,  4000$ & $0.000605$ & $\pm 0.000058$ & $\pm 0.000060$ & $\pm 0.000016$ & $0.94$ & $0.92$ \\ 
$ 4000, 10000$ & $0.000106$ & $\pm 0.000013$ & $\pm 0.000011$ & $\pm 0.000006$ & $0.91$ & $0.92$ \\ 
$10000, 20000$ & $9.3\cdot 10^{-6}$ & $\pm 2.4\cdot 10^{-6}$ & $\pm 3.2\cdot 10^{-5}$ & $_{-1.0}^{+1.1}\cdot 10^{-6}$ & $0.92$ & $0.92$ \\ 
$20000, 88000$ & $7.4\cdot 10^{-8}$ & $\pm 7.4\cdot 10^{-8}$ & $\pm 5.8\cdot 10^{-8}$ & $\pm 1.1\cdot 10^{-8}$ & $0.93$ & $0.88$ \\ 
\hline
\multicolumn{7}{||c||}{unpolarized - three jets} \\
\hline
$  500,  1000$ & $0.00038$ & $\pm 0.00020$ & $\pm 0.00017$ & $\pm 0.00006$ & $0.97$ & $0.78$ \\ 
$ 1000,  2000$ & $0.000156$ & $\pm 0.000055$ & $\pm 0.000013$ & $_{-0.000013}^{+0.000014}$ & $0.91$ & $0.80$ \\ 
$ 2000,  4000$ & $0.000065$ & $\pm 0.000021$ & $\pm 0.000008$ & $_{-0.000002}^{+0.000003}$ & $0.91$ & $0.80$ \\ 
$ 4000, 10000$ & $9.2\cdot 10^{-6}$ & $\pm 4.4\cdot 10^{-6}$ & $\pm 4.1\cdot 10^{-6}$ & $\pm 0.7\cdot 10^{-6}$ & $0.88$ & $0.79$ \\ 
$10000, 20000$ & $4.6\cdot 10^{-7}$ & $\pm 4.6\cdot 10^{-7}$ & $\pm 3.1\cdot 10^{-7}$ & $\pm 0.6\cdot 10^{-7}$ & $1.00$ & $0.82$ \\ 
\hline
    \end{tabular}}
 \caption{
   Differential unpolarized inclusive-jet, dijet and three-jet
   cross-sections $d\sigma/d\q2$ in $e^+p$ collisions. Other
   details as in the caption to Table~\ref{tabseven}.}
 \label{tabtwelve}
\end{center}
\end{table}

\clearpage
\newpage
\begin{table}
\begin{center}
    \begin{tabular}{||c|cccc||c||c||}
\hline
  $x$ bin
& $d\sigma/dx$ (pb)
& $\delta_{\rm stat}$
& $\delta_{\rm syst}$
& $\delta_{\rm ES}$
& $C_{\rm QED}$
& $C_{\rm had}$\\
\hline
\multicolumn{7}{||c||}{unpolarized - inclusive jets} \\
\hline
$0.006 , 0.025$ & $13.12$ & $\pm 0.50$ & $\pm 0.74$ & $_{-0.82}^{+0.86}$ & $0.95$ & $1.01$ \\ 
$0.025 , 0.063$ & $35.82$ & $\pm 0.85$ & $\pm 1.13$ & $_{-0.69}^{+0.70}$ & $0.98$ & $1.00$ \\ 
$0.063 , 0.16$ & $49.26$ & $\pm 0.98$ & $\pm 0.09$ & $\pm 0.14$ & $0.97$ & $0.99$ \\ 
$0.16 , 0.40$ & $31.81$ & $\pm 0.77$ & $\pm 1.30$ & $_{-0.54}^{+0.53}$ & $0.97$ & $0.99$ \\ 
$0.40 , 1.0$ & $ 3.11$ & $\pm 0.23$ & $\pm 0.30$ & $\pm 0.23$ & $0.95$ & $0.98$ \\ 
\hline
    \end{tabular}
 \caption{
   Differential unpolarized inclusive-jet
   cross-sections $d\sigma/dx$ in $e^-p$ collisions. Other
   details as in the caption to Table~\ref{tabseven}.}
 \label{tabthirteen}
\end{center}
\end{table}

\begin{table}
\begin{center}
    \begin{tabular}{||c|cccc||c||c||}
\hline
  $x$ bin
& $d\sigma/dx$ (pb)
& $\delta_{\rm stat}$
& $\delta_{\rm syst}$
& $\delta_{\rm ES}$
& $C_{\rm QED}$
& $C_{\rm had}$\\
\hline
\multicolumn{7}{||c||}{unpolarized - inclusive jets} \\
\hline
$0.006 , 0.025$ & $ 9.49$ & $\pm 0.47$ & $\pm 0.41$ & $_{-0.53}^{+0.54}$ & $0.93$ & $1.01$ \\ 
$0.025 , 0.063$ & $21.40$ & $\pm 0.71$ & $\pm 0.76$ & $\pm 0.24$ & $0.96$ & $1.00$ \\ 
$0.063 , 0.16$ & $21.64$ & $\pm 0.69$ & $\pm 0.76$ & $_{-0.23}^{+0.24}$ & $0.95$ & $0.99$ \\ 
$0.16 , 0.40$ & $ 8.55$ & $\pm 0.41$ & $\pm 0.48$ & $\pm 0.21$ & $0.93$ & $0.98$ \\ 
$0.40 , 1.0$ & $ 0.439$ & $\pm 0.090$ & $\pm 0.084$ & $\pm 0.032$ & $0.91$ & $0.98$ \\ 
\hline
    \end{tabular}
 \caption{
   Differential unpolarized inclusive-jet
   cross-sections $d\sigma/dx$ in $e^+p$ collisions. Other
   details as in the caption to Table~\ref{tabseven}.}
 \label{tabfourteen}
\end{center}
\end{table}

\clearpage
\newpage
\begin{table}
\begin{center}
\scalebox{0.95}{
    \begin{tabular}{|c||c|c|c|c|c|c|c|}
\hline
  lepton/
& $\sigma_{\rm jets}$ (pb)
& $\delta_{\rm stat}$ (pb)
& $\delta_{\rm syst}$ (pb)
& $\delta_{\rm ES}$ (pb)
& \multicolumn{3}{c|}{QCD predictions (pb)}\\ \cline{6-8}
  jet multiplicity
& 
& 
& 
& 
& ZEUS-S & CTEQ6 & MRST \\
\hline\hline
$e^-$/inclusive jet & $56.18$ & $0.68$ & $0.53$ & $_{-0.32}^{+0.34}$ &
$54.47\pm 0.75$ & $54.05$ & $54.56$\\
$e^+$/inclusive jet & $26.88$ & $0.51$ & $0.82$ & $_{-0.17}^{+0.18}$ &
$26.77\pm 0.45$ & $25.85$ & $26.49$\\
\hline
$e^-$/dijet & $10.87$ & $0.34$ & $0.80$ & $_{-0.23}^{+0.24}$ &
$9.14\pm 0.35$ & $9.05$ & $9.26$\\
$e^+$/dijet & $5.83$ & $0.29$ & $0.45$ & $_{-0.12}^{+0.13}$ & 
$4.57\pm 0.19$ & $4.38$ & $4.55$\\
\hline
$e^-$/three jet & $1.52$ & $0.15$ & $0.09$ & $\pm 0.06$ &
$0.79\pm 0.22$ & $0.79$ & $0.82$\\
$e^+$/three jet & $0.563$ & $0.110$ & $0.037$ & $_{-0.022}^{+0.025}$ &
$0.397\pm 0.118$ & $0.386$ & $0.409$\\
\hline
    \end{tabular}}
 \caption{
   Integrated unpolarized jet cross-sections $\sigma_{\rm jets}$
   for jets of hadrons in the laboratory frame selected with the
   longitudinally invariant $\kt$ cluster algorithm. The statistical,
   uncorrelated systematic and energy-scale ({\rm ES}) uncertainties
   are shown separately. The predictions of QCD as given by the 
   {\sc Mepjet} calculations using the ZEUS-S PDFs are shown at NLO
   for the inclusive-jet and dijet cross sections and at LO for the
   three-jet cross sections, together with the total theoretical
   uncertainty. Also shown are the total cross sections predicted by
   QCD using the CTEQ6 or MRST PDF sets.}
 \label{tabtwo}
\end{center}
\end{table}

\clearpage
\newpage
\begin{table}
\begin{center}
    \begin{tabular}{||c|cccc||c||c||}
\hline
  $\mj$ bin (GeV)
& $d\sigma/d\mj$ (pb/GeV)
& $\delta_{\rm stat}$
& $\delta_{\rm syst}$
& $\delta_{\rm ES}$
& $C_{\rm QED}$
& $C_{\rm had}$\\
\hline
\multicolumn{7}{||c||}{unpolarized - dijets} \\
\hline
$ 10,  15$ & $0.090$ & $\pm 0.011$ & $\pm 0.010$ & $\pm 0.001$ & $1.02$ & $0.87$ \\ 
$ 15,  20$ & $0.207$ & $\pm 0.018$ & $\pm 0.011$ & $\pm 0.002$ & $0.99$ & $0.88$ \\ 
$ 20,  30$ & $0.284$ & $\pm 0.016$ & $\pm 0.032$ & $_{-0.004}^{+0.005}$ & $0.97$ & $0.89$ \\ 
$ 30,  45$ & $0.250$ & $\pm 0.014$ & $\pm 0.028$ & $_{-0.007}^{+0.006}$ & $0.96$ & $0.90$ \\ 
$ 45,  65$ & $0.0960$ & $\pm 0.0083$ & $\pm 0.0191$ & $_{-0.0038}^{+0.0051}$ & $0.95$ & $0.94$ \\ 
$ 65,  90$ & $0.0385$ & $\pm 0.0060$ & $\pm 0.0058$ & $_{-0.0032}^{+0.0023}$ & $0.94$ & $0.96$ \\ 
$ 90, 120$ & $0.0132$ & $\pm 0.0046$ & $\pm 0.0035$ & $_{-0.0016}^{+0.0023}$ & $0.97$ & $0.96$ \\ 
\hline
\hline
  $\m3j$ bin (GeV)
& $d\sigma/d\m3j$ (pb/GeV)
& $\delta_{\rm stat}$
& $\delta_{\rm syst}$
& $\delta_{\rm ES}$
& $C_{\rm QED}$
& $C_{\rm had}$\\
\hline
\multicolumn{7}{||c||}{unpolarized - three jets} \\
\hline
$ 20,  30$ & $0.0069$ & $\pm 0.0024$ & $\pm 0.0017$ & $\pm 0.0001$ & $0.99$ & $0.83$ \\ 
$ 30,  45$ & $0.0253$ & $\pm 0.0045$ & $\pm 0.0009$ & $_{-0.0008}^{+0.0006}$ & $0.99$ & $0.76$ \\ 
$ 45,  65$ & $0.0307$ & $\pm 0.0049$ & $\pm 0.0012$ & $_{-0.0011}^{+0.0012}$ & $0.97$ & $0.77$ \\ 
$ 65,  90$ & $0.0150$ & $\pm 0.0034$ & $\pm 0.0054$ & $_{-0.0012}^{+0.0014}$ & $0.94$ & $0.79$ \\ 
$ 90, 120$ & $0.0036$ & $\pm 0.0022$ & $\pm 0.0001$ & $_{-0.0003}^{+0.0004}$ & $0.94$ & $0.85$ \\ 
\hline
    \end{tabular}
 \caption{
   Differential unpolarized dijet and three-jet cross-sections
   $d\sigma/d\mj$ and $d\sigma/d\m3j$ in $e^-p$ collisions. Other
   details as in the caption to Table~\ref{tabseven}.}
 \label{tabfifteen}
\end{center}
\end{table}

\clearpage
\newpage
\begin{table}
\begin{center}
    \begin{tabular}{||c|cccc||c||c||}
\hline
  $\mj$ bin (GeV)
& $d\sigma/d\mj$ (pb/GeV)
& $\delta_{\rm stat}$
& $\delta_{\rm syst}$
& $\delta_{\rm ES}$
& $C_{\rm QED}$
& $C_{\rm had}$\\
\hline
\multicolumn{7}{||c||}{unpolarized - dijets} \\
\hline
$ 10,  15$ & $0.069$ & $\pm 0.011$ & $\pm 0.007$ & $\pm 0.001$ & $0.96$ & $0.90$ \\ 
$ 15,  20$ & $0.146$ & $\pm 0.017$ & $\pm 0.010$ & $_{-0.002}^{+0.001}$ & $0.96$ & $0.90$ \\ 
$ 20,  30$ & $0.171$ & $\pm 0.015$ & $\pm 0.022$ & $\pm 0.003$ & $0.95$ & $0.91$ \\ 
$ 30,  45$ & $0.108$ & $\pm 0.010$ & $\pm 0.013$ & $\pm 0.003$ & $0.95$ & $0.93$ \\ 
$ 45,  65$ & $0.0522$ & $\pm 0.0077$ & $\pm 0.0103$ & $_{-0.0025}^{+0.0023}$ & $0.94$ & $0.95$ \\ 
$ 65,  90$ & $0.0169$ & $\pm 0.0052$ & $\pm 0.0026$ & $_{-0.0011}^{+0.0015}$ & $0.94$ & $0.97$ \\ 
$ 90, 120$ & $0.0020$ & $\pm 0.0020$ & $\pm 0.0002$ & $_{-0.0002}^{+0.0003}$ & $0.92$ & $0.99$ \\ 
\hline
\hline
  $\m3j$ bin (GeV)
& $d\sigma/d\m3j$ (pb/GeV)
& $\delta_{\rm stat}$
& $\delta_{\rm syst}$
& $\delta_{\rm ES}$
& $C_{\rm QED}$
& $C_{\rm had}$\\
\hline
\multicolumn{7}{||c||}{unpolarized - three jets} \\
\hline
$ 20,  30$ & $0.0051$ & $\pm 0.0024$ & $\pm 0.0013$ & $\pm 0.0001$ & $0.91$ & $0.84$ \\ 
$ 30,  45$ & $0.0124$ & $\pm 0.0037$ & $\pm 0.0006$ & $_{-0.0003}^{+0.0005}$ & $0.95$ & $0.75$ \\ 
$ 45,  65$ & $0.0076$ & $\pm 0.0029$ & $\pm 0.0007$ & $\pm 0.0004$ & $0.94$ & $0.78$ \\ 
$ 65,  90$ & $0.0049$ & $\pm 0.0024$ & $\pm 0.0018$ & $_{-0.0004}^{+0.0005}$ & $0.90$ & $0.81$ \\ 
\hline
    \end{tabular}
 \caption{
   Differential unpolarized dijet and three-jet cross-sections
   $d\sigma/d\mj$ and $d\sigma/d\m3j$ in $e^+p$ collisions. Other
   details as in the caption to Table~\ref{tabseven}.}
 \label{tabsixteen}
\end{center}
\end{table}

\newpage
\clearpage

\newpage
\clearpage

\newpage
\clearpage

\newpage
\clearpage

\newpage
\clearpage

\newpage
\clearpage

\newpage
\clearpage

\newpage
\clearpage

\newpage
\clearpage

\newpage
\clearpage

\newpage
\clearpage

\newpage
\clearpage

\newpage
\clearpage

\newpage
\clearpage

\newpage
\clearpage

\end{document}